\documentstyle[12pt,dina4p]{article}

\title{
\bf
       Construction
       of Nonlinear
       Symplectic      \\
       Six-Dimensional
       Thin\,-\,Lens
       Maps               \\
       by Exponentiation
                       }
\author{K. Heinemann, G. Ripken,
                   F. Schmidt\thanks{CERN, SL-Division, Geneva,
Switzerland}}
\begin{document}
\maketitle

\def\gga{(1+a\gamma)}
\def\Kvec{{\vec{F}}^{self}}
\def\K{F^{self}}
\def\beq{\begin{equation}}
\def\eeq{\end{equation}}
%
\newcounter{subfig}
\newcommand{\firstsubfig}%
{\renewcommand{\thefigure}{\arabic{figure}\alph{subfig}}%
\setcounter{subfig}{1}}
\newcommand{\nextsubfig}%
{\addtocounter{figure}{-1}\addtocounter{subfig}{1}}
\newcounter{INDEX}
\renewcommand
{\theequation}{\mbox{\thesection .\arabic{equation}\alph{INDEX}}}

\begin{abstract}
The aim of this paper is to construct
six\,-\,dimensional symplectic
thin--lens transport maps
for the tracking program
SIXTRACK
\cite{SIXTR},
continuing an earlier report
\cite{RS}
by using another method
which consistes in applying
Lie series and exponentiation
as described by
 W. Gr\"obner
\cite{Groeb}
and for canonical systems by
 A.J. Dragt
\cite{DRAGT}.
As in Ref.
\cite{RS}
we firstly use an approximate
Hamiltonian obtained by a series expansion
of the square root\,
\begin{eqnarray*}
                           \left\{1-
          \frac{
                [p_{x}+H\cdot z]^{2}+
                [p_{z}-H\cdot x]^{2}
                                                   }
               {
                \left[1+
                   f(p_{\sigma})\right]^{2}
                                }
                                   \right\}^{1/2}
\end{eqnarray*}
up to first order in terms of the quantity
\begin{eqnarray*}
          \frac{
                [p_{x}+H\cdot z]^{2}+
                [p_{z}-H\cdot x]^{2}
                                      }
               {
                \left[1+
                   f(p_{\sigma})\right]^{2}
                                }\ .
\end{eqnarray*}
Furthermore, nonlinear crossing terms
due to the curvature in bending magnets
are neglected.
An improved Hamiltonian,
excluding solenoids,
is introduced in Appendix A
by using the unexpanded square root
mentioned above,
but neglecting again nonlinear crossing terms
in bending magnets.
It is shown
that the thin\,-\,lens maps remain unchanged
and that
the corrections
due to the new Hamiltonian are fully absorbed
into the drift spaces.
Finally a symplectic treatment
of the crossing terms
appearing in bending magnets
is presented in Appendix B,
taking into account only the lowest order.
The equations derived are valid for arbitrary particle velocity,
i.e. below and above transition energy
and shall be incorporated into
the tracking code SIXTRACK
\cite{SIXTR}.
\end{abstract}
\newpage

\tableofcontents
\newpage

\section{Introduction}

\ \ \ \ \
Continuing an earlier report
\cite{RS},
in this paper
we show
how to solve
the nonlinear
canonical equations of motion
in the framework of the fully
six--dimensional formalism
for various kinds of magnets
(bending magnets, quadrupoles, synchrotron magnets,
skew quadrupoles, sextupoles, octupoles, solenoids),
using an approach different from Ref.
\cite{RS}, namely
by applying Lie series and exponentiation
to a kick approximation
\cite{Groeb,DRAGT}
\footnote{There is a vast literature on solving
differential equations
by Lie series.
A nice treatment is given by Ref.
\cite{Groeb}.}.

In addition to Ref.
\cite{RS}
we also study the thin\,-\,lens formalism
for an improved Hamiltonian
which is exact outside solenoids and bending magnets.
It is shown
that the thin\,-\,lens maps obtained earlier
remain unchanged
and that
the corrections
due to the new Hamiltonian only appear
in the drift spaces.

The equations derived are valid for arbitrary particle velocity,
i.e. below and above transition energy
and shall be incorporated into
the tracking code SIXTRACK
\cite{SIXTR}.

The paper is organized as follows\,:

In chapter 2 the general canonical equations of motion are
derived.
The thin\,-\,lens method using Lie series and exponentiation
is described in chapter 3.
Using the thin--lens approximation the equations of motion
are solved for each element in chapter
4.
The improved Hamiltonian is introduced in Appendix A.
In addition to Ref.
\cite{RS}
              the influence of
nonlinear $''$crossing terms$''$
resulting from the curvature in bending magnets
is investigated
in Appendix B
and
a superposition
of a solenoid and a quadrupole
in Appendix C.
Finally
a summary of the results is presented in chapter 5.
\bigskip

\section{The Canonical Equations
            of Motion
                           }

\subsection{Notation}

\ \ \ \ \
The formalism and notation in this paper will be identical to that
used in Ref.
\cite{RS}.
Thus we will begin by simply stating the canonical equations
of motion already used in this earlier
paper and refer the reader to the latter for details.

\subsection{The Hamiltonian in Machine Coordinates}

\ \ \ \ \
The Hamiltonian for orbital motion in storage rings reads as
\cite{RS}\,:
\setcounter{INDEX}{1}
\begin{eqnarray}
    {\cal{H}}(x,p_{x},z,p_{z},\sigma,p_{\sigma};s)
               &=&
            p_{\sigma}
               -\left[1+
                   f(p_{\sigma})\right]
                 \cdot
                        [1+K_{x}\cdot x+K_{z}\cdot z]\times
                                    \nonumber  \\
 & &
\hspace*{1.5cm}
                           \left\{1-
          \frac{
                [p_{x}+H\cdot z]^{2}+
                [p_{z}-H\cdot x]^{2}
                                                   }
               {
                \left[1+
                   f(p_{\sigma})\right]^{2}
                                }
                                   \right\}^{1/2}
                                    \nonumber  \\
 & &
                  +\frac{1}{2}\cdot
                              [1+K_{x}\cdot{x}+K_{z}\cdot{z}]^{2}
                  -\frac{1}{2}
                              \cdot
                              g\cdot{(z^{2}-x^{2})}
                  -N
                    \cdot{xz}
\nonumber   \\
                            & &+
    \frac{\lambda}{6}\cdot
    (x^{3}-3\,xz^{2})
\nonumber   \\
       & &+
    \frac{\mu}{24}\cdot
    (z^{4}-6\,x^{2}\,z^{2}+x^{4})
\nonumber   \\
                            & &+
                \frac{1}{\beta_{0}^{2}}\cdot
                        \frac{L}{2\pi\cdot h}\cdot
                        \frac{eV(s)}{E_{0}}\cdot\cos
                       \left[
             h\cdot\frac{2\pi}{L}\cdot\sigma+\varphi
                       \right]
\end{eqnarray}
with
\begin{eqnarray}
   \addtocounter{equation}{-1}
\addtocounter{INDEX}{1}
            f(p_{\sigma})
                       &=&
                        \frac{1}{\beta_{0}}\,
           \sqrt{
                             \left(1+
                                 \beta_{0}^{2}\cdot p_{\sigma}
                                    \right)^{2}
                                         -
             \left(\frac{m_{0} c^{2}}{E_{0}}\right)^{2}
                           }\,
               -1
\end{eqnarray}
\setcounter{INDEX}{0}
($g,\, N,\, H,\, K_{x},\, K_{z},\,
\lambda$,\, and\,
$\mu$\,
are defined in Ref.
\cite{RS}\,).
\bigskip

Since
\begin{eqnarray*}
        |p_{x}+H\cdot z|
                   &\ll&{1} \ ;      \\
        |p_{z}-H\cdot x|
                   &\ll&{1}
\end{eqnarray*}
the square root
\begin{eqnarray*}
                           \left[1-
          \frac{
                [p_{x}+H\cdot z]^{2}+
                [p_{z}-H\cdot x]^{2}
                                                   }
               {
                \left[1+
                   f(p_{\sigma})\right]^{2}
                                }
                                   \right]^{1/2}
\end{eqnarray*}
in (2.1)
may be expanded in a series\,:
\\
\begin{eqnarray}
& &
                           \left[1-
          \frac{
                [p_{x}+H\cdot z]^{2}+
                [p_{z}-H\cdot x]^{2}
                                                   }
               {
                \left[1+
                   f(p_{\sigma})\right]^{2}
                                }
                                   \right]^{1/2}
                                                =
\nonumber   \\
\nonumber   \\
& &  \hspace*{2.5cm}
     1-
       \frac{1}{2}\cdot
          \frac{
                [p_{x}+H\cdot z]^{2}+
                [p_{z}-H\cdot x]^{2}
                                                   }
               {
                \left[1+
                   f(p_{\sigma})\right]^{2}
                                }
        +
      \cdot
      \cdot
      \cdot\ .
\end{eqnarray}
The power at which the series is truncated defines the order
of the approximation to the particle motion.
\bigskip

In the following we will use
(as in Ref.
\cite{RS})
the approximation\,:
\begin{eqnarray}
     {\cal{H}}
              &=&
          \frac{1}{2}\cdot
          \frac{
                [p_{x}+H\cdot z]^{2}+
                [p_{z}-H\cdot x]^{2}
                                                   }
               {
                \left[1+
                   f(p_{\sigma})\right]
                                }
                        +
                                    \nonumber  \\
                                    \nonumber  \\
 & &
            p_{\sigma}
           -[1+K_{x}\cdot x+K_{z}\cdot z]
                 \cdot
            f(p_{\sigma})
                        +
                                    \nonumber  \\
                                    \nonumber  \\
 & &
                   \frac{1}{2}\,
                              [K_{x}^{2}+g]
                              \cdot
                              x^{2}
                  +\frac{1}{2}\,
                              [K_{z}^{2}-g]
                              \cdot
                              z^{2}
                  -N
                    \cdot{x\,z}
                        +
\nonumber   \\
\nonumber   \\
                            & &
    \frac{\lambda}{6}\cdot
    (x^{3}-3\,xz^{2})
          +
    \frac{\mu}{24}\cdot
    (z^{4}-6\,x^{2}\,z^{2}+x^{4})
                        +
\nonumber   \\
\nonumber   \\
                            & &
                \frac{1}{\beta_{0}^{2}}\cdot
                        \frac{L}{2\pi\cdot h}\cdot
                        \frac{eV(s)}{E_{0}}\cdot\cos
                       \left[
             h\cdot\frac{2\pi}{L}\cdot\sigma+\varphi
                       \right]\ .
\end{eqnarray}
An improved Hamiltonian is introduced in
Appendix A.
\bigskip

The
canonical equations
corresponding to the Hamiltonian (2.3)
take the form\,:
\begin{eqnarray}
    \frac{d}{ds}\
    \vec{y}
                   &=&
   -\underline{S}\cdot
                \frac{
                      \partial{{\cal{H}}}
                                                  }{\partial
                          \vec{y}
                                 }
\end{eqnarray}
with
\begin{eqnarray*}
       \vec{y}^{\,T}
         &=&
                   \left(
                       y_{1},\,
                       y_{2},\,
                       y_{3},\,
                       y_{4},\,
                       y_{5},\,
                       y_{6}
                       \right)
\\
         &\equiv&
                   \left(
                       x,\,
                       p_{x},\,
                       z,\,
                       p_{z},\,
                       \sigma,\,
                       p_{\sigma}
                       \right)
\end{eqnarray*}
and
\begin{eqnarray}
   \underline{S}=\left( \begin{array}{rrr}
     \underline{S}_{\,2} & \underline{0} & \underline{0}      \\
     \underline{0}     & \underline{S}_{\,2} & \underline{0}  \\
     \underline{0}     & \underline{0}     & \underline{S}_{\,2}
              \end{array}
       \right)\ ; \ \ \
   \underline{S}_{\,2}=\left( \begin{array}{rr}
     0                 & -1         \\
    +1                 &  0
              \end{array}
       \right)
\end{eqnarray}
or, written in components\,:
\setcounter{INDEX}{1}
\begin{eqnarray}
    \frac{d}{ds}\,x
                    &=&
               +\,
                \frac{\partial{
                              {\cal{H}}
                                       }
                                       }
                                        {\partial
                          {p_{x}}}
\nonumber    \\
\nonumber    \\
                    &=&
      \frac{
           p_{x}+H\cdot z}
               {
                \left[1+
                   f(p_{\sigma})\right]
                                }\ ;
 \\       \nonumber
 \\
   \addtocounter{equation}{-1}
\addtocounter{INDEX}{1}
    \frac{d}{ds}
          \,p_{x}
                        &=&
               -\frac{\partial{{\cal{H}}}}{\partial
                          {x}}
\nonumber    \\
\nonumber    \\
                    &=&
                 +\frac{
                  [p_{z}-H\cdot x]
                                 }
               {\left[1+
                   f(p_{\sigma})\right]
                                }
                                  \cdot H
                   -
                [K_{x}^{2}+g]\cdot x
                   +
                          N\cdot z
                   +
                          K_{x}\cdot f(p_{\sigma})
\nonumber    \\
                        & &
\hspace*{0.5cm}
                       -
    \frac{\lambda}{2}\cdot
    (x^{2}-z^{2})
                       -
    \frac{\mu}{6}\cdot
    (x^{3}-3\,x\,z^{2})\ ;
 \\       \nonumber
 \\
   \addtocounter{equation}{-1}
\addtocounter{INDEX}{1}
    \frac{d}{ds}\,z
                   &=&
               +\frac{\partial{
                              {\cal{H}}
                                       }
                                       }
                                        {\partial
                          {p_{z}}}
\nonumber    \\
\nonumber    \\
                    &=&
        \frac{
           p_{z}-H\cdot x}
               {
                \left[1+
                   f(p_{\sigma})\right]
                                }\ ;
 \\       \nonumber
 \\
   \addtocounter{equation}{-1}
\addtocounter{INDEX}{1}
    \frac{d}{ds}
          \,p_{z}
                       &=&
               -\frac{\partial{{\cal{H}}}}{\partial
                          {z}}
\nonumber    \\
\nonumber    \\
                    &=&
                 -\frac{[p_{x}+H\cdot z]
                                 }
               {
                \left[1+
                   f(p_{\sigma})\right]
                                }
                                        \cdot H
                   -
                [K_{z}^{2}-g]\cdot z
                   +
                          N\cdot x
                   +
                          K_{z}\cdot f(p_{\sigma})
\nonumber    \\
                       & &
\hspace*{0.5cm}
                       +
    \lambda\cdot x z
                       -
    \frac{\mu}{6}\cdot
    (z^{3}-3\,x^{2}\,z)\ ;
 \\       \nonumber
 \\
   \addtocounter{equation}{-1}
\addtocounter{INDEX}{1}
    \frac{d}{ds}\,\sigma
                        &=&
               +\frac{\partial{
                              {\cal{H}}
                                       }
                                       }
                                        {\partial
                          {p_{\sigma}}}
\nonumber    \\
\nonumber    \\
                    &=&
              1-[1+K_{x}\cdot x+
                 K_{z}\cdot z]
                \cdot f'(p_{\sigma})
                                                      \nonumber\\
                 & &
\hspace*{0.5cm}
         -\frac{1}{2}\cdot
          \frac{
                [p_{x}+H\cdot z]^{2}+
                [p_{z}-H\cdot x]^{2}
                                                   }
               {
                \left[1+
                   f(p_{\sigma})\right]^{2}
                                }
               \cdot f'(p_{\sigma})
\nonumber    \\
\nonumber    \\
                    &=&
              1-[1+K_{x}\cdot x+
                 K_{z}\cdot z]
                \cdot f'(p_{\sigma})
                                                      \nonumber\\
                 & &
\hspace*{0.5cm}
         -\frac{1}{2}\cdot
                [(x')^{2}+(z')^{2}]
                \cdot f'(p_{\sigma})\ ;
 \\       \nonumber
 \\
   \addtocounter{equation}{-1}
\addtocounter{INDEX}{1}
    \frac{d}{ds}\,p_{\sigma}
                      &=&
               -\frac{\partial{{\cal{H}}}}{\partial
                          {\sigma}}
\nonumber    \\
\nonumber    \\
                    &=&
              \frac{1}{\beta_{0}^{2}}
                         \cdot\frac{eV(s)}{E_{0}}\cdot
                   \sin\left[
                        h\cdot\frac{2\pi}{L}\cdot\sigma
                              +
                             \varphi
                                    \right]\ .
\end{eqnarray}

\setcounter{INDEX}{0}
In detail, one has\,:
\bigskip
\begin{eqnarray*}
\begin{array}{llll}
         \mbox{a})&   K_{x}^{2}+K_{z}^{2}\neq{0};& \ g=N
                            =\lambda=\mu=H=V
                                    =0:
                                    & \mbox{bending magnet} ;\\
         \mbox{b})&   g\neq{0};&
                         \ K_{x}=K_{z}=N
                            =\lambda=\mu=H=V
                                            =0:&
                                              \mbox{quadrupole};\\
         \mbox{c})&   N\neq{0};& \ K_{x}=K_{z}=g
                            =\lambda=\mu=H=V
                        =0:&
                              \mbox{skew quadrupole} ;\\
         \mbox{d})&   \lambda\neq{0};
                                                 & \
                       K_{x}=K_{z}=g=
                               N=\mu=H=V
                                    =0:
                      & \mbox{sextupole} ;\\
         \mbox{e})&   \mu\neq{0};
                                                 & \
                       K_{x}=K_{z}=g=
                               N=\lambda=H=V
                                    =0:
                      & \mbox{octupole} ;\\
         \mbox{f})&   H\neq{0};& \ K_{x}=K_{z}=g=N
                            =\lambda=\mu=V
                                             =0:
                                            & \mbox{solenoid} ;\\
         \mbox{g})&   V\neq{0};& \ K_{x}=K_{z}=g=
                               N
                            =\lambda=\mu=H
                                  =0:
                                    & \mbox{cavity} .\\
\end{array}
\end{eqnarray*}
\bigskip
\setcounter{equation}{0}

\section{Description of the
            Thin\,-\,Lens Method}
\subsection{Thin\,-\,Lens Approximation}

\ \ \ \ \
The equations of motion (2.6)
have the general form\,:
\setcounter{INDEX}{1}
\begin{eqnarray}
    \frac{d\ }{ds}\,
       y_{i}
         &=&
    \vartheta_{i}\left(
                       y_{1},\,
                  y_{2},\,
                  y_{3},\,
                  y_{4},\,
                  y_{5},\,
                  y_{6};\,
                  s
                       \right)\ ;
\end{eqnarray}
\ \ \ \ \ \ \ \ \ \ \ \ \ \ \ \ \
\ \ \ \ \ \ \ \ \ \ \ \ \ \ \ \ \
\ \ \ \ \ \ \ \ \ \ \ \ \ \ \ \ \
   $(i= 1, 2, 3, 4, 5, 6)$
\newline
or
\begin{eqnarray}
   \addtocounter{equation}{-1}
\addtocounter{INDEX}{1}
    \frac{d\ }{ds}\,
       \vec{y}
         &=&
    \vec{\vartheta}\left(
                         \vec{y};\,s
                       \right)
\end{eqnarray}
with
\setcounter{INDEX}{0}
\begin{eqnarray*}
    \vec{\vartheta}^{\,T}
         &=&
                   \left(
                         \vartheta_{1},\,
                         \vartheta_{2},\,
                         \vartheta_{3},\,
                         \vartheta_{4},\,
                         \vartheta_{5},\,
                         \vartheta_{6}
                       \right)
\end{eqnarray*}
and
\setcounter{INDEX}{1}
\begin{eqnarray}
    \vartheta_{1}
         (\vec{y}; s)
                    &=&
                 +\frac{p_{x}
                                 }
               {\left[1+
                   f(p_{\sigma})\right]
                                }
                 +
                 \frac{
                  H(s)
                   \cdot z
                                 }
               {\left[1+
                   f(p_{\sigma})\right]
                                }\ ;
 \\       \nonumber
 \\
   \addtocounter{equation}{-1}
\addtocounter{INDEX}{1}
    \vartheta_{2}
         (\vec{y}; s)
                        &=&
                 +\frac{p_{z}}
               {\left[1+
                   f(p_{\sigma})\right]
                                }
                   \cdot
                         H(s)
                 -\frac{
                  H(s)
                   \cdot x
                                 }
               {\left[1+
                   f(p_{\sigma})\right]
                                }
                                  \cdot H(s)
                   +\
                          K_{x}(s)\cdot f(p_{\sigma})
\nonumber    \\
                        & &
\hspace*{0.5cm}
                  -
               [
                 K_{x}^{2}(s)
                        +g(s)
                                           ]\cdot x
                  +
               N(s)\cdot z
\nonumber    \\
                        & &
\hspace*{0.5cm}
                       -
    \frac{\lambda(s)}{2}\cdot
    (x^{2}-z^{2})
                       -
    \frac{\mu(s)}{6}\cdot
    (x^{3}-3\,x\,z^{2})\ ;
 \\       \nonumber
 \\
   \addtocounter{equation}{-1}
\addtocounter{INDEX}{1}
    \vartheta_{3}
         (\vec{y}; s)
                   &=&
       +\frac{p_{z}
                            }
               {
                \left[1+
                   f(p_{\sigma})\right]
                                }
                 -
        \frac{
                 H(s)\cdot x}
               {
                \left[1+
                   f(p_{\sigma})\right]
                                }\ ;
 \\       \nonumber
 \\
   \addtocounter{equation}{-1}
\addtocounter{INDEX}{1}
    \vartheta_{4}
         (\vec{y}; s)
                       &=&
                 -\frac{p_{x}}
               {\left[1+
                   f(p_{\sigma})\right]
                                }
                   \cdot H(s)
                 -\frac{H(s)\cdot z
                                 }
               {
                \left[1+
                   f(p_{\sigma})\right]
                                }
                                        \cdot H(s)
                   +
                          K_{z}(s)\cdot f(p_{\sigma})
\nonumber    \\
                       & &
\hspace*{0.5cm}
                   -
                [K_{z}(s)^{2}-g(s_{0})]\cdot z
                   +
                          N(s)\cdot x
\nonumber    \\
                       & &
\hspace*{0.5cm}
                       +
    \lambda(s)\cdot x z
                       -
    \frac{\mu(s)}{6}\cdot
    (z^{3}-3\,x^{2}\,z)\ ;
 \\       \nonumber
 \\
   \addtocounter{equation}{-1}
\addtocounter{INDEX}{1}
    \vartheta_{5}
         (\vec{y}; s)
                        &=&
                         1
               -[K_{x}(s)\cdot x+
                 K_{z}(s)\cdot z]
                \cdot f'(p_{\sigma})
                         -
          \frac{1}{2}\cdot
          \frac{p_{x}^{2}+p_{z}^{2}
                                      }
               {
                \left[1+
                   f(p_{\sigma})\right]^{2}
                                }
               \cdot f'(p_{\sigma})
                                                      \nonumber\\
                 & &
\hspace*{0.5cm}
         -\frac{1}{2}\cdot
          \frac{
                H(s)^{2}\cdot[x^{2}+
                            z^{2}]
                                                   }
               {
                \left[1+
                   f(p_{\sigma})\right]^{2}
                                }
               \cdot f'(p_{\sigma})
         -
          \frac{
                     H(s)
               \cdot
              [p_{x}\cdot z-p_{z}\cdot x]
                                                   }
               {
                \left[1+
                   f(p_{\sigma})\right]^{2}
                                }
               \cdot f'(p_{\sigma})\ ;
 \\       \nonumber
 \\
   \addtocounter{equation}{-1}
\addtocounter{INDEX}{1}
    \vartheta_{6}
         (\vec{y}; s)
                      &=&
              \frac{1}{\beta_{0}^{2}}
                         \cdot\frac{eV(s)}{E_{0}}\cdot
                   \sin\left[
                        h\cdot\frac{2\pi}{L}\cdot\sigma
                              +
                             \varphi
                                    \right]\ .
\end{eqnarray}
\setcounter{INDEX}{0}

Equation (3.1) represents a system of
differential equations
the solutions of which
can be written in the form\,:
\begin{eqnarray}
    \vec{y}(s_{f})
       &=&
    \underline{T}(s_{f}, s_{i})\,
    \vec{y}(s_{i})
\end{eqnarray}
by defining a transport operator\,
   $\underline{T}(s_{f}, s_{i})$\,
connecting
the final vector\,
   $\vec{y}(s_{f})$\,
at position\,
   $s_{f}$\,
with the initial vector
   $\vec{y}(s_{i})$\,
at position\,
   $s_{i}$.

The aim of this chapter is now to calculate
the transport map\,
   $\underline{T}(s_{f}, s_{i})$\,
(approximately)
by using symplectic kicks.
\vspace*{0.5ex}

We achieve that in two steps\,:
\vspace*{0.5ex}

In a first step
we decompose the r.h.s. of (3.1)
into two components\,:
\begin{eqnarray}
    \vec{\vartheta}
       &=&
    \vec{\vartheta}_{D}
        +
    \vec{\vartheta}_{L}
\end{eqnarray}
gathering in\,
       $\vec{\vartheta}_{L}$\,
all terms of\,
       $\vec{\vartheta}$\,
containing the external electric and magnetic fields
(expressed by the lens functions
      $V$,\,
      $g$,\,
      $N$,\,
      $H$,\,
      $K_{x}$,\,
      $K_{z}$,\,
      $\lambda$,\,
      $\mu$
      ).
\bigskip

As a result, the component\,
   $\vec{\vartheta}_{D}$\,
in (3.4) then corresponds to
the Hamiltonian
\begin{eqnarray}
     {\cal{H}}_{D}
              &=&
          \frac{1}{2}\cdot
          \frac{
                 p_{x}^{2}+
                 p_{z}^{2}
                                                   }
               {
                \left[1+
                   f(p_{\sigma})\right]
                                }
                        +
            p_{\sigma}
           -
            f(p_{\sigma})
\end{eqnarray}
(to be obtained from (2.3) by neglecting
all external fields)
leading to the (canonical) equations of motion
for a pure drift space\,:
\setcounter{INDEX}{1}
\begin{eqnarray}
    \frac{d}{ds}\,x
                    &=&
               +\frac{\partial
                                      }{\partial
                          {p_{x}}}\,
      {\cal{H}}_{D}(x,p_{x},z,p_{z},\sigma,p_{\sigma})
\nonumber    \\
                    &=&
      \frac{
           p_{x}}
               {
                \left[1+
                   f(p_{\sigma})\right]
                                }\ ;
 \\       \nonumber
 \\
   \addtocounter{equation}{-1}
\addtocounter{INDEX}{1}
    \frac{d}{ds}
          \,p_{x}
                        &=&
               -\frac{\partial
                                      }{\partial
                         {x}}\,
      {\cal{H}}_{D}(x,p_{x},z,p_{z},\sigma,p_{\sigma})
\nonumber    \\
                    &=&
       0
                      \ \ \
         \Longrightarrow\ \ \
         p_{x}\ =\ const\ ;
 \\       \nonumber
 \\
   \addtocounter{equation}{-1}
\addtocounter{INDEX}{1}
    \frac{d}{ds}\,z
                   &=&
               +\frac{\partial
                                      }{\partial
                          {p_{z}}}\,
      {\cal{H}}_{D}(x,p_{x},z,p_{z},\sigma,p_{\sigma})
\nonumber    \\
                    &=&
        \frac{
           p_{z}}
               {
                \left[1+
                   f(p_{\sigma})\right]
                                }\ ;
 \\       \nonumber
 \\
   \addtocounter{equation}{-1}
\addtocounter{INDEX}{1}
    \frac{d}{ds}
          \,p_{z}
                       &=&
               -\frac{\partial
                                      }{\partial
                          {z}}\,
      {\cal{H}}_{D}(x,p_{x},z,p_{z},\sigma,p_{\sigma})
\nonumber    \\
                    &=&
                        0
                      \ \ \
         \Longrightarrow\ \ \
         p_{z}\ =\ const\ ;
 \\       \nonumber
 \\
   \addtocounter{equation}{-1}
\addtocounter{INDEX}{1}
    \frac{d}{ds}\,\sigma
                        &=&
               +\frac{\partial
                                      }{\partial
                          {p_{\sigma}}}
      {\cal{H}}_{D}(x,p_{x},z,p_{z},\sigma,p_{\sigma})
\nonumber    \\
                    &=&
              1-
                      f'(p_{\sigma})
         -\frac{1}{2}\cdot
                [\,(p_{x})^{2}+(p_{z})^{2}\,]
                \cdot
      \frac{
                      f'(p_{\sigma})
                                  }
               {
                \left[1+
                   f(p_{\sigma})\right]^{2}
                                }\ ;
 \\       \nonumber
 \\
   \addtocounter{equation}{-1}
\addtocounter{INDEX}{1}
    \frac{d}{ds}\,p_{\sigma}
                      &=&
               -\frac{\partial
                                      }{\partial
                          {\sigma}}\,
      {\cal{H}}_{D}(x,p_{x},z,p_{z},\sigma,p_{\sigma})
\nonumber    \\
                    &=&
                       0
                      \ \ \
         \Longrightarrow\ \ \
         p_{\sigma}\ =\ const
\end{eqnarray}
\setcounter{INDEX}{0}
(see also eqn. (2.6)\,).
The solutions
for a drift of length
$l$ are\,:

\setcounter{INDEX}{1}
\begin{eqnarray}
                  x^{f}
                    &=&
                  x^{i}
                     +
      \frac{
           p_{x}^{\,i}}
               {
                \left[1+
                   f(p_{\sigma}^{\,i})\right]
                                }
          \cdot l\ ;
 \\       \nonumber
 \\
   \addtocounter{equation}{-1}
\addtocounter{INDEX}{1}
            p_{x}^{f}
                        &=&
            p_{x}^{\,i}\ ;
 \\       \nonumber
 \\
   \addtocounter{equation}{-1}
\addtocounter{INDEX}{1}
                  z^{f}
                   &=&
                  z^{i}
                    +
        \frac{
           p_{z}^{\,i}}
               {
                \left[1+
                   f(p_{\sigma}^{\,i})\right]
                                }
               \cdot l\ ;
 \\       \nonumber
 \\
   \addtocounter{equation}{-1}
\addtocounter{INDEX}{1}
            p_{z}^{f}
                       &=&
            p_{z}^{\,i}\ ;
 \\       \nonumber
 \\
   \addtocounter{equation}{-1}
\addtocounter{INDEX}{1}
    \sigma^{f}
                        &=&
    \sigma^{i}
              +
    \left\{
              1-
                      f'(p_{\sigma}^{\,i})
         -\frac{1}{2}\cdot
                [
          (p_{x}^{\,i})^{2}
                +
          (p_{z}^{\,i})^{2}
                                  ]
                \cdot
      \frac{
                      f'(p_{\sigma}^{\,i})
                                  }
               {
                \left[1+
                   f(p_{\sigma}^{\,i})\right]^{2}
                                }
                         \right\}
          \cdot l\ ;
 \\       \nonumber
 \\
   \addtocounter{equation}{-1}
\addtocounter{INDEX}{1}
                 p_{\sigma}^{f}
                      &=&
                 p_{\sigma}^{\,i}
\end{eqnarray}
\setcounter{INDEX}{0}

The second component
       $\vec{\vartheta}_{L}$\,
corresponds
to the Hamiltonian\,
\begin{eqnarray}
         {\cal{H}}_{L}
              &=&
          \frac{H}
               {
                \left[1+
                   f(p_{\sigma})\right]
                                }
          \cdot
          \left[p_{x}\cdot z-p_{z}\cdot x\right]
                  +
          \frac{1}{2}\cdot
          \frac{H^{2}}
               {
                \left[1+
                   f(p_{\sigma})\right]
                                }
          \cdot
          \left[x^{2}+z^{2}\right]
                        -
                                    \nonumber  \\
                                    \nonumber  \\
 & &
           -[K_{x}\cdot x+K_{z}\cdot z]
                 \cdot
            f(p_{\sigma})
                        +
                                    \nonumber  \\&&
                                    \nonumber  \\
 & &
                   \frac{1}{2}\,
                              [K_{x}^{2}+g]
                              \cdot
                              x^{2}
                  +\frac{1}{2}\,
                              [K_{z}^{2}-g]
                              \cdot
                              z^{2}
                  -N
                    \cdot{x\,z}
                        +
\nonumber   \\
\nonumber   \\
                            & &
    \frac{\lambda}{6}\cdot
    (x^{3}-3\,xz^{2})
          +
    \frac{\mu}{24}\cdot
    (z^{4}-6\,x^{2}\,z^{2}+x^{4})
                        +
\nonumber   \\
\nonumber   \\
                            & &
                \frac{1}{\beta_{0}^{2}}\cdot
                        \frac{L}{2\pi\cdot h}\cdot
                        \frac{eV(s)}{E_{0}}\cdot\cos
                       \left[
             h\cdot\frac{2\pi}{L}\cdot\sigma+\varphi
                       \right]
\end{eqnarray}
containing the remaining terms
in eqn. (2.3).
In particular there are no
       $p_{x}^{2}$
or
       $p_{z}^{2}$
terms.

Thus we have
\begin{eqnarray*}
         {\cal{H}}
              &=&
         {\cal{H}}_{L}
               +
         {\cal{H}}_{D}
\end{eqnarray*}
and
\begin{eqnarray*}
    \vec{\vartheta}_{D}
                   &=&
   -\underline{S}\cdot
                \frac{
                      \partial
                                                  }{\partial
                          \vec{y}
                                 }\,
                               {\cal{H}}_{D}\ ;
\\
    \vec{\vartheta}_{L}
                   &=&
   -\underline{S}\cdot
                \frac{
                      \partial
                                                  }{\partial
                          \vec{y}
                                 }\,
                               {\cal{H}}_{L}\ ,
\end{eqnarray*}
where the matrix $\underline{S}$ is given by eqn. (2.5).
\bigskip

In the second step
we replace the function
   $\vec{\vartheta}$\,
in (3.4)
for a thin lens of length
      $\Delta s$
at position
       $s_{0}$
by
\cite{RS}
\begin{eqnarray}
    \vec{\vartheta}_{mod}
                       (\vec{y}; s)
        &=&
    \vec{\vartheta}_{D}
                       (\vec{y})
         +
    \vec{\vartheta}_{L}
                       (\vec{y}; s)
        \cdot
   \Delta s
        \cdot
   \delta(s-s_{0})
\nonumber    \\
        &=&
    \vec{\vartheta}_{D}
                       (\vec{y})
         +
    \vec{\vartheta}_{L}
                       (\vec{y}; s_{0})
        \cdot
   \Delta s
        \cdot
   \delta(s-s_{0})
\end{eqnarray}
with
\begin{eqnarray}
    \vec{\vartheta}_{L}
              (\vec{y}; s_{0})
                   &=&
   -\underline{S}\cdot
                \frac{
                      \partial
                                                  }{\partial
                          \vec{y}
                                 }\,
                               \hat{\cal{H}}_{L}
\end{eqnarray}
and
\begin{eqnarray}
         \hat{\cal{H}}_{L}
                   &\equiv&
             {\cal{H}}_{L}
              (\vec{y}; s_{0})
\nonumber     \\
                   &=&
          \frac
               {H(s_{0})}
               {
                \left[1+
                   f(p_{\sigma})\right]
                                }
          \cdot
          \left[p_{x}\cdot z-p_{z}\cdot x\right]
                  +
          \frac{1}{2}\cdot
          \frac{
               H^{2}(s_{0})
                            }
               {
                \left[1+
                   f(p_{\sigma})\right]
                                }
          \cdot
          \left[x^{2}+z^{2}\right]
                        -
                                    \nonumber  \\
                                    \nonumber  \\
 & &
           [K_{x}(s_{0})\cdot x+K_{z}(s_{0})\cdot z]
                 \cdot
            f(p_{\sigma})
                        +
                                    \nonumber  \\&&
                                    \nonumber  \\
 & &
                   \frac{1}{2}\,
                              [K_{x}^{2}(s_{0})+g(s_{0})]
                              \cdot
                              x^{2}
                  +\frac{1}{2}\,
                              [K_{x}^{2}(s_{0})-g(s_{0})]
                              \cdot
                              z^{2}
                  -N(s_{0})
                    \cdot{x\,z}
                        +
\nonumber   \\
\nonumber   \\
                            & &
    \frac{\lambda(s_{0})}{6}\cdot
    (x^{3}-3\,xz^{2})
          +
    \frac{\mu(s_{0})}{24}\cdot
    (z^{4}-6\,x^{2}\,z^{2}+x^{4})
                        +
\nonumber   \\
\nonumber   \\
                            & &
                \frac{1}{\beta_{0}^{2}}\cdot
                        \frac{L}{2\pi\cdot h}\cdot
                        \frac{eV(s_{0})}{E_{0}}\cdot\cos
                       \left[
             h\cdot\frac{2\pi}{L}\cdot\sigma+\varphi
                       \right]\ ,
\end{eqnarray}
whereby the new function\,
   $\vec{\vartheta}_{mod}$\,
in (3.9) results from the modified Hamiltonian
\begin{eqnarray}
             {\cal{H}}_{mod}
               &=&
             {\cal{H}}_{D}
                +
             \hat{\cal{H}}_{L}
        \cdot
   \Delta s
        \cdot
   \delta(s-s_{0})\ .
\end{eqnarray}

In order to solve eqn. (3.1)
using
the modified function\,
   $\vec{\vartheta}_{mod}$\,
in (3.9),
we then have to decompose the region
\begin{eqnarray*}
       s_{0}-\frac{\Delta s}{2}\,
          \leq\,
            s\,
          \leq\,
       s_{0}+\frac{\Delta s}{2}
\end{eqnarray*}
of the lens
into three parts\,:
\setcounter{INDEX}{1}
\begin{eqnarray}
      {\rm region}\;\;\; I
                          &:&
       s_{0}-\frac{\Delta s}{2}\,
          \leq\,
            s\,
          \leq\,
       s_{0}-\epsilon\ ;
 \\
   \addtocounter{equation}{-1}
\addtocounter{INDEX}{1}
      {\rm region}\;\; II
                         &:&
       s_{0}-\epsilon\,
          \leq\,
            s\,
          \leq\,
       s_{0}+\epsilon\ ;
 \\
   \addtocounter{equation}{-1}
\addtocounter{INDEX}{1}
      {\rm region}\; III
                        &:&
       s_{0}+\epsilon
          \leq\,
            s\,
          \leq\,
       s_{0}+\frac{\Delta s}{2}\ ;
\end{eqnarray}
\setcounter{INDEX}{0}
\begin{eqnarray*}
     (0\,<\,\epsilon\,
        \rightarrow\,
             0)\ .
\end{eqnarray*}

For region I and III we obtain a drift space
of length\,
     ${\displaystyle
       l\,=\,\frac{\Delta s}{2}
                               }$\,,
described by the differential equation
\begin{eqnarray}
    \frac{d\ }{ds}\,
       \vec{y}(s)
         &=&
    \vec{\vartheta}_{D}
\end{eqnarray}
the solution of which is given by
eqn. (3.7)
and may be expressed by a transport operator\,
       $\underline{T}_{D}(l)$.
\bigskip

The equation of motion for the central region II
reads as\,:
\begin{eqnarray}
    \frac{d\ }{ds}\,
       \vec{y}(s)
         &=&
    \vec{\vartheta}_{L}(\vec{y}; s_{0})
        \cdot
   \Delta s
        \cdot
   \delta(s-s_{0})
\end{eqnarray}
with
\begin{eqnarray*}
   (\vec{\vartheta}_{L})^{\,T}
         &=&
                   \left(
                         \vartheta_{L1},\,
                         \vartheta_{L2},\,
                         \vartheta_{L3},\,
                         \vartheta_{L4},\,
                         \vartheta_{L5},\,
                         \vartheta_{L6}
                       \right)
\end{eqnarray*}
and
\setcounter{INDEX}{1}
\begin{eqnarray}
    \vartheta_{L1}
         (\vec{y}; s_{0})
                    &=&
               +\frac{\partial
                                      }{\partial
                          {p_{x}}}\,
  \hat{\cal{H}}_{L}(x,p_{x},z,p_{z},\sigma,p_{\sigma})
\nonumber   \\
                    &=&
     +\frac{
                 H(s_{0})
                  \cdot z}
               {
                \left[1+
                   f(p_{\sigma})\right]
                                }\ ;
 \\       \nonumber
 \\
   \addtocounter{equation}{-1}
\addtocounter{INDEX}{1}
    \vartheta_{L2}
         (\vec{y}; s_{0})
                    &=&
               -\frac{\partial
                                      }{\partial
                          {x}}\,
  \hat{\cal{H}}_{L}(x,p_{x},z,p_{z},\sigma,p_{\sigma})
\nonumber   \\
                        &=&
                 +\frac{p_{z}}
               {\left[1+
                   f(p_{\sigma})\right]
                                }
                   \cdot
                         H(s_{0})
                 -\frac{
                  H(s_{0})
                   \cdot x
                                 }
               {\left[1+
                   f(p_{\sigma})\right]
                                }
                                  \cdot H(s_{0})
\nonumber    \\
                        & &
\hspace*{0.5cm}
                   +
                          K_{x}(s_{0})\cdot f(p_{\sigma})
                  -
               [
                 K_{x}^{2}(s_{0})
                        +g(s_{0})
                                           ]\cdot x
                  +
               N(s_{0})\cdot z
\nonumber    \\
                        & &
\hspace*{0.5cm}
                       -
    \frac{\lambda(s_{0})}{2}\cdot
    (x^{2}-z^{2})
                       -
    \frac{\mu(s_{0})}{6}\cdot
    (x^{3}-3\,x\,z^{2})\ ;
 \\       \nonumber
 \\
   \addtocounter{equation}{-1}
\addtocounter{INDEX}{1}
    \vartheta_{L3}
         (\vec{y}; s_{0})
                    &=&
               +\frac{\partial
                                      }{\partial
                          {p_{z}}}\,
  \hat{\cal{H}}_{L}(x,p_{x},z,p_{z},\sigma,p_{\sigma})
\nonumber   \\
                   &=&
       -\frac{
                 H(s_{0})\cdot x}
               {
                \left[1+
                   f(p_{\sigma})\right]
                                }\ ;
 \\       \nonumber
 \\
   \addtocounter{equation}{-1}
\addtocounter{INDEX}{1}
    \vartheta_{L4}
         (\vec{y}; s_{0})
                    &=&
               -\frac{\partial
                                      }{\partial
                          {z}}\,
  \hat{\cal{H}}_{L}(x,p_{x},z,p_{z},\sigma,p_{\sigma})
\nonumber   \\
                       &=&
                 -\frac{p_{x}}
               {\left[1+
                   f(p_{\sigma})\right]
                                }
                   \cdot H(s_{0})
                 -\frac{H(s_{0})\cdot z
                                 }
               {
                \left[1+
                   f(p_{\sigma})\right]
                                }
                                        \cdot H(s_{0})
\nonumber    \\
                       & &
\hspace*{0.5cm}
                   +
                          K_{z}(s_{0})\cdot f(p_{\sigma})
                   -
                [K_{z}^{2}(s_{0})-g(s_{0})]\cdot z
                   +
                          N(s_{0})\cdot x
\nonumber    \\
                       & &
\hspace*{0.5cm}
                       +
    \lambda(s_{0})\cdot x z
                       -
    \frac{\mu(s_{0})}{6}\cdot
    (z^{3}-3\,x^{2}\,z)\ ;
 \\       \nonumber
 \\
   \addtocounter{equation}{-1}
\addtocounter{INDEX}{1}
    \vartheta_{L5}
         (\vec{y}; s_{0})
                    &=&
               +\frac{\partial
                                      }{\partial
                          {p_{\sigma}}}\,
  \hat{\cal{H}}_{L}(x,p_{x},z,p_{z},\sigma,p_{\sigma})
\nonumber   \\
                        &=&
               -[K_{x}(s_{0})\cdot x+
                 K_{z}(s_{0})\cdot z]
                \cdot f'(p_{\sigma})
                                                      \nonumber\\
                 & &
\hspace*{0.5cm}
         -\frac{1}{2}\cdot
          \frac{
                H(s_{0})^{2}\cdot[x^{2}+
                            z^{2}]
                                                   }
               {
                \left[1+
                   f(p_{\sigma})\right]^{2}
                                }
               \cdot f'(p_{\sigma})
         -
          \frac{
                     H(s_{0})
               \cdot
              [p_{x}\cdot z-p_{z}\cdot x]
                                                   }
               {
                \left[1+
                   f(p_{\sigma})\right]^{2}
                                }
               \cdot f'(p_{\sigma})\ ;
 \\       \nonumber
 \\
   \addtocounter{equation}{-1}
\addtocounter{INDEX}{1}
    \vartheta_{L6}
         (\vec{y}; s_{0})
                    &=&
               -\frac{\partial
                                      }{\partial
                          {\sigma}}\,
  \hat{\cal{H}}_{L}(x,p_{x},z,p_{z},\sigma,p_{\sigma})
\nonumber   \\
                      &=&
              \frac{1}{\beta_{0}^{2}}
                         \cdot\frac{eV(s_{0})}{E_{0}}\cdot
                   \sin\left[
                        h\cdot\frac{2\pi}{L}\cdot\sigma
                              +
                             \varphi
                                    \right]
\end{eqnarray}
\setcounter{INDEX}{0}
(see eqns. (3.10) and (3.11)\,)
and determines the transport map\,
\begin{eqnarray}
     \underline{T}_{L}
          &\equiv&
     \underline{T}(s_{0}+0, s_{0}-0)
\end{eqnarray}
of region II.

Finally
the transport map of the whole lens
takes the form\,:
\begin{eqnarray}
     \underline{T}(s_{0}+\Delta s/2,\,
                   s_{0}-\Delta s/2)
                &=&
     \underline{T}_{D}(\Delta s/2)
                \cdot
     \underline{T}_{L}
                \cdot
     \underline{T}_{D}(\Delta s/2)
\end{eqnarray}
corresponding to the decomposition
of the length\,
     $\Delta s$\,
into three parts
(see eqn. (3.13)\,).

Note that the (nonlinear) transport maps
    $\underline{T}_{D}$
and
    $\underline{T}_{L}$
corresponding to (3.14) and (3.15)
and thus also\,
    $\underline{T}(s_{0}+\Delta s/2,\,
                   s_{0}-\Delta s/2)$
in (3.18)
are automatically symplectic
for an arbitrary\,
       $\Delta s$\,
due to the canonical structure
of the equations of motion
(see also Ref.
\cite{DRAGT}
and Appendix A in Ref.
\cite{RS}).

In the limit
\begin{eqnarray*}
       \Delta s\
       \longrightarrow\  0
\end{eqnarray*}
one obtains the exact solution
of the canonical equations of motion
corresponding to the
starting Hamiltonian (2.3).
\bigskip

Since\,
    $\underline{T}_{D}$\,
is already known from eqn. (3.7)\,
we are left with the problem of calculating
the transport map
    $\underline{T}_{L}$\,
by solving eqn. (3.15).
This is done in the next section,
using a Lie series and exponentiation.
\bigskip

\subsection{Integration by Lie\,-\,Series}

\subsubsection{General Autonomous Case}

\ \ \ \ \
In the thin\,-\,lens approximation
the equations to be solved are not autonomous
but the
    $s$\,-\,dependence
is trivial
which
reduces the calculation
to an autonomous system.

An autonomous system of
differential equations
of the form\,:
\begin{eqnarray}
    \frac{d\ }{ds}\,
       y_{i}
         &=&
    \tilde{\vartheta}_{i}\left(
                       y_{1},\,
                  y_{2},\,
                  \dots,\,
                  y_{n}
                       \right)\ ;
\ \
         \frac{\partial\ }
              {\partial{s}}\,
  \tilde{\vartheta}_{i}
            \ =\
              0\ ;
\end{eqnarray}
\ \ \ \ \ \ \ \ \ \ \ \ \ \ \ \ \
\ \ \ \ \ \ \ \ \ \ \ \ \ \ \ \ \
\ \ \ \ \ \ \ \ \ \ \ \ \ \ \ \ \
   $(i= 1,\, 2,\, \dots,\,
                    n)$
\newline
\\
(no explicit $s$ dependence)
where the terms
   $\tilde{\vartheta}_{i}\left(y_{1},\,
                  y_{2},\,
                  \dots,\,
                  y_{n}\right)$
represent analytical functions, can be solved by
Lie-series
\cite{Groeb}\,:
\setcounter{INDEX}{1}
\begin{eqnarray}
       y_{i}(s)
               &=&
                e^{
          \left[
          (s-s_{0})D
                       \right]
                      }\,
                   \hat{y}_{i}
\end{eqnarray}
 with
\begin{eqnarray}
   \addtocounter{equation}{-1}
\addtocounter{INDEX}{1}
       D
        &=&
    \tilde{\vartheta}_{1}\left(
                       \hat{y}_{1},\,
                       \hat{y}_{2},\,
                  \dots,\,
                       \hat{y}_{n}
                                  \right)\cdot
                \frac{\partial\ }
                     {\partial{\hat{y}_{1}}}
          +
    \tilde{\vartheta}_{2}\left(\hat{y}_{1},\,
                       \hat{y}_{2},\,
                  \dots,\,
                       \hat{y}_{n}\right)\cdot
                \frac{\partial\ }
                     {\partial{\hat{y}_{2}}}
                    +
                  \dots
          +
    \tilde{\vartheta}_{n}\left(\hat{y}_{1},\,
                       \hat{y}_{2},\,
                  \dots\,
                       \hat{y}_{n}\right)\cdot
                \frac{\partial\ }
                     {\partial{\hat{y}_{n}}}
\nonumber    \\
\end{eqnarray}
and
\begin{eqnarray}
   \addtocounter{equation}{-1}
\addtocounter{INDEX}{1}
        {y}_{i}(s_{0})
           &\equiv&
               \hat{y}_{i}\ .
\end{eqnarray}
\setcounter{INDEX}{0}

Applying this result to the canonical equations
of motion\,:
\begin{eqnarray*}
    \tilde{\vartheta}_{1}
               &=&
               +\frac{\partial
                                      }{\partial
                          {p_{x}}}\,
  {\cal{H}}(x,p_{x},z,p_{z},\sigma,p_{\sigma})\ ;
\\
    \tilde{\vartheta}_{2}
               &=&
               -\frac{\partial
                                      }{\partial
                          {x}}\,
  {\cal{H}}(x,p_{x},z,p_{z},\sigma,p_{\sigma})\ ;
\\
    \tilde{\vartheta}_{3}
               &=&
               +\frac{\partial
                                      }{\partial
                          {p_{z}}}\,
  {\cal{H}}(x,p_{x},z,p_{z},\sigma,p_{\sigma})\ ;
\\
    \tilde{\vartheta}_{4}
               &=&
               -\frac{\partial
                                      }{\partial
                          {z}}\,
  {\cal{H}}(x,p_{x},z,p_{z},\sigma,p_{\sigma})\ ;
\\
    \tilde{\vartheta}_{5}
               &=&
               +\frac{\partial
                                      }{\partial
                          {p_{\sigma}}}\,
  {\cal{H}}(x,p_{x},z,p_{z},\sigma,p_{\sigma})\ ;
\\
    \tilde{\vartheta}_{6}
               &=&
               -\frac{\partial
                                      }{\partial
                          {\sigma}}\,
  {\cal{H}}(x,p_{x},z,p_{z},\sigma,p_{\sigma})
\end{eqnarray*}
we obtain\,:
\setcounter{INDEX}{1}
\begin{eqnarray}
    y_{1}(s)
            &\equiv\
    x(s)
             &=\
                e^{
          (s-s_{0})\, D}\,
    \hat{x}\ ;
\\       \nonumber
\\
   \addtocounter{equation}{-1}
\addtocounter{INDEX}{1}
    y_{2}(s)
            &\equiv\
    p_{x}(s)
             &=\
                e^{
          (s-s_{0})\, D}\,
    \hat{p}_{x}\ ;
\\       \nonumber
\\
   \addtocounter{equation}{-1}
\addtocounter{INDEX}{1}
    y_{3}(s)
            &\equiv\
    z(s)
             &=\
                e^{
          (s-s_{0})\, D}\,
    \hat{z}\ ;
\\       \nonumber
\\
   \addtocounter{equation}{-1}
\addtocounter{INDEX}{1}
    y_{4}(s)
            &\equiv\
    p_{z}(s)
             &=\
                e^{
          (s-s_{0})\, D}\,
    \hat{p}_{z}\ ;
\\       \nonumber
\\
   \addtocounter{equation}{-1}
\addtocounter{INDEX}{1}
    y_{5}(s)
            &\equiv\
    \sigma(s)
             &=\
                e^{
          (s-s_{0})\, D}\,
    \hat{\sigma}\ ;
\\       \nonumber
\\
   \addtocounter{equation}{-1}
\addtocounter{INDEX}{1}
    y_{6}(s)
            &\equiv\
    p_{\sigma}(s)
             &=\
                e^{
          (s-s_{0})\, D}\,
    \hat{p}_{\sigma}
\end{eqnarray}
\setcounter{INDEX}{0}
with
\begin{eqnarray}
         D
          &=&
    \left[
               \frac{\partial\ }
                   {\partial
                   {\hat{p}_{x}}}\,
       {\cal{H}}
                (\vec{\hat{y}})
       \right]\,
               \frac{\partial\ }
                   {\partial
                   {\hat{x}}}
                    -
    \left[
               \frac{\partial\ }
                   {\partial
                   {\hat{x}}}\,
       {\cal{H}}
                (\vec{\hat{y}})
       \right]\,
               \frac{\partial\ }
                   {\partial
                   {\hat{p}_{x}}}
\nonumber    \\
          &+&
    \left[
               \frac{\partial\ }
                   {\partial
                   {\hat{p}_{z}}}\,
       {\cal{H}}
                (\vec{\hat{y}})
       \right]\,
               \frac{\partial\ }
                   {\partial
                   {\hat{z}}}
                    -
    \left[
               \frac{\partial\ }
                   {\partial
                   {\hat{z}}}\,
       {\cal{H}}
                (\vec{\hat{y}})
       \right]\,
               \frac{\partial\ }
                   {\partial
                   {\hat{p}_{z}}}
\nonumber    \\
          &+&
    \left[
               \frac{\partial\ }
                   {\partial
                   {\hat{p}_{\sigma}}}\,
       {\cal{H}}
                (\vec{\hat{y}})
       \right]\,
               \frac{\partial\ }
                   {\partial
                   {\hat{\sigma}}}
                    -
    \left[
               \frac{\partial\ }
                   {\partial
                   {\hat{\sigma}}}\,
       {\cal{H}}
                (\vec{\hat{y}})
       \right]\,
               \frac{\partial\ }
                   {\partial
                   {\hat{p}_{\sigma}}}
\end{eqnarray}
and
\begin{eqnarray}
         \hat{x}&\equiv& x(s_{0})\ ;\ \
         \hat{p_{x}}\ \equiv\ p_{x}(s_{0})\ ;
\nonumber    \\
         \hat{z}&\equiv& z(s_{0})\ ;\ \
         \hat{p_{z}}\ \equiv\ p_{z}(s_{0})\ ;
\nonumber    \\
         \hat{\sigma}&\equiv& \sigma(s_{0})\ ;\ \
         \hat{p_{\sigma}}\ \equiv\ p_{\sigma}(s_{0})\ .
\end{eqnarray}
\\
\underline{Remarks:}
                    \newline

\vspace{-0.3ex}
1) Using the notation of Ref.
\cite{DRAGT},
eqn. (3.20a) may also be written in the form\,:
\begin{eqnarray*}
       y_{i}(s)
               &=&
                e^{
          :\,
          (s-s_{0})
         {\cal{H}}
                   \,:
                      }\,
                   \hat{y}_{i}
\end{eqnarray*}
if the autonomous equations of motion
result from an Hamiltonian\,
        ${\cal{H}}$.
So when the approach in Ref.
\cite{Groeb}
is restricted to canonical systems
it is identical to the Lie Algebra
method introduced by Dragt.
\\

2) Since the equations of motion (3.6)
for a drift space
represent an autonomous system of differential equations,
eqns. (3.20a, b, c) can be used to determine
the transport map\,
       $\underline{T}_{D}$\,
of a drift space.

In this case we get by comparing (3.19) with (3.6)\,:
\setcounter{INDEX}{1}
\begin{eqnarray}
    \tilde{\vartheta}_{1}
                    &=&
      \frac{
           y_{2}
                }
               {
                \left[1+
                   f(y_{6})\right]
                                }\ ;
 \\       \nonumber
 \\
   \addtocounter{equation}{-1}
\addtocounter{INDEX}{1}
    \tilde{\vartheta}_{2}
                        &=&
       0\ ;
 \\       \nonumber
 \\
   \addtocounter{equation}{-1}
\addtocounter{INDEX}{1}
    \tilde{\vartheta}_{3}
                   &=&
      \frac{
           y_{4}
                }
               {
                \left[1+
                   f(y_{6})\right]
                                }\ ;
 \\       \nonumber
 \\
   \addtocounter{equation}{-1}
\addtocounter{INDEX}{1}
    \tilde{\vartheta}_{4}
                       &=&
                        0\ ;
 \\       \nonumber
 \\
   \addtocounter{equation}{-1}
\addtocounter{INDEX}{1}
    \tilde{\vartheta}_{5}
                        &=&
              1-
                      f'(y_{6})
         -\frac{1}{2}\cdot
                [y_{2}^{2}+y_{4}^{2}]
                \cdot
                \frac{
                      f'(y_{6})
                                    }
               {
                \left[1+
                   f(y_{6})\right]^{2}
                                }\ ;
 \\       \nonumber
 \\
   \addtocounter{equation}{-1}
\addtocounter{INDEX}{1}
    \tilde{\vartheta}_{6}
                       &=&
                       0\ .
\end{eqnarray}
\setcounter{INDEX}{0}
This leads to
\setcounter{INDEX}{1}
\begin{eqnarray}
       y_{i}(s_{0}+l)&=&
                e^{
           l\cdot D}\,
                   \hat{y}_{i}
\end{eqnarray}
 with
\begin{eqnarray}
   \addtocounter{equation}{-1}
\addtocounter{INDEX}{1}
       D
        &=&
    \tilde{\vartheta}_{1}\left(
                       \vec{\hat{y}}
                                  \right)\cdot
                \frac{\partial\ }
                     {\partial{\hat{y}_{1}}}
          +
    \tilde{\vartheta}_{3}\left(
                       \vec{\hat{y}}
                                  \right)\cdot
                \frac{\partial\ }
                     {\partial{\hat{y}_{3}}}
                    +
    \tilde{\vartheta}_{5}\left(
                       \vec{\hat{y}}
                                  \right)\cdot
                \frac{\partial\ }
                     {\partial{\hat{y}_{5}}}
\end{eqnarray}
and
\begin{eqnarray}
   \addtocounter{equation}{-1}
\addtocounter{INDEX}{1}
        \hat{y}_{i}
           &\equiv&
                    y_{i}(s_{0})\ .
\end{eqnarray}
\setcounter{INDEX}{0}

We then have\,:
\setcounter{INDEX}{1}
\begin{eqnarray}
    D\,
    \hat{y}_{1}
             &=&
      \frac{
           \hat{y}_{2}
                }
               {
                \left[1+
                   f(\hat{y}_{6})\right]
                                }\ ;
\\       \nonumber
\\
   \addtocounter{equation}{-1}
\addtocounter{INDEX}{1}
    D\,
    \hat{y}_{2}
             &=&
              0\ ;
\\       \nonumber
\\
   \addtocounter{equation}{-1}
\addtocounter{INDEX}{1}
    D\,
    \hat{y}_{3}
        &=&
      \frac{
           \hat{y}_{4}
                }
               {
                \left[1+
                   f(\hat{y}_{6})\right]
                                }\ ;
\\       \nonumber
\\
   \addtocounter{equation}{-1}
\addtocounter{INDEX}{1}
    D\,
    \hat{y}_{4}
             &=&
              0\ ;
\\       \nonumber
\\
   \addtocounter{equation}{-1}
\addtocounter{INDEX}{1}
    D\,
    \hat{y}_{5}
            &=&
              1-
                      f'(\hat{y}_{6})
         -\frac{1}{2}\cdot
                [\hat{y}_{2}^{2}+\hat{y}_{4}^{2}]
                \cdot
                \frac{
                      f'(\hat{y}_{6})
                                    }
               {
                \left[1+
                   f(\hat{y}_{6})\right]^{2}
                                }\ ;
\\       \nonumber
\\
   \addtocounter{equation}{-1}
\addtocounter{INDEX}{1}
    D\,
    \hat{y}_{6}
             &=&
              0
\end{eqnarray}
\setcounter{INDEX}{0}
and
\begin{eqnarray*}
           D^{\,\nu}\,
       \vec{\hat{y}}
        &=&
       \vec{0}\ \
       \mbox{for}\ \
       \nu\;>\;1\ .
\end{eqnarray*}

Thus\,:
\begin{eqnarray}
      \vec{y}(s_{0}+l)
          &=&
      \left[
            1+l\cdot D
                       \right]\,
      \vec{\hat{y}}\ .
\end{eqnarray}

Putting (3.26) into (3.27),
we regain eqn. (3.7).
\\

3) The method
for calculating thin\,-\,lens transport maps
described in this paper
works also in the presence of
nonsymplectic terms
resulting for instance from radiation damping
\cite{BHKMR}.
One must simply include these terms in\,
        $D$\,
before expanding\,
              $\exp{
                    \left[
                            {\hat{D}}
                             \right]
                               }$.
\bigskip

\subsubsection{Calculation of the Thin\,-\,Lens Transport Map
         for the Central Region.
                                }

\ \ \ \ \
In order to determine the transport map\,
         $\underline{T}_{L}$\,
for the central region,
we investigate the special case\,:
\begin{eqnarray}
    \tilde{\vartheta}_{i}\left(
                       \vec{y}\,
                       \right)
         &=&
    \delta(s-s_{0})
       \cdot
    F_{i}\left(
                       \vec{y}
                       \right)\ ;\ \
    \frac{\partial\ }
       {\partial{s}}\,
    F_{i}
        \ =\
          0\ .
\end{eqnarray}

Replacing
the $\delta$\,-\,function\,
  $\delta(s-s_{0})$\,
in (3.28)
by a step function of height\,
   $(1/2\epsilon)$\,
and length\,
   $(2\epsilon)$,
we obtain in this case
from eqn. (3.20)\,:
\setcounter{INDEX}{1}
\begin{eqnarray*}
     \vec{y}(s)
           &=&
  \left\{
     \exp{\left[(s-[s_{0}-\epsilon])
           \cdot
           \frac{1}{2\,\epsilon}\,
           \hat{D}\right]}
                \right\}
     \vec{\hat{y}}\ \
\\
\\
& &
      \mbox{for}\ \
     (s_{0}-\epsilon\,\leq\,s\,\leq\,s_{0}+\epsilon)
\end{eqnarray*}
with
\begin{eqnarray*}
     \vec{\hat{y}}
           &\equiv&
     \vec{y}(s_{0}-\epsilon)
\end{eqnarray*}
and
\setcounter{INDEX}{1}
\begin{eqnarray}
       \hat{D}
        &=&
    F_{1}(
               \vec{\hat{y}}
                                  )\cdot
                \frac{\partial\ }
                     {\partial{\hat{y}_{1}}}
          +
    F_{2}(
               \vec{\hat{y}}
                                  )\cdot
                \frac{\partial\ }
                     {\partial{\hat{y}_{2}}}
                    +
                  \cdot
                  \cdot
                  \cdot\,
          +
    F_{6}(
               \vec{\hat{y}}
                                  )\cdot
                \frac{\partial\ }
                     {\partial{\hat{y}_{6}}}\ .
\end{eqnarray}

In particular
by putting\,
      $s\,=\,s_{0}+\epsilon$\,
we have\,:
\begin{eqnarray*}
     \vec{y}(s_{0}+\epsilon)
           &=&
  \left\{
     \exp{\left[([s_{0}+\epsilon]
           -[s_{0}-\epsilon])
           \cdot
           \frac{1}{2\,\epsilon}\,
           \hat{D}\right]}
                \right\}
     \vec{\hat{y}}
\nonumber     \\
           &=&
  \left\{
     \exp{[
           \hat{D}]}
                \right\}
     \vec{\hat{y}}\ ,
\end{eqnarray*}
which leads
in the limit\,
      $\epsilon\,\longrightarrow\,0$\,
to\,:
\begin{eqnarray*}
     \vec{y}(s_{0}+0)
           &=&
  \left\{
     \exp{[
           \hat{D}]}
                \right\}
     \vec{\hat{y}}
\end{eqnarray*}
with
\begin{eqnarray*}
     \vec{\hat{y}}
           &\equiv&
     \vec{y}(s_{0}-0)\ .
\end{eqnarray*}

Then
by choosing
the functions\,
   $F_{i}(\vec{y})$
appearing in (3.29a) as\,:
\begin{eqnarray}
   \addtocounter{equation}{-1}
\addtocounter{INDEX}{1}
    F_{i}(\vec{y})
        &=&
    \vartheta_{Li}(\vec{y}; s_{0})
        \cdot
   \Delta s\ ,
\end{eqnarray}
with\,
   $\vartheta_{Li}(\vec{y}; s_{0})$\,
given by (3.16),
one just gets the transport map\,
   $\underline{T}_{L}$
corresponding to eqn. (3.15)
in the form\,:
\begin{eqnarray}
   \addtocounter{equation}{-1}
\addtocounter{INDEX}{1}
   \underline{T}_{L}
        &=&
     \exp{[
           \hat{D}]}
\end{eqnarray}
\setcounter{INDEX}{0}
as may be seen by comparing (3.28)
with the r.h.s.
of (3.15).
\\
\\
\\
\underline{Remark:}
                    \newline

\vspace{-0.3ex}
The relation (3.29c)
for\,
  $\underline{T}_{L}$\,
can also be derived by solving
the differential equation\,:
\begin{eqnarray}
    \frac{d\ }{ds}\,
      \vec{y}
             &=&
    \vec{\vartheta}_{L}(\vec{y}; s_{0})
            \ \equiv\
   \frac{1}{\Delta s}\cdot
    \vec{F}(\vec{y})\ ,
\end{eqnarray}
which does not contain
the $\delta$\,-\,function\,
  $\delta(s-s_{0})$\,
anymore.
Writing the solution
of (3.30) in the form
\begin{eqnarray}
      \vec{y}(s)
         &=&
      \underline{\tilde{T}}(s,\, s_{0})\,
      \vec{y}(s_{0})\ ,
\end{eqnarray}
one then obtains\,:
\begin{eqnarray}
      \underline{T}_{L}
         &\equiv&
      \underline{\tilde{T}}(s_{0}+\Delta s,\, s_{0})\ ,
\end{eqnarray}
as can be verified by compararing (3.29) with (3.20)
and using (3.19).
\bigskip

By (3.30) we see in fact
that for the central region II
the problem reduces to an autonomous one.

For an example see Remark 2) at the end of
Appendix C.
\bigskip

\setcounter{equation}{0}
\section{Thin\,-\,Lens Approximation
         for Various Kinds of Magnets
         and for Cavities
                                    }

\ \ \ \ \
In this section
the thin\,-\,lens transport map
corresponding to the central region II
(see eqn. (3.13)\,)
is calculated for cavities
and for various kinds of magnets.
\bigskip

\subsection{Quadrupole}

\subsubsection{Exponentiation}

\ \ \ \ \
For a quadrupole we have\,:
\begin{eqnarray*}
       g&\neq&0
\end{eqnarray*}
and
\begin{eqnarray*}
          K_{x}&=&K_{z}\ =\ N\ =\ \lambda\ =\ \mu\ =\ H\ =\ V
                                          \ =\ 0\ .
\end{eqnarray*}

Then we obtain from (3.16) and (3.29b)\,:
\begin{eqnarray*}
    F_{1}
         (\vec{y})
                    &=&
      0\ ;
\\      \nonumber
 \\
    F_{2}
         (\vec{y})
                    &=&
      -g(s_{0})\cdot\Delta s\cdot x\ ;
\\      \nonumber
 \\
    F_{3}
         (\vec{y})
                    &=&
      0\ ;
\\      \nonumber
 \\
    F_{4}
         (\vec{y})
                    &=&
      +g(s_{0})\cdot\Delta s\cdot z  ;
\\      \nonumber
 \\
    F_{5}
         (\vec{y})
                    &=&
      0\ ;
\\      \nonumber
 \\
    F_{6}
         (\vec{y})
                    &=&
      0\ .
\end{eqnarray*}

Thus\,:
\begin{eqnarray}
       \hat{D}
        &=&
    F_{2}
         (\vec{\hat{y}})
         \cdot
                \frac{\partial\ }
                     {\partial{\hat{y}_{2}}}
          +
    F_{4}
         (\vec{\hat{y}})
         \cdot
                \frac{\partial\ }
                     {\partial{\hat{y}_{4}}}
\end{eqnarray}
and
\begin{eqnarray}
       \hat{D}\,
       \vec{\hat{y}}
        &=&
                 \left( \begin{array}{c}
                   0      \\
                   F_{2}(\vec{\hat{y}})  \\
                   0      \\
                   F_{4}(\vec{\hat{y}})  \\
                   0      \\
                   0
              \end{array}
       \right)
       \ =\
       \underline{\hat{A}}\,
       \vec{\hat{y}}
\end{eqnarray}
with
\begin{eqnarray}
   \underline{\hat{A}}
               &=&
   \Delta s
       \cdot
                 \left( \begin{array}{cccccc}
                   0  &  0
                            & 0                 &  0  &  0  &  0  \\
   -g                 &  0
                            &  0  & 0 &  0  &  0      \\
    0                 &  0
                            &  0  &  0  &  0  &  0  \\
                   0  &  0
       &  +g              &  0    &  0  &  0        \\
   0               &  0    &  0       &  0 &  0  & 0               \\
                0  &  0  &  0  &  0 &0 & 0
              \end{array}
       \right)\ .
\end{eqnarray}

The transfer matrix\,
      $\underline{M}$\,
defined by
\begin{eqnarray*}
      \vec{y}(s_{0}+0)
         &=&
      \underline{M}\,
      \vec{y}(s_{0}-0)
        \ \equiv\
      \underline{M}\,
      \vec{\hat{y}}
\end{eqnarray*}
reads as\,:
\begin{eqnarray}
   \underline{M}
               &=&
               \exp{
                    \left[
                  \underline{\hat{A}}
                             \right]
                               }
\nonumber   \\
               &=&
                  \underline{1}
                       +
                  \underline{\hat{A}}
\end{eqnarray}
since
\begin{eqnarray*}
       \hat{D}\,
   \underline{\hat{A}}
               &=&
   \underline{\hat{A}}\,
       \hat{D}
\ \ \Longrightarrow\ \
       \hat{D}^{\,\nu}\,
       \vec{\hat{y}}
       \ =\
       \underline{\hat{A}}^{\,\nu}\,
       \vec{\hat{y}}
\ \ \Longrightarrow\ \
         \left\{
                \exp{
                    \left[
                            {\hat{D}}
                             \right]
                               }
                                  \right\}\,
                                  \vec{\hat{y}}
       \ =\
         \left\{
               \exp{
                    \left[
                  \underline{\hat{A}}
                             \right]
                               }
                                  \right\}\,
                                  \vec{\hat{y}}
\end{eqnarray*}
and
\begin{eqnarray*}
       \underline{\hat{A}}^{\,\nu}
               &=&
       \underline{0}\ \ \
       \mbox{for}\ \
       \nu\;>\;1\ .
\end{eqnarray*}
\bigskip

\subsubsection{Thin\,-\,Lens Transport Map}

\ \ \ \ \
{}From (4.4) we obtain
\footnote{See also
section A.2.2
in Appendix A,
where a
superposition of quadrupoles,
                       skew quadrupoles,
                       bending magnets,
                       sextupoles and
                       octupoles
is investigated.}:
\begin{eqnarray*}
    x^{f}
            &=&
               x^{i}\ ;
\\      \nonumber
 \\
    p_{x}^{\,f}
            &=&
    p_{x}^{\,i}
                -g(s_{0})
                     \cdot\Delta s
                          \cdot x^{i}\ ;
\\      \nonumber
 \\
    z^{f}
            &=&
               z^{i}\ ;
\\      \nonumber
 \\
    p_{z}^{\,f}
            &=&
    p_{z}^{\,i}
                +g(s_{0})
                     \cdot\Delta s
                          \cdot z^{i}\ ;
\\      \nonumber
 \\
    \sigma^{f}
            &=&
               \sigma^{i}\ ;
\\      \nonumber
 \\
    p_{\sigma}^{\,f}
            &=&
    p_{\sigma}^{\,i}
\end{eqnarray*}
with
\begin{eqnarray*}
    y^{i}
            &\equiv&
    y(s_{0}-0)\ ;
\\
    y^{f}
            &\equiv&
    y(s_{0}+0)\ ;
\end{eqnarray*}
\begin{eqnarray*}
         (y&=&x,\, p_{x},\, z,\, p_{z},\, \sigma,\, p_{\sigma})\ .
\end{eqnarray*}

\bigskip

\subsection{Skew Quadrupole}

\subsubsection{Exponentiation}
\ \ \ \ \
For a skew quadrupole we have\,:
\begin{eqnarray*}
       N&\neq&0
\end{eqnarray*}
and
\begin{eqnarray*}
          K_{x}&=&K_{z}\ =\ g\ =\ \lambda\ =\ \mu\ =\ H\ =\ V
                                          \ =\ 0\ .
\end{eqnarray*}

Thus we get from (3.16) and (3.29b)\,:
\bigskip
\begin{eqnarray*}
    F_{1}
         (\vec{y})
                    &=&
      0\ ;
\\      \nonumber
 \\
    F_{2}
         (\vec{y})
                    &=&
      +N(s_{0})\cdot\Delta s\cdot z\ ;
\\      \nonumber
 \\
    F_{3}
         (\vec{y})
                    &=&
      0\ ;
\\      \nonumber
 \\
    F_{4}
         (\vec{y})
                    &=&
      +N(s_{0})\cdot\Delta s\cdot x  ;
\\      \nonumber
 \\
    F_{5}
         (\vec{y})
                    &=&
      0\ ;
\\      \nonumber
 \\
    F_{6}
         (\vec{y})
                    &=&
      0\ .
\end{eqnarray*}

Thus\,:
\begin{eqnarray}
       \hat{D}
        &=&
    F_{2}
         (\vec{\hat{y}})
         \cdot
                \frac{\partial\ }
                     {\partial{\hat{y}_{2}}}
          +
    F_{4}
         (\vec{\hat{y}})
         \cdot
                \frac{\partial\ }
                     {\partial{\hat{y}_{4}}}
\end{eqnarray}
and
\begin{eqnarray}
       \hat{D}\,
       \vec{\hat{y}}
        &=&
                 \left( \begin{array}{c}
                   0      \\
                   F_{2}(\vec{\hat{y}})  \\
                   0      \\
                   F_{4}(\vec{\hat{y}})  \\
                   0      \\
                   0
              \end{array}
       \right)
       \ =\
       \underline{\hat{A}}\;
       \vec{\hat{y}}
\end{eqnarray}
with
\begin{eqnarray}
   \underline{\hat{A}}
               &=&
   \Delta s
       \cdot
                 \left( \begin{array}{cccccc}
                   0  &  0
                            & 0                 &  0  &  0  &  0  \\
    0                 &  0
                            &  N  & 0 &  0  &  0      \\
    0                 &  0
                            &  0  &  0  &  0  &  0  \\
                   N  &  0
       &   0              &  0    &  0  &  0        \\
   0               &  0    &  0       &  0 &  0  & 0               \\
                0  &  0  &  0  &  0 &0 & 0
              \end{array}
       \right)\ .
\end{eqnarray}

The transfer matrix reads as\,:
\begin{eqnarray}
   \underline{M}
               &=&
               \exp{
                    \left[
                  \underline{\hat{A}}
                             \right]
                               }
\nonumber   \\
               &=&
                  \underline{1}
                       +
                  \underline{\hat{A}}
\end{eqnarray}
since
\begin{eqnarray*}
       \hat{D}\,
   \underline{\hat{A}}
               &=&
   \underline{\hat{A}}\,
       \hat{D}
\end{eqnarray*}
and
\begin{eqnarray*}
       \underline{\hat{A}}^{\,\nu}
               &=&
       \underline{0}\ \
       \mbox{for}\ \
       \nu\;>\;1\ .
\end{eqnarray*}
\bigskip

\subsubsection{Thin\,-\,Lens Transport Map}
\bigskip

\ \ \ \ \
{}From (4.8) we obtain\,:
\begin{eqnarray*}
    x^{f}
            &=&
               x^{i}\ ;
\\      \nonumber
 \\
    p_{x}^{\,f}
            &=&
    p_{x}^{\,i}
                +N(s_{0})
                     \cdot\Delta s
                          \cdot z^{i}\ ;
\\      \nonumber
 \\
    z^{f}
            &=&
               z^{i}\ ;
\\      \nonumber
 \\
    p_{z}^{\,f}
            &=&
    p_{z}^{\,i}
                +N(s_{0})
                     \cdot\Delta s
                          \cdot x^{i}\ ;
\\      \nonumber
 \\
    \sigma^{f}
            &=&
               \sigma^{i}\ ;
\\      \nonumber
 \\
    p_{\sigma}^{\,f}
            &=&
    p_{\sigma}^{\,i}\ .
\end{eqnarray*}
\bigskip

\subsection{Bending Magnet}

\subsubsection{Exponentiation}

\ \ \ \ \
For a bending magnet we have\,:
\begin{eqnarray*}
       K_{x}^{2}+K_{z}^{2}&\neq&0\,;\ \
       K_{x}\cdot K_{z}\ =\ 0
\end{eqnarray*}
and
\begin{eqnarray*}
          g&=&N\ =\ \lambda\ =\ \mu\  =\ H\ =\ V
                                    \ =\ 0\ .
\end{eqnarray*}

{}From (3.16) and (3.29b) we thus obtain\,:
\bigskip
\begin{eqnarray*}
    F_{1}
         (\vec{y})
                    &=&
      0\ ;
\\      \nonumber
 \\
    F_{2}
         (\vec{y})
                    &=&
      -[K_{x}(s_{0})]^{2}
               \cdot\Delta s\cdot x
      +K_{x}(s_{0})
               \cdot\Delta s
               \cdot f(p_{\sigma})\ ;
\\      \nonumber
 \\
    F_{3}
         (\vec{y})
                    &=&
      0\ ;
\\      \nonumber
 \\
    F_{4}
         (\vec{y})
                    &=&
      -[K_{z}(s_{0})]^{2}
               \cdot\Delta s\cdot z
      +K_{z}(s_{0})
               \cdot\Delta s
               \cdot f(p_{\sigma})\ ;
\\      \nonumber
 \\
    F_{5}
         (\vec{y})
            &=&
             -
    \left[
       K_{x}(s_{0})\cdot x
             +
       K_{z}(s_{0})\cdot z
               \right]
               \cdot\Delta s
               \cdot f'(p_{\sigma})\ ;
\\      \nonumber
 \\
    F_{6}
         (\vec{y})
                    &=&
      0\ .
\end{eqnarray*}

Thus\,:
\begin{eqnarray}
       \hat{D}
        &=&
    F_{2}
        (\vec{\hat{y}})
         \cdot
                \frac{\partial\ }
                     {\partial{\hat{y}_{2}}}
          +
    F_{4}
        (\vec{\hat{y}})
         \cdot
                \frac{\partial\ }
                     {\partial{\hat{y}_{4}}}
          +
    F_{5}
        (\vec{\hat{y}})
         \cdot
                \frac{\partial\ }
                     {\partial{\hat{y}_{5}}}
\end{eqnarray}
and
\begin{eqnarray}
       \hat{D}\,
       \vec{\hat{y}}
        &=&
                 \left( \begin{array}{c}
                   0      \\
                   F_{2}(\vec{\hat{y}})  \\
                   0      \\
                   F_{4}(\vec{\hat{y}})  \\
                   F_{5}(\vec{\hat{y}})  \\
                   0
              \end{array}
       \right)
       \ =\
       \underline{\hat{A}}\,
       \vec{\hat{y}}
\end{eqnarray}
with
\begin{eqnarray}
   \underline{\hat{A}}
               &=&
   \Delta s
       \cdot
                 \left( \begin{array}{cccccc}
                   0  &  0
                            & 0                 &  0  &  0  &  0
                                                                  \\
   -K_{x}^{\,2}
                      &  0
                            &  0  & 0 &  0  &
                            K_{x}\cdot f(\hat{p}_{\sigma})
                                                      \\
    0                 &  0
                            &  0  &  0  &  0  &  0
                                                    \\
                   0  &  0
       &
          -K_{z}^{\,2}
                          &  0    &  0  &
                            K_{z}\cdot f(\hat{p}_{\sigma})
                                                    \\
          -K_{x}\cdot f'(\hat{p}_{\sigma})
                   &  0    &
          -K_{z}\cdot f'(\hat{p}_{\sigma})
                                      &  0 &  0  & 0
                                                                   \\
                0  &  0  &  0  &  0 &0 & 0
              \end{array}
       \right)\ .
\end{eqnarray}

The transfer matrix reads as\,:
\begin{eqnarray}
   \underline{M}
               &=&
               \exp{
                    \left[
                  \underline{\hat{A}}
                             \right]
                               }
\nonumber   \\
               &=&
                  \underline{1}
                       +
                  \underline{\hat{A}}
\end{eqnarray}
since
\begin{eqnarray*}
       \hat{D}\,
   \underline{\hat{A}}
               &=&
   \underline{\hat{A}}\,
       \hat{D}
\end{eqnarray*}
and
\begin{eqnarray*}
       \underline{\hat{A}}^{\,\nu}
               &=&
       \underline{0}\ \
       \mbox{for}\ \
       \nu\;>\;1\ .
\end{eqnarray*}
\bigskip

\subsubsection{Thin\,-\,Lens Transport Map}
\bigskip

\ \ \ \ \
{}From (4.12) we obtain\,:
\begin{eqnarray*}
    x^{f}
            &=&
               x^{i}\ ;
\\      \nonumber
 \\
    p_{x}^{\,f}
            &=&
    p_{x}^{\,i}
                -\left[K_{x}(s_{0})\right]^{2}
                     \cdot\Delta s
                          \cdot x^{i}
                         +K_{x}(s_{0})
                     \cdot\Delta s
                               \cdot f(p_{\sigma}^{\,i})\ ;
\\      \nonumber
 \\
    z^{f}
            &=&
               z^{i}\ ;
\\      \nonumber
 \\
    p_{z}^{\,f}
            &=&
    p_{z}^{\,i}
                -\left[K_{z}(s_{0})\right]^{2}
                     \cdot\Delta s
                          \cdot z^{i}
                         +K_{z}(s_{0})
                     \cdot\Delta s
                               \cdot f(p_{\sigma}^{\,i})\ ;
\\      \nonumber
 \\
    \sigma^{f}
            &=&
               \sigma^{i}
               -[K_{x}\cdot x^{i}+
                 K_{z}\cdot z^{i}]
                     \cdot\Delta s
                \cdot f'(p_{\sigma}^{\,i})\ ;
\\      \nonumber
 \\
    p_{\sigma}^{\,f}
            &=&
    p_{\sigma}^{\,i}\ .
\end{eqnarray*}
\bigskip

\subsection{Sextupole}

\subsubsection{Exponentiation}

\ \ \ \ \
For a sextupole we have\,:
\begin{eqnarray*}
       \lambda&\neq&0
\end{eqnarray*}
and
\begin{eqnarray*}
           K_{x}&=&K_{z}\ =\ g\ =\ N\ =\ \mu\ =\ H\ =\ V
                                          \ =\ 0\ .
\end{eqnarray*}

{}From (3.16) and (3.29b) we then get\,:
\begin{eqnarray*}
    F_{1}
         (\vec{y})
                    &=&
      0\ ;
\\      \nonumber
 \\
    F_{2}
         (\vec{y})
                    &=&
      -\frac{1}{2}\,
       \lambda(s_{0})\cdot\Delta s
       \cdot
       \left[x^{2}-z^{2}\right]\ ;
\\      \nonumber
 \\
    F_{3}
         (\vec{y})
                    &=&
      0\ ;
\\      \nonumber
 \\
    F_{4}
         (\vec{y})
                    &=&
       \lambda(s_{0})\cdot\Delta s
       \cdot x\,z\ ;
\\      \nonumber
 \\
    F_{5}
         (\vec{y})
                    &=&
      0\ ;
\\      \nonumber
 \\
    F_{6}
         (\vec{y})
                    &=&
      0\ .
\end{eqnarray*}

Thus\,:
\begin{eqnarray}
       \hat{D}
        &=&
    F_{2}
         (\vec{\hat{y}})
         \cdot
                \frac{\partial\ }
                     {\partial{\hat{y}_{2}}}
          +
    F_{4}
         (\vec{\hat{y}})
         \cdot
                \frac{\partial\ }
                     {\partial{\hat{y}_{4}}}
\end{eqnarray}
and
\begin{eqnarray}
       \hat{D}\,
       \vec{\hat{y}}
        &=&
                 \left( \begin{array}{c}
                   0      \\
                   F_{2}(\vec{\hat{y}})   \\
                   0      \\
                   F_{4}(\vec{\hat{y}})   \\
                   0      \\
                   0
              \end{array}
       \right)\ ;\ \ \
       \hat{D}^{\,\nu}\,
       \vec{\hat{y}}
       \ =\
                 \left( \begin{array}{c}
                   0      \\
                   0      \\
                   0      \\
                   0      \\
                   0      \\
                   0
              \end{array}
       \right)
          \ \ \
       \mbox{for}\ \
       \nu\;>\;1
\end{eqnarray}
\begin{eqnarray}
   \Longrightarrow\ \
        \left\{
               \exp{
                    \left[
                            {\hat{D}}
                             \right]
                               }
                       \right\}\,
              \vec{\hat{y}}
               &=&
              \vec{\hat{y}}
                +
              \hat{D}\,
              \vec{\hat{y}}\ .
\end{eqnarray}
\bigskip

\subsubsection{Thin\,-\,Lens Transport Map}

\ \ \ \ \
{}From (4.15) we obtain\,:
\begin{eqnarray*}
    x^{f}
            &=&
               x^{i}\ ;
\\      \nonumber
 \\
    p_{x}^{\,f}
            &=&
    p_{x}^{\,i}
                -\,
                \frac{1}{2}\,
                 \lambda(s_{0})
                     \cdot\Delta s
                          \cdot
                 [(x^{i})^{2}
                 -(z^{i})^{2}]\ ;
\\      \nonumber
 \\
    z^{f}
            &=&
               z^{i}\ ;
\\      \nonumber
 \\
    p_{z}^{\,f}
            &=&
    p_{z}^{\,i}
                +\,
                 \lambda(s_{0})
                     \cdot\Delta s
                          \cdot x^{i}\,
                   z^{i}\ ;
\\      \nonumber
 \\
    \sigma^{f}
            &=&
               \sigma^{i}\ ;
\\      \nonumber
 \\
    p_{\sigma}^{\,f}
            &=&
    p_{\sigma}^{\,i}\ .
\end{eqnarray*}
\bigskip

\subsection{Octupole}

\subsubsection{Exponentiation}

\ \ \ \ \
For an octupole we have\,:
\begin{eqnarray*}
       \mu&\neq&0
\end{eqnarray*}
and
\begin{eqnarray*}
          K_{x}&=&K_{z}\ =\ g\ =\ N\  =\ \lambda\ =\ H\ =\ V
                                          \ =\ 0\ .
\end{eqnarray*}

Then we obtain from (3.16) and (3.29b)\,:
\begin{eqnarray*}
    F_{1}
         (\vec{y})
                    &=&
      0\ ;
\\      \nonumber
 \\
    F_{2}
         (\vec{y})
                    &=&
      -\frac{1}{6}\,
       \mu(s_{0})\cdot\Delta s
       \cdot
       \left[x^{3}-3\,xz^{2}\right]\ ;
\\      \nonumber
 \\
    F_{3}
         (\vec{y})
                    &=&
      0\ ;
\\      \nonumber
 \\
    F_{4}
         (\vec{y})
                    &=&
      -\frac{1}{6}\,
       \mu(s_{0})\cdot\Delta s
       \cdot
       \left[z^{3}-3\,x^{2}z\right]\ ;
\\      \nonumber
 \\
    F_{5}
         (\vec{y})
                    &=&
      0\ ;
\\      \nonumber
 \\
    F_{6}
         (\vec{y})
                    &=&
      0\ .
\end{eqnarray*}

Thus\,:
\begin{eqnarray}
       \hat{D}
        &=&
    F_{2}
         (\vec{\hat{y}})
         \cdot
                \frac{\partial\ }
                     {\partial{\hat{y}_{2}}}
          +
    F_{4}
         (\vec{\hat{y}})
         \cdot
                \frac{\partial\ }
                     {\partial{\hat{y}_{4}}}
\end{eqnarray}
and
\begin{eqnarray}
       \hat{D}\,
       \vec{\hat{y}}
        &=&
                 \left( \begin{array}{c}
                   0      \\
                   F_{2}(\vec{\hat{y}})   \\
                   0      \\
                   F_{4}(\vec{\hat{y}})   \\
                   0      \\
                   0
              \end{array}
       \right)\ ;\ \ \
       \hat{D}^{\,\nu}\,
       \vec{\hat{y}}
       \ =\
                 \left( \begin{array}{c}
                   0      \\
                   0      \\
                   0      \\
                   0      \\
                   0      \\
                   0
              \end{array}
       \right)
          \ \ \
       \mbox{for}\ \
       \nu\;>\;1
\end{eqnarray}
\begin{eqnarray}
   \Longrightarrow\ \
        \left\{
               \exp{
                    \left[
                            {\hat{D}}
                             \right]
                               }
                       \right\}\,
              \vec{\hat{y}}
               &=&
              \vec{\hat{y}}
                +
              \hat{D}\,
              \vec{\hat{y}}\ .
\end{eqnarray}
\bigskip

\subsubsection{Thin\,-\,Lens Transport Map}

\ \ \ \ \
{}From (4.18) we obtain\,:
\begin{eqnarray*}
    x^{f}
            &=&
               x^{i}\ ;
\\      \nonumber
 \\
    p_{x}^{\,f}
            &=&
    p_{x}^{\,i}
                -\,
                \frac{1}{6}\,
                \mu(s_{0})
                     \cdot\Delta s
          \cdot\left[
                (x^{i})^{3}
                -3\,(x^{i})\,(z^{i})^{2}\right]\ ;
\\      \nonumber
 \\
    z^{f}
            &=&
               z^{i}\ ;
\\      \nonumber
 \\
    p_{z}^{\,f}
            &=&
    p_{z}^{\,i}
                -\,
                \frac{1}{6}\,
                \mu(s_{0})
                     \cdot\Delta s
          \cdot\left[
                (z^{i})^{3}
                -3\,(x^{i})^{2}\,(z^{i})\right]\ ;
\\      \nonumber
 \\
    \sigma^{f}
            &=&
               \sigma^{i}\ ;
\\      \nonumber
 \\
    p_{\sigma}^{\,f}
            &=&
    p_{\sigma}^{\,i}\ .
\end{eqnarray*}
\bigskip

\subsection{Synchrotron\,-\,Magnet}

\subsubsection{Exponentiation}

\ \ \ \ \
For a synchrotron magnet
we have\,:
\begin{eqnarray} \
      g&\neq&0\,; \ \
       K_{x}^{2}+K_{z}^{2}\ \neq\ 0\ \ \mbox{with}\ \
       K_{x}\cdot K_{z}\ =\ 0
\end{eqnarray}
and
\begin{eqnarray*}
              N&=&\lambda\ =\ \mu\ =\ H\ =\ V
                                    \ =\ 0\ .
\end{eqnarray*}

We thus obtain from (3.16) and (3.29b)\,:
\bigskip
\begin{eqnarray*}
    F_{1}
         (\vec{y})
                    &=&
      0\ ;
\\      \nonumber
 \\
    F_{2}
         (\vec{y})
                    &=&
      -G_{1}(s_{0})
               \cdot\Delta s\cdot x
      +K_{x}(s_{0})
               \cdot\Delta s
               \cdot f(p_{\sigma})\ ;
\\      \nonumber
 \\
    F_{3}
         (\vec{y})
                    &=&
      0\ ;
\\      \nonumber
 \\
    F_{4}
         (\vec{y})
                    &=&
      -G_{2}(s_{0})
               \cdot\Delta s\cdot z
      +K_{z}(s_{0})
               \cdot\Delta s
               \cdot f(p_{\sigma})\ ;
\\      \nonumber
 \\
    F_{5}
         (\vec{y})
                    &=&
    \left[
       K_{x}(s_{0})\cdot x
             +
       K_{z}(s_{0})\cdot z
               \right]
               \cdot\Delta s
               \cdot f'(p_{\sigma})\ ;
\\      \nonumber
 \\
    F_{6}
         (\vec{y})
                    &=&
      0\ ;
 \\      \nonumber
 \\      \nonumber
& &
\hspace*{1.0cm}
(\,G_{1}\,=\,K_{x}^{2}+g\,;\;
 G_{2}\,=\,K_{z}^{2}-g\,)\ .
\end{eqnarray*}

Thus\,:
\begin{eqnarray}
       \hat{D}
        &=&
    F_{2}
        (\vec{\hat{y}})
         \cdot
                \frac{\partial\ }
                     {\partial{\hat{y}_{2}}}
          +
    F_{4}
        (\vec{\hat{y}})
         \cdot
                \frac{\partial\ }
                     {\partial{\hat{y}_{4}}}
          +
    F_{5}
        (\vec{\hat{y}})
         \cdot
                \frac{\partial\ }
                     {\partial{\hat{y}_{5}}}
\end{eqnarray}
and
\begin{eqnarray}
       \hat{D}\,
       \vec{\hat{y}}
        &=&
                 \left( \begin{array}{c}
                   0      \\
                   F_{2}(\vec{\hat{y}})  \\
                   0      \\
                   F_{4}(\vec{\hat{y}})  \\
                   F_{5}(\vec{\hat{y}})  \\
                   0
              \end{array}
       \right)
       \ =\
       \underline{\hat{A}}\,
       \vec{\hat{y}}
\end{eqnarray}
with
\begin{eqnarray}
   \underline{\hat{A}}
               &=&
   \Delta s
       \cdot
                 \left( \begin{array}{cccccc}
                   0  &  0
                            & 0                 &  0  &  0  &  0
                                                                  \\
   -G_{1}
                      &  0
                            &  0  & 0 &  0  &
                            K_{x}\cdot f(\hat{p}_{\sigma})
                                                      \\
    0                 &  0
                            &  0  &  0  &  0  &  0
                                                    \\
                   0  &  0
       &
          -G_{2}
                          &  0    &  0  &
                            K_{z}\cdot f(\hat{p}_{\sigma})
                                                    \\
           K_{x}\cdot f'(\hat{p}_{\sigma})
                   &  0    &
           K_{z}\cdot f'(\hat{p}_{\sigma})
                                      &  0 &  0  & 0
                                                                   \\
                0  &  0  &  0  &  0 &0 & 0
              \end{array}
       \right)\ .
\end{eqnarray}

The transfer matrix reads as\,:
\begin{eqnarray}
   \underline{M}
               &=&
               \exp{
                    \left[
                  \underline{\hat{A}}
                             \right]
                               }
\nonumber   \\
               &=&
                  \underline{1}
                       +
                  \underline{\hat{A}}
\end{eqnarray}
since
\begin{eqnarray*}
       \hat{D}\,
   \underline{\hat{A}}
               &=&
   \underline{\hat{A}}\,
       \hat{D}
\end{eqnarray*}
and
\begin{eqnarray*}
       \underline{\hat{A}}^{\,\nu}
               &=&
       \underline{0}\ \
       \mbox{for}\ \
       \nu\;>\;1\ .
\end{eqnarray*}
\bigskip

\subsubsection{Thin\,-\,Lens Transport Map}

\ \ \ \ \
{}From (4.23) we obtain\,:
\begin{eqnarray*}
    x^{f}
            &=&
               x^{i}\ ;
\\      \nonumber
 \\
    p_{x}^{\,f}
            &=&
    p_{x}^{\,i}
                -G_{1}(s_{0})
                     \cdot\Delta s
                          \cdot x^{i}
                         +K_{x}(s_{0})
                     \cdot\Delta s
                               \cdot f(p_{\sigma}^{\,i})\ ;
\\      \nonumber
 \\
    z^{f}
            &=&
               z^{i}\ ;
\\      \nonumber
 \\
    p_{z}^{\,f}
            &=&
    p_{z}^{\,i}
                -G_{2}(s_{0})
                     \cdot\Delta s
                          \cdot z^{i}
                         +K_{z}(s_{0})
                     \cdot\Delta s
                               \cdot f(p_{\sigma}^{\,i})\ ;
\\      \nonumber
 \\
    \sigma^{f}
            &=&
               \sigma^{i}
               -[K_{x}(s_{0}\cdot x+
                 K_{z}(s_{0})\cdot z]
                     \cdot\Delta s
                \cdot f'(p_{\sigma}^{\,i})\ ;
\\      \nonumber
 \\
    p_{\sigma}^{\,f}
            &=&
    p_{\sigma}^{\,i}\ .
\end{eqnarray*}
\bigskip

\subsection{Solenoid}

\subsubsection{Exponentiation}

\ \ \ \ \
For a solenoid we have\,:
\begin{eqnarray*}
      H&\neq&0
\end{eqnarray*}
and
\begin{eqnarray*}
          K_{x}&=&K_{z}\ =\ g\ =\ N\ =\ \lambda\ =\ \mu\ =\ V
                                          \ =\ 0\ .
\end{eqnarray*}

Using then (3.16) and (3.29b)
we obtain\,:
\setcounter{INDEX}{1}
\begin{eqnarray}
    F_{1}
         (\vec{y})
                    &=&
     +\frac{
                H(s_{0})
              \cdot\Delta s
                            }
              {
               \left[1+
                  f(p_{\sigma})\right]
                               }
                         \cdot z\ ;
 \\       \nonumber
 \\
   \addtocounter{equation}{-1}
\addtocounter{INDEX}{1}
    F_{2}
         (\vec{y})
                        &=&
     +\frac{
                 H(s_{0})
               \cdot\Delta s
                             }
               {
                \left[1+
                   f(p_{\sigma})\right]
                                }
               \cdot
        \left[p_{z}-H(s_{0})\cdot x\right]\ ;
 \\       \nonumber
 \\
   \addtocounter{equation}{-1}
\addtocounter{INDEX}{1}
    F_{3}
         (\vec{y})
                   &=&
     -\frac{
                 H(s_{0})
               \cdot\Delta s
                             }
               {
                \left[1+
                   f(p_{\sigma})\right]
                                }
                         \cdot x\ ;
 \\       \nonumber
 \\
   \addtocounter{equation}{-1}
\addtocounter{INDEX}{1}
    F_{4}
         (\vec{y})
                       &=&
     -\frac{
                 H(s_{0})
               \cdot\Delta s
                             }
               {
                \left[1+
                   f(p_{\sigma})\right]
                                }
               \cdot
        \left[p_{x}+H(s_{0})\cdot z\right]\ ;
 \\       \nonumber
 \\
   \addtocounter{equation}{-1}
\addtocounter{INDEX}{1}
    F_{5}
         (\vec{y})
                        &=&
     -\frac{
                 H(s_{0})
               \cdot\Delta s
                             }
               {
                \left[1+
                   f(p_{\sigma})\right]
                                }
       \cdot
      \frac{
            f'(p_{\sigma})
                            }
               {
                \left[1+
                   f(p_{\sigma})\right]
                                }
      \cdot
      \left\{
          \frac{1}{2}\,
          H(s_{0})
                  \cdot[x^{2}+
                            z^{2}]
             +
              [p_{x}\cdot z-p_{z}\cdot x]
                        \right\}\ ;
 \\       \nonumber
 \\
   \addtocounter{equation}{-1}
\addtocounter{INDEX}{1}
    F_{6}
         (\vec{y})
                      &=&
                       0\ .
\end{eqnarray}
\setcounter{INDEX}{0}

Thus\,:
\begin{eqnarray}
       \hat{D}
        &=&
    F_{1}
         (\vec{\hat{y}})
         \cdot
                \frac{\partial\ }
                     {\partial{\hat{y}_{1}}}
          +
    F_{2}
         (\vec{\hat{y}})
         \cdot
                \frac{\partial\ }
                     {\partial{\hat{y}_{2}}}
          +
    F_{3}
         (\vec{\hat{y}})
         \cdot
                \frac{\partial\ }
                     {\partial{\hat{y}_{3}}}
          +
    F_{4}
         (\vec{\hat{y}})
         \cdot
                \frac{\partial\ }
                     {\partial{\hat{y}_{4}}}
          +
    F_{5}
         (\vec{\hat{y}})
         \cdot
                \frac{\partial\ }
                     {\partial{\hat{y}_{5}}}
\end{eqnarray}
and
\setcounter{INDEX}{1}
\begin{eqnarray}
       \hat{D}\,
                 \left( \begin{array}{c}
                   \hat{x}         \\
                   \hat{p}_{x}     \\
                   \hat{z}         \\
                   \hat{p}_{z}
              \end{array}
       \right)
        &=&
                 \left( \begin{array}{c}
                   F_{1}(\vec{\hat{y}})  \\
                   F_{2}(\vec{\hat{y}})  \\
                   F_{3}(\vec{\hat{y}})  \\
                   F_{4}(\vec{\hat{y}})
              \end{array}
       \right)
       \ =\
       \underline{\hat{A}}_{\,0}\,
                 \left( \begin{array}{c}
                   \hat{x}         \\
                   \hat{p}_{x}     \\
                   \hat{z}         \\
                   \hat{p}
              \end{array}
       \right)\ ;
 \\         \nonumber
 \\         \nonumber
 \\
   \addtocounter{equation}{-1}
\addtocounter{INDEX}{1}
       \hat{D}\,
       \hat{\sigma}
        &=&
       F_{5}(\vec{\hat{y}})\ ;
 \\         \nonumber
 \\
   \addtocounter{equation}{-1}
\addtocounter{INDEX}{1}
       \hat{D}\,
       \hat{p}_{\sigma}
        &=&
         0
    \ \ \Longrightarrow\ \
         \left\{
               \exp{
                    \left[
                            {\hat{D}}
                             \right]
                               }
                      \right\}\,
       \hat{p}_{\sigma}
       \ =\
       \hat{p}_{\sigma}
\end{eqnarray}
\setcounter{INDEX}{0}
with
\begin{eqnarray}
   \underline{\hat{A}}_{\,0}
               &=&
   \Delta s
       \cdot
          \frac{1}
               {
                \left[1+
                   f(\hat{p}_{\sigma})\right]
                                }
       \cdot
                 \left( \begin{array}{cccc}
                   0  &  0
                            & +H                 &  0   \\
   -H^{2}             &  0
                            &  0  & +H       \\
   -H                 &  0
                            &  0  &  0   \\
                   0  &  -H
       &  -H^{2}             &  0            \\
              \end{array}
       \right)\ .
\end{eqnarray}

We decompose\,
the matrix\,
  $\underline{\hat{A}}_{\,0}$\,
into the components\,
  $\underline{\hat{A}}_{\,01}$\,
and\,
  $\underline{\hat{A}}_{\,02}$\,:
\begin{eqnarray}
   \underline{\hat{A}}_{\,0}
          &=&
   \underline{\hat{A}}_{\,01}
           +
   \underline{\hat{A}}_{\,02}
\end{eqnarray}
with
\setcounter{INDEX}{1}
\begin{eqnarray}
   \underline{\hat{A}}_{\,01}
          &=&
   \Delta s
       \cdot
          \frac{H^{2}}
               {
                \left[1+
                   f(\hat{p}_{\sigma})\right]
                                }
       \cdot
                 \left( \begin{array}{cccc}
                   0  &  0
                            & 0                 &  0    \\
   -1                 &  0
                            &  0  & 0       \\
   0                  &  0
                            &  0  &  0    \\
                   0  &  0
       &  -1                 &  0
              \end{array}
       \right)
\end{eqnarray}
and
\begin{eqnarray}
   \addtocounter{equation}{-1}
\addtocounter{INDEX}{1}
   \underline{\hat{A}}_{\,02}
          &=&
   \Delta s
       \cdot
          \frac{H}
               {
                \left[1+
                   f(\hat{p}_{\sigma})\right]
                                }
       \cdot
                 \left( \begin{array}{cccccc}
                   0  &  0
                            & +1                 &  0    \\
   0                  &  0
                            &  0  & +1      \\
   -1                 &  0
                            &  0  &  0    \\
                   0  &  -1
       &  0                  &  0
              \end{array}
       \right)\ .
\end{eqnarray}
\setcounter{INDEX}{0}

The transfer matrix
for
\begin{eqnarray}
       \vec{y}_{0}
           &=&
                 \left( \begin{array}{c}
                       {x}         \\
                       {p}_{x}     \\
                       {z}         \\
                       {p}
              \end{array}
       \right)
\end{eqnarray}
reads as\,:
\begin{eqnarray}
   \underline{M}_{0}
               &=&
               \exp{
                    \left[
                  \underline{\hat{A}}_{\,0}
                             \right]
                               }
\end{eqnarray}
since
\begin{eqnarray*}
       \hat{D}\,
   \underline{\hat{A}}_{\,0}
               &=&
   \underline{\hat{A}}_{\,0}\,
       \hat{D}
\ \ \Longrightarrow\ \
       \hat{D}^{\,\nu}\,
       \vec{\hat{y}}_{0}
       \ =\
       \underline{\hat{A}}^{\,\nu}
       \vec{\hat{y}}_{0}
\ \ \Longrightarrow\ \
         \left\{
               \exp{
                    \left[
                            {\hat{D}}
                             \right]
                               }
                      \right\}
       \vec{\hat{y}}_{0}
       \ =\
         \left\{
               \exp{
                    \left[
                  \underline{\hat{A}}_{0}
                             \right]
                               }
                      \right\}
       \vec{\hat{y}}_{0}\ .
\end{eqnarray*}
\bigskip

Using the relations\,:
\begin{eqnarray*}
   \underline{\hat{A}}_{\,01}
          \cdot
   \underline{\hat{A}}_{\,02}
            &=&
   \underline{\hat{A}}_{\,02}
          \cdot
   \underline{\hat{A}}_{\,01}
\ \ \Longrightarrow\ \
               \exp{
                    \left[
                  \underline{\hat{A}}
                             \right]
                                    }
                     \ =\
               \exp{
                    \left[
                  \underline{\hat{A}}_{\,01}
                             \right]
                                    }
               \cdot
               \exp{
                    \left[
                  \underline{\hat{A}}_{\,02}
                             \right]
                                    }
\end{eqnarray*}
and
\begin{eqnarray*}
    \left[
       \underline{\hat{A}}_{01}
                             \right]
                               ^{\,\nu}
               &=&
       \underline{0}\ \
       \mbox{for}\ \ \
       \nu\;>\;1
\ \ \Longrightarrow\ \
               \exp{
                    \left[
                  \underline{\hat{A}}_{\,01}
                             \right]
                                    }
                     \ =\
                  \underline{1}
                       +
                  \underline{\hat{A}}_{\,01}
\end{eqnarray*}
as well as
\begin{eqnarray*}
& &
                 \left( \begin{array}{cccccc}
                   0  &  0
                            & 1                 &  0    \\
   0                  &  0
                            &  0  & 1      \\
   -1                 &  0
                            &  0  &  0    \\
                   0  &  -1
       &  0                  &  0
              \end{array}
       \right)^{\,2n}
            \ =\
       (-1)^{n}
       \cdot
       \underline{1}\ ;
\\
\\
& &
                 \left( \begin{array}{cccccc}
                   0  &  0
                            & 1                 &  0    \\
   0                  &  0
                            &  0  & 1      \\
   -1                 &  0
                            &  0  &  0    \\
                   0  &  -1
       &  0                  &  0
              \end{array}
       \right)^{\,2n+1}
            \ =\
       (-1)^{n}
       \cdot
                 \left( \begin{array}{cccccc}
                   0  &  0
                            & 1                 &  0    \\
   0                  &  0
                            &  0  & 1      \\
   -1                 &  0
                            &  0  &  0    \\
                   0  &  -1
       &  0                  &  0
              \end{array}
       \right)
\end{eqnarray*}
we get\,:
\begin{eqnarray*}
& &
               \exp{
                    \left[
                  \underline{\hat{A}}_{\,02}
                             \right]
                                    }
                     \ =\
\\
\\
& &
        \sum_{n,\,=\,0}^{\infty}\,
     \frac{1}{(2n)!}\cdot
     (-1)^{n}\cdot
     (\Delta\Theta)^{2n}
           \cdot
                  \underline{1}
\\
\\
& &
\hspace*{1.0cm}
                       +
        \sum_{n,\,=\,0}^{\infty}\,
     \frac{1}{(2n+1)!}\cdot
     (-1)^{n}\cdot
     (\Delta\Theta)^{2n+1}
           \cdot
                 \left( \begin{array}{cccccc}
                   0  &  0
                            & +1                 &  0    \\
   0                  &  0
                            &  0  & +1      \\
   -1                 &  0
                            &  0  &  0    \\
                   0  &  -1
       &  0                  &  0
              \end{array}
       \right)
\\
\\
& &
           \ =\
          \underline{1}
     \cdot
     \cos(\Delta\Theta)
                       +
                 \left( \begin{array}{cccccc}
                   0  &  0
                            & +1                 &  0    \\
   0                  &  0
                            &  0  & +1      \\
   -1                 &  0
                            &  0  &  0    \\
                   0  &  -1
       &  0                  &  0
              \end{array}
       \right)
           \cdot
     \sin(\Delta\Theta)
\\
\\
& &
           \ =\
                 \left( \begin{array}{cccccc}
     \cos(\Delta\Theta)
                      &  0
                            &
                              +\sin(\Delta\Theta)
                                                 &  0    \\
   0                  &
     \cos(\Delta\Theta)
                            &  0  &
                              +\sin(\Delta\Theta)
                                            \\
   -\sin(\Delta\Theta)
                      &  0
                            &
     \cos(\Delta\Theta)
                                  &  0    \\
                   0  &
                         -\sin(\Delta\Theta)
       &  0                  &
     \cos(\Delta\Theta)
              \end{array}
       \right)
\end{eqnarray*}
with
\begin{eqnarray}
          \Delta\Theta
          &=&
          \frac{
                H
               \cdot\Delta s
                             }
               {
                \left[1+
                   f(\hat{p}_{\sigma})\right]
                                }\ .
\end{eqnarray}
Therefore\,:
\begin{eqnarray}
                  \underline{M}_{0}
                      &=&
        \left[
                  \underline{1}
                       +
                  \underline{\hat{A}}_{\,01}
                        \right]
             \cdot
                 \left( \begin{array}{cccccc}
     \cos(\Delta\Theta)
                      &  0
                            &
                              +\sin(\Delta\Theta)
                                                 &  0    \\
   0                  &
     \cos(\Delta\Theta)
                            &  0  &
                              +\sin(\Delta\Theta)
                                            \\
   -\sin(\Delta\Theta)
                      &  0
                            &
     \cos(\Delta\Theta)
                                  &  0    \\
                   0  &
                         -\sin(\Delta\Theta)
       &  0                  &
     \cos(\Delta\Theta)
              \end{array}
       \right)\ .
\end{eqnarray}

For the variable\,
      $\sigma$\,
we get from (4.26b)\,:
\begin{eqnarray*}
   {\hat{D}}^{2}\,
    \hat{\sigma}
        &=&
           {\hat{D}}\,
       F_{5}(\vec{\hat{y}})
 \\
 \\
        &=&
  \left\{
    F_{1}
         (\vec{\hat{y}})
         \cdot
                \frac{\partial\ }
                     {\partial{\hat{y}_{1}}}
          +
    F_{2}
         (\vec{\hat{y}})
         \cdot
                \frac{\partial\ }
                     {\partial{\hat{y}_{2}}}
          +
    F_{3}
         (\vec{\hat{y}})
         \cdot
                \frac{\partial\ }
                     {\partial{\hat{y}_{3}}}
          +
    F_{4}
         (\vec{\hat{y}})
         \cdot
                \frac{\partial\ }
                     {\partial{\hat{y}_{4}}}
               \right\}\,
       F_{5}(\vec{\hat{y}})
 \\
 \\
        &=&
          \frac{
                H
               \cdot\Delta s
                             }
               {
                \left[1+
                   f(\hat{p}_{\sigma})\right]
                                }\,
    \left\{
    \hat{y}_{3}
         \cdot
                \frac{\partial\ }
                     {\partial{\hat{y}_{1}}}
          -
    \left[H\cdot
                 \hat{y}_{1}
                 -
                 \hat{y}_{4}
                            \right]
         \cdot
                \frac{\partial\ }
                     {\partial{\hat{y}_{2}}}
                 -
    \hat{y}_{1}
         \cdot
                \frac{\partial\ }
                     {\partial{\hat{y}_{3}}}
          -
    \left[H\cdot
                 \hat{y}_{3}
                 +
                 \hat{y}_{2}
                            \right]
         \cdot
                \frac{\partial\ }
                     {\partial{\hat{y}_{4}}}
                                           \right\}
 \\
 \\
& &
\hspace*{2.0cm}
       \left\{
          \frac{
                (-H)
               \cdot f'(\hat{p}_{\sigma})
               \cdot\Delta s
                             }
               {
                \left[1+
                   f(\hat{p}_{\sigma})\right]^{2}
                                }\,
       \left[
              \frac{1}{2}\,H\cdot
              (
               \hat{y}_{1}^{2}
                      +
               \hat{y}_{3}^{2}
                              )
                      +
       \left(
             \hat{y}_{2}\cdot
             \hat{y}_{3}
                      -
             \hat{y}_{4}\cdot
             \hat{y}_{1}
                               \right)
                          \right]
                          \right\}
 \\
 \\
                      &=&
          \frac{
                H
               \cdot\Delta s
                             }
               {
                \left[1+
                   f(\hat{p}_{\sigma})\right]
                                }\cdot
          \frac{
                (-H)
               \cdot f'(\hat{p}_{\sigma})
               \cdot\Delta s
                             }
               {
                \left[1+
                   f(\hat{p}_{\sigma})\right]^{2}
                                }
 \\
 \\
& &
         \times\,
       \left\{
               \hat{y}_{3}
                          \cdot
              [
               H\cdot\hat{y}_{1}
                      -
               \hat{y}_{4}
                              ]
                      -
              [
               H\cdot\hat{y}_{1}
                      -
               \hat{y}_{4}
                              ]
                          \cdot
               \hat{y}_{3}
                      -
               \hat{y}_{1}
                          \cdot
              [
               H\cdot\hat{y}_{3}
                      +
               \hat{y}_{2}
                              ]
                      +
              [
               H\cdot\hat{y}_{3}
                      +
               \hat{y}_{2}
                              ]
                          \cdot
               \hat{y}_{1}
                          \right\}
 \\
 \\
                      &=&
                       0\ ;
\end{eqnarray*}
\begin{eqnarray}
        \Longrightarrow\ \
         \left\{
               \exp{
                    \left[
                            {\hat{D}}
                             \right]
                               }
                      \right\}
       \hat{\sigma}
        &=&
       \hat{\sigma}
           +
       F_{5}(\vec{\hat{y}})
           +
       \frac{1}{2!}\cdot
           {\hat{D}}\,
       F_{5}(\vec{\hat{y}})
           +
      \cdot
      \cdot
      \cdot
 \nonumber    \\
 \nonumber    \\
        &=&
       \hat{\sigma}
           +
       F_{5}(\vec{\hat{y}})\ .
\hspace*{7.0cm}
\end{eqnarray}
\vspace*{3.0ex}
\underline{Remarks:}

1) We have used a separation of the matrix\,
      $\underline{\hat{A}}_{0}$
into the components\,
      $\underline{\hat{A}}_{01}$\,
and\,
      $\underline{\hat{A}}_{02}$\,
with
\begin{eqnarray*}
       \underline{\hat{A}}_{01}
              \cdot
       \underline{\hat{A}}_{02}
              &=&
       \underline{\hat{A}}_{02}
              \cdot
       \underline{\hat{A}}_{01}
\end{eqnarray*}
and
\begin{eqnarray*}
       \underline{\hat{A}}_{01}\ \
      \mbox{nilpotent}\ .
\end{eqnarray*}
Since in addition\,
      $\underline{\hat{A}}_{02}$\,
is diagonalizable,
eqn. (4.28) represents an additive
Jordan decomposition.
\\

2) The decomposition (4.28) factors
       $\underline{M}_{0}$
into a rotation map and a focussing map.
\bigskip

\subsubsection{Thin\,-\,Lens Transport Map}

Equations (4.31), (4.34) and (4.26c)
finally lead to\,:
\setcounter{INDEX}{1}
\begin{eqnarray}
        \vec{y}_{0}^{\;f}
             &=&
        \underline{M}_{0}\,
        \vec{y}_{0}^{\;i}\ ;
 \\   \nonumber
 \\
   \addtocounter{equation}{-1}
\addtocounter{INDEX}{1}
          \sigma^{f}
             &=&
                   {\sigma}^{i}
         -\frac{f'({p}_{\sigma}^{\,i})}
               {
                \left[1+
                   f({p}_{\sigma}^{\,i})\right]
                                }
                    \cdot
                \Delta\Theta
\nonumber     \\
& &
\hspace*{1.0cm}
      \times
  \left\{
          \frac{1}{2}\,
                   H(s_{0})
                   \cdot
           \left[
                ({x}^{i})^{2}
               +({z}^{i})^{2}
                                \right]
                   +
    \left[
    {p}_{x}^{\,i}
                   \cdot
            {z}^{i}
                 -
    {p}_{z}^{\,i}
                   \cdot
            {x}^{i}
                      \right]
                 \right\}\ ;
 \\
   \addtocounter{equation}{-1}
\addtocounter{INDEX}{1}
          p_{\sigma}^{\,f}
             &=&
          p_{\sigma}^{\,i}
\end{eqnarray}
with\,
\begin{eqnarray}
\addtocounter{INDEX}{1}
   \addtocounter{equation}{-1}
      \Delta\Theta
           &=&
      \frac{
               H(s_{0})\cdot\Delta s
             }
               {
                \left[1+
                   f({p}_{\sigma}^{\,i})\right]
                                }
\end{eqnarray}
and
       $\underline{M}_{0}$\,
given by (4.33),
where the matrix\,
  $\underline{\hat{A}}_{\,01}$\,
appearing in (4.33) takes the form\,:
\begin{eqnarray}
\addtocounter{INDEX}{1}
   \addtocounter{equation}{-1}
   \underline{\hat{A}}_{\,01}
          &=&
   \Delta\Theta
       \cdot
          H(s_{0})
       \cdot
                 \left( \begin{array}{cccc}
                   0  &  0
                            & 0                 &  0    \\
   -1                 &  0
                            &  0  & 0       \\
   0                  &  0
                            &  0  &  0    \\
                   0  &  0
       &  -1                 &  0
              \end{array}
       \right)
\end{eqnarray}
\setcounter{INDEX}{0}
(see eqn. (4.29a)\,).
\bigskip

In Appendix B the superposition
of a solenoid with a quadrupole
is investigated.
\bigskip

\subsection{Cavity}

\subsubsection{Exponentiation}

\ \ \ \ \
For a cavity we have\,:
\begin{eqnarray*}
       V&\neq&0
\end{eqnarray*}
and
\begin{eqnarray*}
      K_{x}&=&K_{z}\ =\ g\ =\ N\ =\ \lambda\ =\ \mu\ =\ H
                                          \ =\ 0\ .
\end{eqnarray*}

Then we obtain from (3.16) and (3.29b)\,:
\begin{eqnarray*}
    F_{1}
         (\vec{y})
                    &=&
      0\ ;
\\      \nonumber
 \\
    F_{2}
         (\vec{y})
                    &=&
      0\ ;
\\      \nonumber
 \\
    F_{3}
         (\vec{y})
                    &=&
      0\ ;
\\      \nonumber
 \\
    F_{4}
         (\vec{y})
                    &=&
      0\ ;
\\      \nonumber
 \\
    F_{5}
         (\vec{y})
                    &=&
      0\ ;
\\      \nonumber
 \\
    F_{6}
         (\vec{y})
                    &=&
              \frac{1}{\beta_{0}^{2}}
                         \cdot\frac{eV(s_{0})}{E_{0}}\cdot
                   \sin\left[
                        h\cdot\frac{2\pi}{L}\cdot\sigma
                              +
                             \varphi
                                    \right]
       \cdot
   \Delta s\ .
\end{eqnarray*}

Thus\,:
\begin{eqnarray}
       \hat{D}
        &=&
    F_{6}
        (\vec{\hat{y}})
         \cdot
                \frac{\partial\ }
                     {\partial{\hat{y}_{6}}}
\end{eqnarray}
and
\begin{eqnarray}
       \hat{D}\,
       \vec{\hat{y}}
        &=&
                 \left( \begin{array}{c}
                   0      \\
                   0      \\
                   0      \\
                   0      \\
                   0      \\
                   F_{6}(\vec{\hat{y}})
              \end{array}
       \right)\ ;\ \ \
       \hat{D}^{\,\nu}\,
       \vec{\hat{y}}
       \ =\
                 \left( \begin{array}{c}
                   0      \\
                   0      \\
                   0      \\
                   0      \\
                   0      \\
                   0
              \end{array}
       \right)
          \ \ \
       \mbox{for}\ \
       \nu\;>\;1
\end{eqnarray}
\begin{eqnarray}
   \Longrightarrow\ \
        \left\{
               \exp{
                    \left[
                            {\hat{D}}
                             \right]
                               }
                       \right\}\,
              \vec{\hat{y}}
               &=&
              \vec{\hat{y}}
                +
              \hat{D}\,
              \vec{\hat{y}}\ .
\end{eqnarray}
\bigskip

\subsubsection{Thin\,-\,Lens Transport Map}
\bigskip

\ \ \ \ \
{}From (4.37) and (4.38)
we obtain\,:
\begin{eqnarray*}
    x^{f}
            &=&
               x^{i}\ ;
\\      \nonumber
 \\
    p_{x}^{\,f}
            &=&
    p_{x}^{\,i}\ ;
\\      \nonumber
 \\
    z^{f}
            &=&
               z^{i}\ ;
\\      \nonumber
 \\
    p_{z}^{\,f}
            &=&
    p_{z}^{\,i}\ ;
\\      \nonumber
 \\
    \sigma^{f}
            &=&
               \sigma^{i}\ ;
\\      \nonumber
 \\
    p_{\sigma}^{\,f}
            &=&
    p_{\sigma}^{\,i}
             +\frac{1}{\beta_{0}^{2}}
                         \cdot\frac{eV(s_{0})}{E_{0}}
                             \cdot
                   \sin\left[
                        h\cdot\frac{2\pi}{L}\cdot\sigma^{i}
                              +
                             \varphi
                                    \right]
                     \cdot\Delta s\ .
\end{eqnarray*}
\bigskip

\section{Summary}

\ \ \ \ \
As a continuation of Ref.
\cite{RS},
we have shown how to solve the nonlinear
canonical equations of motion
in the framework of the fully
six--dimensional formalism
using Lie series and exponentiation.
Various kinds of magnets
(bending magnets, quadrupoles, synchrotron magnets,
skew quadrupoles, sextupoles, octupoles, solenoids)
and cavities are treated,
taking into account the energy dependence
of the focusing strength.

In Appendix A
we have introduced an improved Hamiltonian (Appendix A).
which is exact outside bending magnets
and solenoids.

Since the equations of motion are canonical,
the transport maps obtained are automatically symplectic.

The equations derived are valid for arbitrary particle velocity,
i.e. below and above transition energy,
and have been incorporated into the computer program
SIXTRACK.
\bigskip

\section*{Acknowledgments}
\addcontentsline{toc}{section}{
          Acknowledgments}
\ \ \ \ \
We wish to thank D.~P.~Barber
for very stimulating and interesting discussions
and for carefully reading
the manuscript.
\bigskip
%
%

\setcounter{section}{1}
\setcounter{subsection}{0}
\setcounter{equation}{0}
\renewcommand{\thesection}{\Alph{section}}
\addcontentsline{toc}{section}{Appendix A:
                      Thin\,-\,Lens Approximation
                      with an Improved Hamiltonian}
\section*{Appendix A:
                      Thin\,-\,Lens Approximation
                      with an Improved Hamiltonian}

\ \ \ \ \
In chapter 4
we have used an approximate
Hamiltonian which is obtained by a series expansion
of the square root\,
\begin{eqnarray*}
                           \left\{1-
          \frac{
                [p_{x}+H\cdot z]^{2}+
                [p_{z}-H\cdot x]^{2}
                                                   }
               {
                \left[1+
                   f(p_{\sigma})\right]^{2}
                                }
                                   \right\}^{1/2}
\end{eqnarray*}
up to first order in terms of the quantity
\begin{eqnarray*}
          \frac{
                [p_{x}+H\cdot z]^{2}+
                [p_{z}-H\cdot x]^{2}
                                      }
               {
                \left[1+
                   f(p_{\sigma})\right]^{2}
                                }\ .
\end{eqnarray*}

The aim of this Appendix is to repeat
the calculations of chapter 4
in the absence of solenoids
with an improved Hamiltonian by using
the unexpanded square root.
The new Hamiltonian is again decomposed into a lens part\,
        ${\cal{H}}_{L}$\,
and a drift component\,
        ${\cal{H}}_{D}$.
Then we
apply the thin\,-\,lens approximation
described in chapter 3 by defining the
modified Hamiltonian
\begin{eqnarray*}
             {\cal{H}}_{mod}
               &=&
             {\cal{H}}_{D}
                +
             \hat{\cal{H}}_{L}
        \cdot
   \Delta s
        \cdot
   \delta(s-s_{0})
\end{eqnarray*}
(see eqn. (3.12)\,)
and solve the equations of motion resulting from the improved
Hamiltonian for the regions I, II, and III
of eqn. (3.13).
It is shown that the thin\,-\,lens maps
for the central region II
obtained earlier
remain valid
and that corrections induced by the new Hamiltonian
only appear in the solutions
of the
       {\it drift}
             spaces (regions I and III).

The improved Hamiltonian introduced in this Appendix
is exact outside the bending magnets.
Inside a bending magnet we neglect nonlinear
crossing terms resulting from the curvature.
These terms are investigated in Appendix B.
\bigskip

\subsection{The Improved Hamiltonian}
\vspace*{0.5ex}

\ \ \ \ \
In the absence of solenoids the
Hamiltonian for orbital motion
in storage rings takes the form\,:
\begin{eqnarray}
    {\cal{H}}(x,p_{x},z,p_{z},\sigma,p_{\sigma};s)
               &=&
            p_{\sigma}
               -\left[1+
                   f(p_{\sigma})\right]
                 \cdot
                        [1+K_{x}\cdot x+K_{z}\cdot z]
                 \cdot
                           \left\{1-
          \frac{
                p_{x}^{2}+
                p_{z}^{2}
                                                   }
               {
                \left[1+
                   f(p_{\sigma})\right]^{2}
                                }
                                   \right\}^{1/2}
                                    \nonumber  \\
 & &
                  +\frac{1}{2}\cdot
                              [1+K_{x}\cdot{x}+K_{z}\cdot{z}]^{2}
                  -\frac{1}{2}
                              \cdot
                              g\cdot{(z^{2}-x^{2})}
                  -N
                    \cdot{xz}
\nonumber   \\
                            & &+
    \frac{\lambda}{6}\cdot
    (x^{3}-3\,xz^{2})
\nonumber   \\
       & &+
    \frac{\mu}{24}\cdot
    (z^{4}-6\,x^{2}\,z^{2}+x^{4})
\nonumber   \\
                            & &+
                \frac{1}{\beta_{0}^{2}}\cdot
                        \frac{L}{2\pi\cdot h}\cdot
                        \frac{eV(s)}{E_{0}}\cdot\cos
                       \left[
             h\cdot\frac{2\pi}{L}\cdot\sigma+\varphi
                       \right]
\end{eqnarray}
(see eqn. (2.1)\,).
\bigskip

If we neglect
nonlinear crossing terms containing the
factors
     $K_{x}\cdot x$
and
     $K_{z}\cdot z$
we may write\,:
\begin{eqnarray*}
& &
           -\left[1+
                   f(p_{\sigma})\right]
                 \cdot
                        [1+K_{x}\cdot x+K_{z}\cdot z]
                 \cdot
              \left\{1-
          \frac{
                p_{x}^{2}+
                p_{z}^{2}
                                                   }
               {
                \left[1+
                   f(p_{\sigma})\right]^{2}
                                }
                                   \right\}^{1/2}
\nonumber  \\
               &\approx&
           -\left[1+
                   f(p_{\sigma})\right]
                 \cdot
                           \left\{1-
          \frac{
                p_{x}^{2}+
                p_{z}^{2}
                                                   }
               {
                \left[1+
                   f(p_{\sigma})\right]^{2}
                                }
                                   \right\}^{1/2}
           -\left[1+
                   f(p_{\sigma})\right]
                 \cdot
                        [K_{x}\cdot x+K_{z}\cdot z]\ .
\end{eqnarray*}

Then we obtain\,:
\begin{eqnarray}
     {\cal{H}}
              &=&
            p_{\sigma}
           -[1+f(p_{\sigma})]
                 \cdot
                           \left\{1-
          \frac{
                 p_{x}^{2}+
                 p_{z}^{2}
                                                   }
               {
                [1+f(p_{\sigma})]^{2}
                                }
                                   \right\}^{1/2}
                                    \nonumber  \\
                                    \nonumber  \\
 & &
                        -\,
            [K_{x}\cdot x+K_{z}\cdot z]
                 \cdot
            f(p_{\sigma})
                                    \nonumber  \\
                                    \nonumber  \\
 & &
                        +\,
                   \frac{1}{2}\,
                              [K_{x}^{2}+g]
                              \cdot
                              x^{2}
                  +\frac{1}{2}\,
                              [K_{z}^{2}-g]
                              \cdot
                              z^{2}
                  -N
                    \cdot{x\,z}
\nonumber   \\
\nonumber   \\
                            & &
                        +\,
    \frac{\lambda}{6}\cdot
    (x^{3}-3\,xz^{2})
          +
    \frac{\mu}{24}\cdot
    (z^{4}-6\,x^{2}\,z^{2}+x^{4})
\nonumber   \\
\nonumber   \\
                            & &
                        +\,
                \frac{1}{\beta_{0}^{2}}\cdot
                        \frac{L}{2\pi\cdot h}\cdot
                        \frac{eV(s)}{E_{0}}\cdot\cos
                       \left[
             h\cdot\frac{2\pi}{L}\cdot\sigma+\varphi
                       \right]
\end{eqnarray}
(constant terms
which have no influence on the motion
have been dropped).
\bigskip

In particular we get for
the Hamiltonian
of a drift space\ :
\begin{eqnarray}
    {\cal{H}}_{D}
                   (x,p_{x},z,p_{z},\sigma,p_{\sigma};s)
               &=&
            p_{\sigma}
           -[1+f(p_{\sigma})]
                 \cdot
                           \left\{1-
          \frac{
                 p_{x}^{2}+
                 p_{z}^{2}
                                                   }
               {
                [1+f(p_{\sigma}]^{2}
                                }
                                   \right\}^{1/2}\ ,
\end{eqnarray}
and the lens component of
   ${\cal{H}}$\,:
\begin{eqnarray*}
    {\cal{H}}_{L}
       &=&
    {\cal{H}}
        -
    {\cal{H}}_{D}
\end{eqnarray*}
then reads\,:
\begin{eqnarray}
     {\cal{H}}_{L}
                   (x,p_{x},z,p_{z},\sigma,p_{\sigma};s)
              &=&
           -\,
            [K_{x}\cdot x+K_{z}\cdot z]
                 \cdot
            f(p_{\sigma})
                                    \nonumber  \\
                                    \nonumber  \\
 & &
                 +\,
                   \frac{1}{2}\,
                              [K_{x}^{2}+g]
                              \cdot
                              x^{2}
                  +\frac{1}{2}\,
                              [K_{z}^{2}-g]
                              \cdot
                              z^{2}
                  -N
                    \cdot{x\,z}
\nonumber   \\
\nonumber   \\
                            & &
                        +\,
    \frac{\lambda}{6}\cdot
    (x^{3}-3\,xz^{2})
          +
    \frac{\mu}{24}\cdot
    (z^{4}-6\,x^{2}\,z^{2}+x^{4})
\nonumber   \\
\nonumber   \\
                            & &
                        +\,
                \frac{1}{\beta_{0}^{2}}\cdot
                        \frac{L}{2\pi\cdot h}\cdot
                        \frac{eV(s)}{E_{0}}\cdot\cos
                       \left[
             h\cdot\frac{2\pi}{L}\cdot\sigma+\varphi
                       \right]\ .
\end{eqnarray}
\bigskip

\subsection{Equations of Motion}

\subsubsection{Drift Space}

\ \ \ \ \
The Hamiltonian (A.3)
for a drift space
leads to the
canonical equations of motion\,:
\setcounter{INDEX}{1}
\begin{eqnarray}
    \frac{d}{ds}\,x
                    &=&
               +\,
                \frac{\partial{
                              {\cal{H}}_{D}
                                       }
                                       }
                                        {\partial
                          {p_{x}}}
\nonumber    \\
\nonumber    \\
                    &=&
           -[1+f(p_{\sigma})]
          \cdot\frac{1}{2}\,
                           \left\{1-
          \frac{
                 p_{x}^{2}+
                 p_{z}^{2}
                                                   }
               {
                [1+f(p_{\sigma}]^{2}
                                }
                                   \right\}^{-1/2}
          \cdot\frac{(-2\,p_{x})}
               {
                [1+f(p_{\sigma})]^{2}
                                }
\nonumber    \\
\nonumber    \\
                    &=&
           +\,
                           \left\{1-
          \frac{
                 p_{x}^{2}+
                 p_{z}^{2}
                                                   }
               {
                [1+f(p_{\sigma})]^{2}
                                }
                                   \right\}^{-1/2}
          \cdot\frac{p_{x}}
               {
                [1+f(p_{\sigma})]
                                }\ ;
\\      \nonumber
\\      \nonumber
 \\
   \addtocounter{equation}{-1}
\addtocounter{INDEX}{1}
    \frac{d}{ds}
          \,p_{x}
                        &=&
               -\frac{\partial{
                               {\cal{H}}_{D}
                                        }}{\partial
                          {x}}
\nonumber    \\
\nonumber    \\
                    &=&
                     0
                      \ \ \
         \Longrightarrow\ \ \
         p_{x}\ =\ const\ ;
\\      \nonumber
 \\
   \addtocounter{equation}{-1}
\addtocounter{INDEX}{1}
    \frac{d}{ds}\,z
                   &=&
               +\frac{\partial{
                              {\cal{H}}_{D}
                                       }
                                       }
                                        {\partial
                          {p_{z}}}
\nonumber    \\
\nonumber    \\
                    &=&
           +\,
                           \left\{1-
          \frac{
                 p_{x}^{2}+
                 p_{z}^{2}
                                                   }
               {
                [1+f(p_{\sigma})]^{2}
                                }
                                   \right\}^{-1/2}
          \cdot\frac{p_{z}}
               {
                [1+f(p_{\sigma})]
                                }\ ;
\\      \nonumber
\\      \nonumber
 \\
   \addtocounter{equation}{-1}
\addtocounter{INDEX}{1}
    \frac{d}{ds}
          \,p_{z}
                       &=&
               -\frac{\partial{
                               {\cal{H}}_{D}
                                        }}{\partial
                          {z}}
\nonumber    \\
\nonumber    \\
                    &=&
          0
                      \ \ \
         \Longrightarrow\ \ \
         p_{z}\ =\ const\ ;
\\      \nonumber
\\      \nonumber
 \\
   \addtocounter{equation}{-1}
\addtocounter{INDEX}{1}
    \frac{d}{ds}\,\sigma
                        &=&
               +\frac{\partial{
                              {\cal{H}}_{D}
                                       }
                                       }
                                        {\partial
                          {p_{\sigma}}}
\nonumber    \\
\nonumber    \\
                    &=&
           1-
                   f'(p_{\sigma})\cdot
                           \left\{1-
          \frac{
                 p_{x}^{2}+
                 p_{z}^{2}
                                                   }
               {
                [1+f(p_{\sigma})]^{2}
                                }
                                   \right\}^{1/2}
\nonumber    \\
\nonumber    \\
& &
           -[1+f(p_{\sigma})]
          \cdot\frac{1}{2}\,
                           \left\{1-
          \frac{
                 p_{x}^{2}+
                 p_{z}^{2}
                                                   }
               {
                [1+f(p_{\sigma})]^{2}
                                }
                                   \right\}^{-1/2}
          \cdot\frac{2\,[p_{x}^{2}+p_{z}^{2}]}
               {
                [1+f(p_{\sigma}])^{3}
                                }
               \cdot
               f'(p_{\sigma})
\nonumber    \\
\nonumber    \\
                    &=&
           1-
                   f'(p_{\sigma})\cdot
                           \left\{1-
          \frac{
                 p_{x}^{2}+
                 p_{z}^{2}
                                                   }
               {
                [1+f(p_{\sigma})]^{2}
                                }
                                   \right\}^{1/2}
\nonumber    \\
\nonumber    \\
& &
           -\,
                           \left\{1-
          \frac{
                 p_{x}^{2}+
                 p_{z}^{2}
                                                   }
               {
                [1+f(p_{\sigma})]^{2}
                                }
                                   \right\}^{-1/2}
          \cdot\frac{[p_{x}^{2}+p_{z}^{2}]}
               {
                [1+f(p_{\sigma})]^{2}
                                }
               \cdot
               f'(p_{\sigma})
\nonumber    \\
\nonumber    \\
                    &=&
           1-
                   f'(p_{\sigma})\cdot
                           \left\{1-
          \frac{
                 p_{x}^{2}+
                 p_{z}^{2}
                                                   }
               {
                [1+f(p_{\sigma})]^{2}
                                }
                                   \right\}^{-1/2}\ ;
\\      \nonumber
\\      \nonumber
 \\
   \addtocounter{equation}{-1}
\addtocounter{INDEX}{1}
    \frac{d}{ds}\,p_{\sigma}
                      &=&
               -\frac{\partial{
                               {\cal{H}}_{D}
                                        }}{\partial
                          {\sigma}}
\nonumber    \\
\nonumber    \\
                    &=&
                     0
                      \ \ \
         \Longrightarrow\ \ \
         p_{\sigma}\ =\ const\ .
\end{eqnarray}
\setcounter{INDEX}{0}

The solution of eqn. (A.5)
for a drift space of length\,$l$\,
reads as\,:
\setcounter{INDEX}{1}
\begin{eqnarray}
    x^{f}
            &=&
               x^{i}
           +
                           \left\{1-
          \frac{
               \left[
                 (p_{x}^{\,i})^{2}+
                 (p_{z}^{\,i})^{2}
                                  \right]
                                                   }
               {
                [1+f(p_{\sigma}^{\,i})]^{2}
                                }
                                   \right\}^{-1/2}
          \cdot\frac{p_{x}^{\,i}}
               {
                [1+f(p_{\sigma}^{\,i})]
                                }
           \cdot l\ ;
\\      \nonumber
 \\
   \addtocounter{equation}{-1}
\addtocounter{INDEX}{1}
    p_{x}^{\,f}
            &=&
    p_{x}^{\,i}\ ;
\\      \nonumber
 \\
   \addtocounter{equation}{-1}
\addtocounter{INDEX}{1}
    z^{f}
            &=&
    z^{i}
           +
                           \left\{1-
          \frac{
               \left[
                 (p_{x}^{\,i})^{2}+
                 (p_{z}^{\,i})^{2}
                                  \right]
                                                   }
               {
                [1+f(p_{\sigma}^{\,i})]^{2}
                                }
                                   \right\}^{-1/2}
          \cdot\frac{p_{z}^{\,i}}
               {
                [1+f(p_{\sigma}^{\,i})]
                                }
           \cdot l\ ;
\\      \nonumber
 \\
   \addtocounter{equation}{-1}
\addtocounter{INDEX}{1}
    p_{z}^{\,f}
            &=&
    p_{z}^{\,i}\ ;
\\      \nonumber
 \\
   \addtocounter{equation}{-1}
\addtocounter{INDEX}{1}
    \sigma^{f}
            &=&
               \sigma^{i}
        +
    \left[
           1-
                   f'(p_{\sigma}^{\,i})\cdot
                           \left\{1-
          \frac{
               \left[
                 (p_{x}^{\,i})^{2}+
                 (p_{z}^{\,i})^{2}
                                  \right]
                                                   }
               {
                [1+f(p_{\sigma}^{\,i})]^{2}
                                }
                                   \right\}^{-1/2}
          \right]
               \cdot l\ ;
\\      \nonumber
 \\
   \addtocounter{equation}{-1}
\addtocounter{INDEX}{1}
    p_{\sigma}^{\,f}
            &=&
    p_{\sigma}^{\,i}\ .
\end{eqnarray}
\\
\\
\underline{Remarks:}
1) Using (A.5a) and (A.5c), eqn. (A.5e)
may also be written in the form\,:
\begin{eqnarray}
    \frac{d}{ds}\,\sigma
                    &=&
           1-
             f'(p_{\sigma})\cdot
           \sqrt{1+(x')^{2}+(z')^{2}}\ .
\end{eqnarray}
This result can be obtained directly
from the defining equation for\,
        $\sigma$\,:
\begin{eqnarray*}
       \sigma
             &=&
                s-v_{0}\cdot t(s)\ ;
\\
\\
   \Longrightarrow\
    \frac{d}{ds}\,\sigma
             &=&
                1-v_{0}\cdot
                            \frac{dt}{ds}
\end{eqnarray*}
with
\begin{eqnarray*}
      dt
             &=&
      \frac{1}{v}\cdot
      \sqrt{ds^{2}+dx^{2}+dz^{2}}
\end{eqnarray*}
and leads to
\begin{eqnarray}
    \frac{d}{ds}\,\sigma
                    &=&
           1-
             \frac{v_{0}}{v}\cdot
           \sqrt{1+(x')^{2}+(z')^{2}}\ ,
\end{eqnarray}
which agrees with eqn. (A.6), since
\begin{eqnarray*}
             f'(p_{\sigma})
                    &=&
             \frac{v_{0}}{v}
\end{eqnarray*}
(see Appendix C in Ref.
\cite{RS}\,).
Since
\begin{eqnarray*}
           \sqrt{1+(x')^{2}+(z')^{2}}
                 &\approx&
           1+
           \frac{1}{2}\,[\,(x')^{2}+(z')^{2}\,]\ ,
\end{eqnarray*}
one may write\,:
\begin{eqnarray*}
    \frac{d}{ds}\,\sigma
                    &\approx&
           1-
             f'(p_{\sigma})\cdot
           \left\{
           1+
           \frac{1}{2}\,[\,(x')^{2}+(z')^{2}\,]\right\}\ .
\end{eqnarray*}
This approximation was used in Ref.
\cite{RS}.
\\

2) As in Ref.
\cite{RS}
one obtains for a drift space\,:
\begin{eqnarray*}
     x'(s)\ =\ const.\ \
  &\Longrightarrow&\ \
     x(s)\  =\ x(s_{0})+x'(s_{0})\cdot(s-s_{0})\ ;
\\
     z'(s)\ =\ const.\ \
  &\Longrightarrow&\ \
     z(s)\  =\ z(s_{0})+z'(s_{0})\cdot(s-s_{0})\ ,
\end{eqnarray*}
(see
eqns.(A.5a,\,b) and (A.5c,\,d)\,)
but the connection betwenn\,
        $y'$\,
and
        $p_{y}$\,
        $(y\,\equiv\,x,\,z)$
is modified
(see
eqns.(A.5a,\,c) and (3.7a,\,c)\,).
\bigskip

\subsubsection{The Central Part}

\ \ \ \ \
The equation of motion for the central part
(region II in eqn. (3.13)\,)
due to the Hamiltonian (A.2)
reads as\,:
\setcounter{INDEX}{1}
\begin{eqnarray}
    \frac{d\ }{ds}\,
       \vec{y}(s)
         &=&
       \vec{F}
              (\vec{y})
        \cdot
   \delta(s-s_{0})
\end{eqnarray}
with
\begin{eqnarray}
   \addtocounter{equation}{-1}
\addtocounter{INDEX}{1}
       \vec{F}
              (\vec{y})
         &=&
    \vec{\vartheta}_{L}
              (\vec{y}; s_{0})
        \cdot
   \Delta s
\end{eqnarray}
\setcounter{INDEX}{0}
and with\,
   $\vec{\vartheta}_{L}
              (\vec{y}; s)$\,
given by\,:
\setcounter{INDEX}{1}
\begin{eqnarray}
    \vartheta_{L1}
        (\vec{y}; s)
                    &=&
               +\frac{\partial
                                      }{\partial
                          {p_{x}}}\,
  {\cal{H}}_{L}(x,p_{x},z,p_{z},\sigma,p_{\sigma}; s)
\nonumber   \\
                    &=&
   0\ ;
 \\       \nonumber
 \\
   \addtocounter{equation}{-1}
\addtocounter{INDEX}{1}
    \vartheta_{L2}
        (\vec{y}; s)
                    &=&
               -\frac{\partial
                                      }{\partial
                          {x}}\,
      {\cal{H}}_{L}(x,p_{x},z,p_{z},\sigma,p_{\sigma}; s)
\nonumber   \\
                        &=&
                   +
                          K_{x}(s)\cdot f(p_{\sigma})
                   -
                [K_{x}^{2}(s)+g(s)]\cdot x
                   +
                          N(s)\cdot z
\nonumber   \\
& &
\hspace*{1.0cm}
                       -
    \frac{\lambda(s)}{2}\cdot
    (x^{2}-z^{2})
                       -
    \frac{\mu(s)}{6}\cdot
    (x^{3}-3\,x\,z^{2})\ ;
 \\       \nonumber
 \\
   \addtocounter{equation}{-1}
\addtocounter{INDEX}{1}
    \vartheta_{L3}
        (\vec{y}; s)
                    &=&
               +\frac{\partial
                                      }{\partial
                          {p_{z}}}\,
      {\cal{H}}_{L}(x,p_{x},z,p_{z},\sigma,p_{\sigma}; s)
\nonumber   \\
                   &=&
     0\ ;
 \\       \nonumber
 \\
   \addtocounter{equation}{-1}
\addtocounter{INDEX}{1}
    \vartheta_{L4}
        (\vec{y}; s)
                    &=&
               -\frac{\partial
                                      }{\partial
                          {z}}\,
      {\cal{H}}_{L}(x,p_{x},z,p_{z},\sigma,p_{\sigma}; s)
\nonumber   \\
                       &=&
                   +
                          K_{z}(s)\cdot f(p_{\sigma})
                   -
                [K_{z}^{2}(s)-g(s)]\cdot z
                   +
                          N(s)\cdot x
\nonumber   \\
& &
\hspace*{1.0cm}
                       +
    \lambda(s)\cdot x z
                       -
    \frac{\mu(s)}{6}\cdot
    (z^{3}-3\,x^{2}\,z)\ ;
 \\       \nonumber
 \\
   \addtocounter{equation}{-1}
\addtocounter{INDEX}{1}
    \vartheta_{L5}
        (\vec{y}; s)
                    &=&
               +\frac{\partial
                                      }{\partial
                          {p_{\sigma}}}\,
      {\cal{H}}_{L}(x,p_{x},z,p_{z},\sigma,p_{\sigma}; s)
\nonumber   \\
                        &=&
               -[K_{x}(s)\cdot x+
                 K_{z}(s)\cdot z]
                \cdot f'(p_{\sigma})\ ;
 \\       \nonumber
 \\
   \addtocounter{equation}{-1}
\addtocounter{INDEX}{1}
    \vartheta_{L6}
        (\vec{y}; s)
                    &=&
               -\frac{\partial
                                      }{\partial
                          {\sigma}}\,
      {\cal{H}}_{L}(x,p_{x},z,p_{z},\sigma,p_{\sigma}; s)
\nonumber   \\
                      &=&
              \frac{1}{\beta_{0}^{2}}
                         \cdot\frac{eV(s)}{E_{0}}\cdot
                   \sin\left[
                        h\cdot\frac{2\pi}{L}\cdot\sigma
                              +
                             \varphi
                                    \right]
\end{eqnarray}
\setcounter{INDEX}{0}
resulting from
the Hamiltonian\,
    ${\cal{H}}_{L}$\,
in eqn. (A.4).
\setcounter{INDEX}{0}
\bigskip

Since the relations (A.10a\,--\,f) coincide
with eqns. (3.16a\,--\,f) in the absence of solenoids
      ($H\,=\,0$),
the thin\,-\,lens transport maps calculated in sections 4.1\,--\,4.7
remain valid also for the Hamiltonian (A.2).
Thus the corrections resulting from the new Hamiltonian
are fully absorbed in the solutions for the
drift space
as can be seen by comparing (A.6) with (3.7).
\bigskip

As an example we consider the
superposition of quadrupoles,
                       skew quadrupoles,
                       bending magnets,
                       sextupoles and
                       octupoles
and obtain from (A.10)\,:
\begin{eqnarray*}
    F_{1}
         (\vec{y})
                    &=&
      0\ ;
\\      \nonumber
 \\
    F_{2}
         (\vec{y})
                    &=&
                   \left\{
                          K_{x}(s_{0})\cdot f(p_{\sigma})
                   -
                 G_{1}(s_{0})\cdot x
                   +
                          N(s_{0})\cdot z
                       -
    \frac{\lambda(s_{0})}{2}\cdot
    (x^{2}-z^{2})
              \right.
\\
& &
\hspace*{7.0cm}
        \left.
                       -
    \frac{\mu_(s_{0})}
                       {6}\cdot
    (x^{3}-3\,x\,z^{2})
                     \right\}
    \cdot\Delta s\ ;
\\      \nonumber
 \\
    F_{3}
         (\vec{y})
                    &=&
      0\ ;
\\      \nonumber
 \\
    F_{4}
         (\vec{y})
                    &=&
                   \left\{
                          K_{z}(s_{0})\cdot f(p_{\sigma})
                   -
                G_{2}(s_{0})\cdot z
                   +
                          N(s_{0})\cdot x
                       +
    \lambda(s_{0})\cdot x z
              \right.
\\
& &
\hspace*{7.0cm}
        \left.
                       -
    \frac{\mu_(s_{0})}{6}\cdot
    (z^{3}-3\,x^{2}\,z)
                     \right\}
    \cdot\Delta s\ ;
\\      \nonumber
 \\
    F_{5}
         (\vec{y})
                    &=&
    \left[
       K_{x}(s_{0})\cdot x
             +
       K_{z}(s_{0})\cdot z
               \right]
               \cdot f'(p_{\sigma})
               \cdot\Delta s\ ;
\\      \nonumber
 \\
    F_{6}
         (\vec{y})
                    &=&0\ ;
 \\      \nonumber
 \\      \nonumber
 \\      \nonumber
& &
\hspace*{1.0cm}
(\,G_{1}\,=\,K_{x}^{2}+g\,;\;
 G_{2}\,=\,K_{z}^{2}-g\,)\ .
\end{eqnarray*}

Thus\,:
\begin{eqnarray}
       \hat{D}
        &=&
    F_{2}
        (\vec{\hat{y}})
         \cdot
                \frac{\partial\ }
                     {\partial{\hat{y}_{2}}}
          +
    F_{4}
        (\vec{\hat{y}})
         \cdot
                \frac{\partial\ }
                     {\partial{\hat{y}_{4}}}
          +
    F_{5}
        (\vec{\hat{y}})
         \cdot
                \frac{\partial\ }
                     {\partial{\hat{y}_{5}}}\ .
\end{eqnarray}
We then have\,:
\bigskip
\setcounter{INDEX}{1}
\begin{eqnarray}
& &
    \hat{D}\,
    \hat{y}_{1}
            \ =\
   0\ ;
\\       \nonumber
\\
   \addtocounter{equation}{-1}
\addtocounter{INDEX}{1}
& &
    \hat{D}\,
    \hat{y}_{2}
            \ =\
                   \left\{
                          K_{x}(s_{0})\cdot f(p_{\sigma})
                   -
                 G_{1}(s_{0})\cdot x
                   +
                          N(s_{0})\cdot z
                       -
    \frac{\lambda(s_{0})}{2}\cdot
    (x^{2}-z^{2})
              \right.
\nonumber    \\
& &
\hspace*{5.0cm}
        \left.
                       -
    \frac{\mu_(s_{0})}
                       {6}\cdot
    (x^{3}-3\,x\,z^{2})
                     \right\}
    \cdot\Delta s\ ;
\\       \nonumber
\\
   \addtocounter{equation}{-1}
\addtocounter{INDEX}{1}
& &
    \hat{D}\,
    \hat{y}_{3}
       \ =\
  0\ ;
\\       \nonumber
\\
   \addtocounter{equation}{-1}
\addtocounter{INDEX}{1}
& &
    \hat{D}\,
    \hat{y}_{4}
            \ =\
                   \left\{
                          K_{z}(s_{0})\cdot f(p_{\sigma})
                   -
                G_{2}(s_{0})\cdot z
                   +
                          N(s_{0})\cdot x
                       +
    \lambda(s_{0})\cdot x z
              \right.
\nonumber    \\
& &
\hspace*{7.0cm}
        \left.
                       -
    \frac{\mu_(s_{0})}{6}\cdot
    (z^{3}-3\,x^{2}\,z)
                     \right\}
    \cdot\Delta s\ ;
\\       \nonumber
\\
   \addtocounter{equation}{-1}
\addtocounter{INDEX}{1}
& &
    \hat{D}\,
    \hat{y}_{5}
           \ =\
    \left[
       K_{x}(s_{0})\cdot x
             +
       K_{z}(s_{0})\cdot z
               \right]
               \cdot f'(p_{\sigma})
               \cdot\Delta s\ ;
\\       \nonumber
\\
   \addtocounter{equation}{-1}
\addtocounter{INDEX}{1}
& &
    \hat{D}\,
    \hat{y}_{6}
            \ =\
                0
\end{eqnarray}
\setcounter{INDEX}{0}
\bigskip
and
\begin{eqnarray*}
       \hat{D}^{\,\nu}\,
       \vec{\hat{y}}
        &=&
       \vec{0}\ \
       \mbox{for}\ \
       \nu\;>\;1\ .
\end{eqnarray*}

Thus\,:
\begin{eqnarray}
      \vec{y}(s_{0}+0)
          &=&
      \vec{\hat{y}}
             +\hat{D}\,
      \vec{\hat{y}}
\end{eqnarray}
with\,
\begin{eqnarray*}
      \vec{\hat{y}}&\equiv&
      \vec{y}(s_{0})
\end{eqnarray*}
and
             $\hat{D}\,
      \vec{\hat{y}}$\,
given by (A.12).
\bigskip

Equation (A.13) contains as special cases
the transport maps of simple
                       quadrupoles,
                       skew quadrupoles,
                       bending magnets,
                       sextupoles and
                       octupoles
which are identical
with those already derived
in section 4.
\\
\\
\\
\underline{Remarks:}
                    \newline

\vspace{-0.3ex}
1) As in chapter 4
the transport maps
\begin{eqnarray*}
       \vec{y}\,\left(s_{0}-\frac{1}{2}\,\Delta s\right)
             \ \ \longrightarrow\ \
       \vec{y}\,\left(s_{0}+\frac{1}{2}\,\Delta s\right)
\end{eqnarray*}
described by a composition of (A.6) and (A.13)
(combining the regions I, II and III
in eqn. (3.13)\,)
are symplectic for an arbitrary
       $\Delta s$
due to the canonical structure
of the equations of motion.
Furthermore one obtains the exact solution
corresponding to the Hamiltonian (A.2) for
      $\Delta s\,
                 \longrightarrow\,
       0$\ .
\\

2) The Hamiltonian (A.2) is exact
for a straight section with
\begin{eqnarray*}
       K_{x}&=&K_{z}\ =\ 0\ ,
\end{eqnarray*}
i.e. outside the bending magnets.
For a bending magnet with
\begin{eqnarray*}
       K_{x}^{\,2}+K_{z}^{\,2}&\neq&0\ ;\ \
       K_{x}\cdot K_{z}\ =\ 0
\end{eqnarray*}
the exact Hamiltonian reads as\,:
\begin{eqnarray}
    {\cal{H}}_{bend}
               &=&
            p_{\sigma}
               -\left[1+
                   f(p_{\sigma})\right]
                 \cdot
                        [1+K_{x}\cdot x+K_{z}\cdot z]
                 \cdot
                           \left\{1-
          \frac{
                p_{x}^{2}+
                p_{z}^{2}
                                                   }
               {
                \left[1+
                   f(p_{\sigma})\right]^{2}
                                }
                                   \right\}^{1/2}
 \nonumber  \\
 & &
                  +\frac{1}{2}\cdot
                              [1+K_{x}\cdot{x}+K_{z}\cdot{z}]^{\,2}
                  -\frac{1}{2}
 \nonumber  \\
 \nonumber  \\
               &=&
    {\cal{H}}_{D}
               -\left[1+
                   f(p_{\sigma})\right]
                 \cdot
                        [K_{x}\cdot x+K_{z}\cdot z]
                 \cdot
                           \left\{1-
          \frac{
                p_{x}^{2}+
                p_{z}^{2}
                                                   }
               {
                \left[1+
                   f(p_{\sigma})\right]^{2}
                                }
                                   \right\}^{1/2}
 \nonumber  \\
 & &
                  +[K_{x}\cdot{x}+K_{z}\cdot{z}]
                  +\frac{1}{2}\,K_{x}^{\,2}\cdot x^{2}
                  +\frac{1}{2}\,K_{z}^{\,2}\cdot z^{2}
 \nonumber  \\
 \nonumber  \\
               &=&
    {\cal{H}}_{D}
 \nonumber  \\
 & &
               -
         \left[1+
                 f(p_{\sigma})\right]
                 \cdot
                        [K_{x}\cdot x+K_{z}\cdot z]
                 \cdot
      \left(
                           \left\{1-
          \frac{
                p_{x}^{2}+
                p_{z}^{2}
                                                   }
               {
                \left[1+
                   f(p_{\sigma})\right]^{2}
                                }
                                   \right\}^{1/2}
                    -1\right)
 \nonumber  \\
 & &
               -
                 f(p_{\sigma})
                 \cdot
                        [K_{x}\cdot x+K_{z}\cdot z]
                  +\frac{1}{2}\,K_{x}^{\,2}\cdot x^{2}
                  +\frac{1}{2}\,K_{z}^{\,2}\cdot z^{2}
\end{eqnarray}
with
\begin{eqnarray}
    {\cal{H}}_{D}
               &=&
            p_{\sigma}
           -[1+f(p_{\sigma})]
                 \cdot
                           \left\{1-
          \frac{
                 p_{x}^{2}+
                 p_{z}^{2}
                                                   }
               {
                [1+f(p_{\sigma}]^{2}
                                }
                                   \right\}^{1/2}\ .
\end{eqnarray}

The drift part\,
   ${\cal{H}}_{D}$\,
of the bending Hamiltonian
coincides with eqn. (A.3).
Thus for the regions I) and III)
we have again the solution (A.6).
\bigskip

For the central part described
by the equations\,:
\setcounter{INDEX}{1}
\begin{eqnarray}
    \vartheta_{L1}
              (\vec{y}; s_{0})
                    &=&
               +\frac{\partial
                                      }{\partial
                          {p_{x}}}\,
  \hat{\cal{H}}_{bend}(x,p_{x},z,p_{z},\sigma,p_{\sigma})\ ;
 \\       \nonumber
 \\
   \addtocounter{equation}{-1}
\addtocounter{INDEX}{1}
    \vartheta_{L2}
              (\vec{y}; s_{0})
                        &=&
               -\frac{\partial
                                      }{\partial
                          {x}}\,
  \hat{\cal{H}}_{bend}(x,p_{x},z,p_{z},\sigma,p_{\sigma})\ ;
 \\       \nonumber
 \\
   \addtocounter{equation}{-1}
\addtocounter{INDEX}{1}
    \vartheta_{L3}
              (\vec{y}; s_{0})
                   &=&
               +\frac{\partial
                                      }{\partial
                          {p_{z}}}\,
  \hat{\cal{H}}_{bend}(x,p_{x},z,p_{z},\sigma,p_{\sigma})\ ;
 \\       \nonumber
 \\
   \addtocounter{equation}{-1}
\addtocounter{INDEX}{1}
    \vartheta_{L4}
              (\vec{y}; s_{0})
                       &=&
               -\frac{\partial
                                      }{\partial
                          {z}}\,
  \hat{\cal{H}}_{bend}(x,p_{x},z,p_{z},\sigma,p_{\sigma})\ ;
 \\       \nonumber
 \\
   \addtocounter{equation}{-1}
\addtocounter{INDEX}{1}
    \vartheta_{L5}
              (\vec{y}; s_{0})
                        &=&
               +\frac{\partial
                                      }{\partial
                          {p_{\sigma}}}\,
  \hat{\cal{H}}_{bend}(x,p_{x},z,p_{z},\sigma,p_{\sigma})\ ;
 \\       \nonumber
 \\
   \addtocounter{equation}{-1}
\addtocounter{INDEX}{1}
    \vartheta_{L6}
              (\vec{y}; s_{0})
                      &=&
               -\frac{\partial
                                      }{\partial
                          {\sigma}}\,
  \hat{\cal{H}}_{bend}(x,p_{x},z,p_{z},\sigma,p_{\sigma})
\end{eqnarray}
\setcounter{INDEX}{0}
(see eqns. (3.10) and (3.11)\,)
we obtain the Hamiltonian\,:
\begin{eqnarray}
    \hat{\cal{H}}_{bend}
               &=&
               -
         \left[1+
                 f(p_{\sigma})\right]
                 \cdot
                 [K_{x}(s_{0})\cdot x+K_{z}(s_{0})\cdot z]
                 \cdot
      \left(
                           \left\{1-
          \frac{
                p_{x}^{2}+
                p_{z}^{2}
                                                   }
               {
                \left[1+
                   f(p_{\sigma})\right]^{2}
                                }
                                   \right\}^{1/2}
                    -1\right)
 \nonumber  \\
 & &
               -
                 f(p_{\sigma})
                 \cdot
                 [K_{x}(s_{0})\cdot x+K_{z}(s_{0})\cdot z]
                  +\frac{1}{2}\,[K_{x}(s_{0})]^{2}\cdot x^{2}
                  +\frac{1}{2}\,[K_{z}(s_{0})]^{2}\cdot z^{2}\ .
\ \ \ \ \ \ \ \ \ \
\end{eqnarray}

Using (A.9b) and (3.29),
we can calculate\,
          $\vec{y}(s_{0}+0)$\,
up to an arbitrary order by a series expansion of\,
    $\exp{[\hat{D}]}$.
This shall be done in
Appendix B where it is shown
how to express the transport map by elementary functions
(without disturbing the symplecticity),
restricting the crossing terms to the lowest order
and using
a special decomposition of\,
   $\hat{\cal{H}}_{bend}$\,
into two parts.
\bigskip
%
%

\setcounter{section}{2}
\setcounter{subsection}{0}
\setcounter{equation}{0}
\addcontentsline{toc}{section}{Appendix B:
                       Bending Magnet
                       with Nonlinear Crossing Terms}
\section*{Appendix B:
                       Bending Magnet
                       with Nonlinear Crossing Terms
                       Resulting from the Curvature
                                                    }
\par
\newcommand{\yr}{{\bf x},s}
\subsection{The Hamiltonian}

\ \ \ \ \
By expanding the Hamiltonian (A.14)
and keeping the nonlinear crossing terms
of lowest order
resulting from the curvature\,
        $K_{x}$\ ,
the whole Hamiltonian
of a horizontal bending magnet
\footnote{A vertical bending magnet
can be dealt with analogously.}
($K_{x}\,\neq\,0;\, K_{z}\,=\,0$)
can be written in the form\,:
\begin{eqnarray}
          {\cal{H}}_{bend}
            &=&
       {\cal{H}}_{D}
             +
       {\cal{H}}_{L}
\end{eqnarray}
with a lens component
\begin{eqnarray}
     {\cal{H}}_{L}
              &=&
           -K_{x}(s)\cdot x
                 \cdot
            f(p_{\sigma})
                        +
                   \frac{1}{2}\,
                              [K_{x}(s)]^{2}
                              \cdot
                              x^{2}
                        +
       \frac{1}{2}\cdot
          \frac
               {K_{x}(s)}
               {
                \left[1+
                   f(p_{\sigma})\right]
                                }
          \cdot x
          \cdot
          \left[p_{x}^{2}
               +p_{z}^{2}
                         \right]
\end{eqnarray}
and a drift component\,
    ${\cal{H}}_{D}$\,
given by (3.5)
or more precisely by (A.3).
\bigskip

In order to obtain
a thin\,-\,lens approximation
of a horizontal bending magnet
due to the Hamiltonian (B.1)
which can be expressed by elementary functions,
we decompose the lens component\,
    ${\cal{H}}_{L}$\,
of eqn. (B.2)
into two parts\,:
\begin{eqnarray}
          {\cal{H}}_{L}
            &=&
       {\cal{H}}_{L}^{(1)}
             +
       {\cal{H}}_{L}^{(2)}
\end{eqnarray}
with
\setcounter{INDEX}{1}
\begin{eqnarray}
     {\cal{H}}_{L}^{(1)}
              &=&
           -K_{x}(s)\cdot x
                 \cdot
            f(p_{\sigma})
                        +
                   \frac{1}{2}\,
                              [K_{x}(s)]^{2}
                              \cdot
                              x^{2}
                        +
       \frac{1}{2}\cdot
          \frac
               {K_{x}(s)}
               {
                \left[1+
                   f(p_{\sigma})\right]
                                }
          \cdot x
          \cdot
                p_{z}^{2}\ ;
\\
   \addtocounter{equation}{-1}
\addtocounter{INDEX}{1}
     {\cal{H}}_{L}^{(2)}
              &=&
       \frac{1}{2}\cdot
          \frac
               {K_{x}(s)}
               {
                \left[1+
                   f(p_{\sigma})\right]
                                }
          \cdot x
          \cdot
                p_{x}^{2}\ .
\end{eqnarray}
\setcounter{INDEX}{0}

For each part we then separately
construct a thin lens
(placing one lens behind the other).
This is achieved
replacing\,
    ${\cal{H}}_{bend}$\,
in (B.1) by
\begin{eqnarray}
     {\cal{H}}_{bend}
           &=&
     {\cal{H}}_{D}
            +
           \delta(s-[s_{0}-0])\cdot
           \Delta s\cdot
     \hat{\cal{H}}_{L}^{(1)}
            +
           \delta(s-[s_{0}+0])\cdot
           \Delta s\cdot
     \hat{\cal{H}}_{L}^{(2)}
\end{eqnarray}
with
\setcounter{INDEX}{1}
\begin{eqnarray}
     \hat{\cal{H}}_{L}^{(1)}
              &=&
           -K_{x}(s_{0})\cdot x
                 \cdot
            f(p_{\sigma})
                        +
                   \frac{1}{2}\,
                              [K_{x}(s_{0})]^{2}
                              \cdot
                              x^{2}
                        +
       \frac{1}{2}\cdot
          \frac
               {K_{x}(s_{0})}
               {
                \left[1+
                   f(p_{\sigma})\right]
                                }
          \cdot x
          \cdot
                p_{z}^{2}\ ;
\\
   \addtocounter{equation}{-1}
\addtocounter{INDEX}{1}
     \hat{\cal{H}}_{L}^{(2)}
              &=&
       \frac{1}{2}\cdot
          \frac
               {K_{x}(s_{0})}
               {
                \left[1+
                   f(p_{\sigma})\right]
                                }
          \cdot x
          \cdot
                p_{x}^{2}\ .
\end{eqnarray}
\setcounter{INDEX}{0}
\bigskip

\subsection{Thin\,-\,Lens Approximation}
\subsubsection{The Component
    $\hat{\cal{H}}_{L}^{(1)}$}

\ \ \ \ \
{}From eqns. (3.29b) and (B.6a) we have\,:
\setcounter{INDEX}{1}
\begin{eqnarray}
    F_{1}
         (\vec{y})
                    &=&
         +\Delta s\cdot
                \frac{\partial
                                      }{\partial
                          {p_{x}}}\,
      \hat{\cal{H}}_{L}^{(1)}(x,p_{x},z,p_{z},\sigma,p_{\sigma})
\nonumber    \\
                    &=&
                     0\ ;
 \\       \nonumber
 \\
   \addtocounter{equation}{-1}
\addtocounter{INDEX}{1}
    F_{2}
         (\vec{y})
                        &=&
         -\Delta s\cdot
                \frac{\partial
                                      }{\partial
                         {x}}\,
      \hat{\cal{H}}_{L}^{(1)}(x,p_{x},z,p_{z},\sigma,p_{\sigma})
\nonumber    \\
                    &=&
          \Delta s\cdot
     \left\{
                        -
           [K_{x}(s_{0})]^{2}
             \cdot
             x
           +
            K_{x}(s_{0})
                 \cdot
            f(p_{\sigma})
                        -
          \frac{1}{2}\cdot
          \frac
               {K_{x}(s_{0})}
               {
                \left[1+
                   f(p_{\sigma})\right]
                                }
          \cdot
          p_{z}^{2}\,
                      \right\};
 \\       \nonumber
 \\
   \addtocounter{equation}{-1}
\addtocounter{INDEX}{1}
    F_{3}
         (\vec{y})
                   &=&
         +\Delta s\cdot
                \frac{\partial
                                      }{\partial
                          {p_{z}}}\,
      \hat{\cal{H}}_{L}^{(1)}(x,p_{x},z,p_{z},\sigma,p_{\sigma})
\nonumber    \\
                    &=&
          \Delta s\cdot
        \frac{
           K_{x}(s_{0})}
               {
                \left[1+
                   f(p_{\sigma})\right]
                                }
        \cdot x
        \cdot p_{z}\ ;
 \\       \nonumber
 \\
   \addtocounter{equation}{-1}
\addtocounter{INDEX}{1}
    F_{4}
         (\vec{y})
                       &=&
         -\Delta s\cdot
                \frac{\partial
                                      }{\partial
                          {z}}\,
      \hat{\cal{H}}_{L}^{(1)}(x,p_{x},z,p_{z},\sigma,p_{\sigma})
\nonumber    \\
                    &=&
                        0\ ;
 \\       \nonumber
 \\
   \addtocounter{equation}{-1}
\addtocounter{INDEX}{1}
    F_{5}
         (\vec{y})
                       &=&
         +\Delta s\cdot
                \frac{\partial
                                      }{\partial
                          {p_{\sigma}}}
      \hat{\cal{H}}_{L}^{(1)}(x,p_{x},z,p_{z},\sigma,p_{\sigma})
\nonumber    \\
                    &=&
     \Delta s\cdot
       \left\{
           -K_{x}(s_{0})\cdot x
                 \cdot
            f'(p_{\sigma})
                  -
          \frac{1}{2}\cdot
          \frac
               {K_{x}(s_{0})}
               {
                \left[1+
                   f(p_{\sigma})\right]
                                }
        \cdot
      \frac{
                      f'(p_{\sigma})
                                  }
               {
                \left[1+
                   f(p_{\sigma})\right]
                                }
          \cdot
          \left[x\cdot p_{z}^{2}\right]\,
                  \right\}\ ;
 \\       \nonumber
 \\
   \addtocounter{equation}{-1}
\addtocounter{INDEX}{1}
    F_{6}
         (\vec{y})
                      &=&
         -\Delta s\cdot
                \frac{\partial
                                      }{\partial
                          {\sigma}}\,
      \hat{\cal{H}}_{L}^{(1)}(x,p_{x},z,p_{z},\sigma,p_{\sigma})
\nonumber    \\
                    &=&
                       0\ .
\end{eqnarray}
\setcounter{INDEX}{0}

Thus\,:
\begin{eqnarray}
       \hat{D}
        &=&
    F_{2}
        (\vec{\hat{y}})
         \cdot
                \frac{\partial\ }
                     {\partial{\hat{y}_{2}}}
          +
    F_{3}
        (\vec{\hat{y}})
         \cdot
                \frac{\partial\ }
                     {\partial{\hat{y}_{3}}}
          +
    F_{5}
        (\vec{\hat{y}})
         \cdot
                \frac{\partial\ }
                     {\partial{\hat{y}_{5}}}
\end{eqnarray}
and
\begin{eqnarray}
       \hat{D}\,
       \vec{\hat{y}}
        &=&
                 \left( \begin{array}{c}
                   0      \\
                   F_{2}(\vec{\hat{y}})  \\
                   F_{3}(\vec{\hat{y}})  \\
                   0      \\
                   F_{5}(\vec{\hat{y}})  \\
                   0
              \end{array}
       \right)\ .
\end{eqnarray}

We then get\,:
\\
\vspace*{0.5ex}

a) For\, $x$\,:
\begin{eqnarray*}
       \hat{D}\,
        \hat{x}
        &=&
         0
\\
\\
    \Longrightarrow\ \
       \hat{D}^{\nu}\,
        \hat{x}
        &=&
      0\ \
       \mbox{for}\ \
       \nu\;>\;0\ .
\end{eqnarray*}
Thus\,:
\begin{eqnarray}
  \left\{
     \exp{[\hat{D}]}
                \right\}
      \hat{x}
         &\equiv&
        \sum_{n\,=\,0}^{\infty}\,
     \frac{1}{n!}\cdot
           \hat{D}^{n}\,
      \hat{x}
        \ =\
      \hat{x}\ .
\end{eqnarray}
\\
\vspace*{+0.5ex}

b) For\, $p_{x}$\,:
\begin{eqnarray*}
       \hat{D}\,
        \hat{p}_{x}
        &=&
       F_{2}(\vec{\hat{y}})
\\
        &=&
          \Delta s\cdot
     \left\{
                        -
           [K_{x}(s_{0})]^{2}
             \cdot
             \hat{x}
           +
            K_{x}(s_{0})
                 \cdot
            f(\hat{p}_{\sigma})
                        -
          \frac{1}{2}\cdot
          \frac
               {K_{x}(s_{0})}
               {
                \left[1+
                   f(\hat{p}_{\sigma})\right]
                                }
          \cdot
          \hat{p}_{z}^{2}\,
                      \right\}
\\
        &=&
          \Delta s\cdot
     \left\{
                        -
           [K_{x}(s_{0})]^{2}
             \cdot
             \hat{y}_{1}
           +
            K_{x}(s_{0})
                 \cdot
            f(\hat{y}_{6})
                        -
          \frac{1}{2}\cdot
          \frac
               {K_{x}(s_{0})}
               {
                \left[1+
                   f(\hat{y}_{6})\right]
                                }
          \cdot
          \hat{y}_{4}^{2}\,
                      \right\}\ ;
\\
\\
       \hat{D}^{2}\,
        \hat{p}_{x}
        &=&
          \Delta s\cdot
       \hat{D}\,
     \left\{
                        -
           [K_{x}(s_{0})]^{2}
             \cdot
             \hat{y}_{1}
           +
            K_{x}(s_{0})
                 \cdot
            f(\hat{y}_{6})
                        -
          \frac{1}{2}\cdot
          \frac
               {K_{x}(s_{0})}
               {
                \left[1+
                   f(\hat{y}_{6})\right]
                                }
          \cdot
          \hat{y}_{4}^{2}\,
                      \right\}
\\
        &=&
         0\ ;
\\
\\
    \Longrightarrow\ \
       \hat{D}^{\nu}\,
        \hat{p}_{x}
        &=&
      0\ \
       \mbox{for}\ \
       \nu\;>\;1\ .
\end{eqnarray*}
Thus\,:
\begin{eqnarray}
  \left\{
     \exp{[\hat{D}]}
                \right\}
      \hat{p}_{x}
         &=&
      \hat{p}_{x}
          +
       \hat{D}\,
        \hat{p}_{x}
\nonumber    \\
         &=&
      \hat{y}_{2}
          +
          \Delta s\cdot
     \left\{
                        -
           [K_{x}(s_{0})]^{2}
             \cdot
             \hat{y}_{1}
           +
            K_{x}(s_{0})
                 \cdot
            f(\hat{y}_{6})
                        -
          \frac{1}{2}\cdot
          \frac
               {K_{x}(s_{0})}
               {
                \left[1+
                   f(\hat{y}_{6})\right]
                                }
          \cdot
          \hat{y}_{4}^{2}\,
                      \right\}
\nonumber    \\
         &\equiv&
      \hat{p}_{x}
            +
          \Delta s\cdot
     \left\{
                        -
           [K_{x}(s_{0})]^{2}
             \cdot
             \hat{x}
           +
            K_{x}(s_{0})
                 \cdot
            f(\hat{p}_{\sigma})
                        -
          \frac{1}{2}\cdot
          \frac
               {K_{x}(s_{0})}
               {
                \left[1+
                   f(\hat{p}_{\sigma})\right]
                                }
          \cdot
          \hat{p}_{z}^{2}\,
                      \right\}\,;
\ \ \
\ \ \
\end{eqnarray}
\\
\vspace*{0.5ex}

c) For\, $z$\,:
\begin{eqnarray*}
       \hat{D}\,
        \hat{z}
        &=&
       F_{3}(\vec{\hat{y}})
       \ =\
          \Delta s\cdot
        \frac{
           K_{x}(s_{0})}
               {
                \left[1+
                   f(p_{\sigma})\right]
                                }
        \cdot
           [\hat{y}_{1}\,
            \hat{y}_{4}]\ ;
\\
\\
       \hat{D}^{2}\,
        \hat{z}
        &=&
          \Delta s\cdot
        \frac{
           K_{x}(s_{0})}
               {
                \left[1+
                   f(p_{\sigma})\right]
                                }
       \cdot
       \hat{D}\,
           [\hat{y}_{1}\,
            \hat{y}_{4}]
       \ =\
         0\ ;
\\
\\
    \Longrightarrow\ \
       \hat{D}^{\nu}\,
        \hat{z}
        &=&
      0\ \
       \mbox{for}\ \
       \nu\;>\;1\ .
\end{eqnarray*}
Thus\,:
\begin{eqnarray}
  \left\{
     \exp{[\hat{D}]}
                \right\}
      \hat{z}
         &=&
      \hat{y}_{3}
          +
          \Delta s\cdot
        \frac{
           K_{x}(s_{0})}
               {
                \left[1+
                   f(p_{\sigma})\right]
                                }
       \cdot
           [\hat{y}_{1}\,
            \hat{y}_{4}]
\nonumber   \\
         &\equiv&
      \hat{z}
          +
          \Delta s\cdot
        \frac{
           K_{x}(s_{0})}
               {
                \left[1+
                   f(p_{\sigma})\right]
                                }
       \cdot
           [\hat{x}\,
            \hat{p}_{z}]\ .
\end{eqnarray}
\\
\vspace*{0.5ex}

d) For\, $p_{z}$\,:
\begin{eqnarray*}
       \hat{D}\,
        \hat{p}_{z}
        &=&
       0\ ;
\\
\\
    \Longrightarrow\ \
       \hat{D}^{\nu}\,
        \hat{p}_{z}
        &=&
      0\ \
       \mbox{for}\ \
       \nu\;>\;0\ .
\end{eqnarray*}
Thus\,:
\begin{eqnarray}
  \left\{
     \exp{[\hat{D}]}
                \right\}
        \hat{p}_{z}
         &=&
        \hat{p}_{z}\ .
\end{eqnarray}
\\
\vspace*{0.5ex}

e) For\, $\sigma$\,:
\begin{eqnarray*}
       \hat{D}\,
        \hat{\sigma}
        &=&
       F_{5}(\vec{\hat{y}})
\\
        &=&
     \Delta s\cdot
       \left\{
           -K_{x}(s_{0})\cdot\hat{x}
                 \cdot
            f'(\hat{p}_{\sigma})
                  -
          \frac{1}{2}\cdot
          \frac
               {K_{x}(s_{0})}
               {
                \left[1+
                   f(\hat{p}_{\sigma})\right]
                                }
        \cdot
      \frac{
                      f'(\hat{p}_{\sigma})
                                  }
               {
                \left[1+
                   f(\hat{p}_{\sigma})\right]
                                }
          \cdot
          \left[\hat{x}\, \hat{p}_{z}^{2}\right]\,
                  \right\}
\\
        &=&
     \Delta s\cdot
       \left\{
           -K_{x}(s_{0})\cdot\hat{y}_{1}
                 \cdot
            f'(\hat{y}_{6})
                  -
          \frac{1}{2}\cdot
          \frac
               {K_{x}(s_{0})}
               {
                \left[1+
                   f(\hat{y}_{6})\right]
                                }
        \cdot
      \frac{
                      f'(\hat{y}_{6})
                                  }
               {
                \left[1+
                   f(\hat{y}_{6})\right]
                                }
          \cdot
          \left[\hat{y}_{1}\, \hat{y}_{4}^{2}\right]\,
                  \right\}\ ;
\\
\\
       \hat{D}^{2}\,
        \hat{\sigma}
        &=&
     \Delta s\cdot
       \hat{D}\,
       \left\{
           -K_{x}(s_{0})\cdot\hat{y}_{1}
                 \cdot
            f'(\hat{y}_{6})
                  -
          \frac{1}{2}\cdot
          \frac
               {K_{x}(s_{0})}
               {
                \left[1+
                   f(\hat{y}_{6})\right]
                                }
        \cdot
      \frac{
                      f'(\hat{y}_{6})
                                  }
               {
                \left[1+
                   f(\hat{y}_{6})\right]
                                }
          \cdot
          \left[\hat{y}_{1}\, \hat{y}_{4}^{2}\right]\,
                  \right\}
\\
        &=&
         0\ ;
\\
\\
    \Longrightarrow\ \
       \hat{D}^{\nu}\,
        \hat{\sigma}
        &=&
      0\ \
       \mbox{for}\ \
       \nu\;>\;1\ .
\end{eqnarray*}
Thus\,:
\begin{eqnarray}
  \left\{
     \exp{[\hat{D}]}
                \right\}
        \hat{\sigma}
         &=&
        \hat{\sigma}
         +
       \hat{D}\,
        \hat{\sigma}
\nonumber   \\
         &=&
      \hat{y}_{5}
         +
     \Delta s\cdot
       \left\{
           -K_{x}(s_{0})\cdot\hat{y}_{1}
                 \cdot
            f'(\hat{y}_{6})
                  -
          \frac{1}{2}\cdot
          \frac
               {K_{x}(s_{0})}
               {
                \left[1+
                   f(\hat{y}_{6})\right]
                                }
        \cdot
      \frac{
                      f'(\hat{y}_{6})
                                  }
               {
                \left[1+
                   f(\hat{y}_{6})\right]
                                }
          \cdot
          \left[\hat{y}_{1}\, \hat{y}_{4}^{2}\right]\,
                  \right\}
\nonumber   \\
         &=&
      \hat{y}_{5}
         +
     \Delta s\cdot
     \hat{y}_{1}\cdot
       \left\{
           -K_{x}(s_{0})
                 \cdot
            f'(\hat{y}_{6})
                  -
          \frac{1}{2}\cdot
          \frac
               {K_{x}(s_{0})}
               {
                \left[1+
                   f(\hat{y}_{6})\right]
                                }
        \cdot
      \frac{
                      f'(\hat{y}_{6})
                                  }
               {
                \left[1+
                   f(\hat{y}_{6})\right]
                                }
          \cdot
          \hat{y}_{4}^{2}
                  \right\}
\nonumber   \\
         &=&
      \hat{\sigma}
         +
     \Delta s\cdot
     \hat{x}\cdot
       \left\{
           -K_{x}(s_{0})
                 \cdot
            f'(\hat{p}_{\sigma})
                  -
          \frac{1}{2}\cdot
          \frac
               {K_{x}(s_{0})}
               {
                \left[1+
                   f(\hat{p}_{\sigma})\right]
                                }
        \cdot
      \frac{
                      f'(\hat{p}_{\sigma})
                                  }
               {
                \left[1+
                   f(\hat{p}_{\sigma})\right]
                                }
          \cdot
          \hat{p}_{z}^{2}
                   \right\}\ .
\nonumber    \\
\end{eqnarray}
\\
\vspace*{0.5ex}

f) For\, $p_{\sigma}$\,:
\begin{eqnarray*}
       \hat{D}\,
        \hat{p}_{\sigma}
        &=&
       0\ ;
\\
\\
    \Longrightarrow\ \
       \hat{D}^{\nu}\,
        \hat{p}_{\sigma}
        &=&
      0\ \
       \mbox{for}\ \
       \nu\;>\;0\ .
\end{eqnarray*}
Thus\,:
\begin{eqnarray}
  \left\{
     \exp{[\hat{D}]}
                \right\}
        \hat{p}_{\sigma}
         &=&
        \hat{p}_{\sigma}\ .
\end{eqnarray}
\bigskip

Equations (B.10\,--\,15)
finally lead to\,:
\setcounter{INDEX}{1}
\begin{eqnarray}
        x^{\,f}
             &=&
          {x}^{\,i}\ ;
 \\   \nonumber
 \\
   \addtocounter{equation}{-1}
\addtocounter{INDEX}{1}
        p_{x}^{\,f}
             &=&
        p_{x}^{\,i}
            +
          \Delta s\cdot
     \left\{
                        -
           [K_{x}(s_{0})]^{2}
             \cdot
             {x}^{\,i}
           +
            K_{x}(s_{0})
                 \cdot
            f(p_{\sigma}^{\,i})
                        -
          \frac{1}{2}\cdot
          \frac
               {K_{x}(s_{0})}
               {
                \left[1+
                   f(p_{\sigma}^{\,i})\right]
                                }
          \cdot
         (p_{z}^{\,i})^{2}\,
                      \right\}\,;
 \\   \nonumber
 \\
   \addtocounter{equation}{-1}
\addtocounter{INDEX}{1}
        z^{\,f}
             &=&
          z^{\,i}
              +
          \Delta s\cdot
        \frac{
           K_{x}(s_{0})}
               {
                \left[1+
                   f(p_{\sigma}^{\,i})\right]
                                }
       \cdot
           [x^{\,i}\,
            p^{\,i}_{z}]\ ;
 \\   \nonumber
 \\
   \addtocounter{equation}{-1}
\addtocounter{INDEX}{1}
        p_{z}^{\,f}
             &=&
        p_{z}^{\,i}\ ;
 \\   \nonumber
 \\
   \addtocounter{equation}{-1}
\addtocounter{INDEX}{1}
          \sigma^{\,f}
             &=&
          \sigma^{\,i}
         +
     \Delta s\cdot
     x^{i}\cdot
       \left\{
           -K_{x}(s_{0})
                 \cdot
            f'(p_{\sigma}^{\,i})
                  -
          \frac{1}{2}\cdot
          \frac
               {K_{x}(s_{0})}
               {
                \left[1+
                   f(p_{\sigma}^{\,i})\right]
                                }
        \cdot
      \frac{
                      f'(p_{\sigma}^{\,i})
                                  }
               {
                \left[1+
                   f(p_{\sigma}^{\,i})\right]
                                }
          \cdot
          (p_{z}^{\,i})^{2}
                   \right\}\ ;
\ \ \ \ \
 \\   \nonumber
 \\
   \addtocounter{equation}{-1}
\addtocounter{INDEX}{1}
          p_{\sigma}^{\,f}
             &=&
          p_{\sigma}^{\,i}.
\end{eqnarray}
\setcounter{INDEX}{0}
\bigskip

\subsubsection{The Component
    $\hat{\cal{H}}_{L}^{(2)}$}

\ \ \ \ \
{}From eqns. (3.29b) and (B.6b) we have\,:
\begin{eqnarray*}
    F_{1}
         (\vec{y})
                    &=&
            +\Delta s\cdot
                \frac{\partial
                                      }{\partial
                          {p_{x}}}\,
  \hat{\cal{H}}_{L}^{(2)}(x,p_{x},z,p_{z},\sigma,p_{\sigma})
\nonumber   \\
                    &=&
      \frac{
             K_{x}(s_{0})
               \cdot\Delta s
                                 }
               {
                \left[1+
                   f(p_{\sigma})\right]
                                }
                         \cdot x
                \cdot
           p_{x}
\nonumber   \\
                    &=&
          A\cdot[y_{1}\,y_{2}]\ ;
\\      \nonumber
 \\
    F_{2}
         (\vec{y})
                    &=&
            -\Delta s\cdot
                \frac{\partial
                                      }{\partial
                          {x}}\,
  \hat{\cal{H}}_{L}^{(2)}(x,p_{x},z,p_{z},\sigma,p_{\sigma})
\nonumber   \\
                    &=&
         -\frac{1}{2}\,
          \frac{
                       K_{x}(s_{0})
               \cdot\Delta s
                                                   }
               {
                \left[1+
                   f(p_{\sigma})\right]
                                }
                     \cdot
                p_{x}^{2}
\nonumber   \\
                    &=&
      -\frac{1}{2}\,A\cdot y_{2}^{2}\ ;
\\      \nonumber
 \\
    F_{3}
         (\vec{y})
                    &=&
            +\Delta s\cdot
                \frac{\partial
                                      }{\partial
                          {p_{z}}}\,
  \hat{\cal{H}}_{L}^{(2)}(x,p_{x},z,p_{z},\sigma,p_{\sigma})
\nonumber   \\
                    &=&
      0\ ;
\\      \nonumber
 \\
    F_{4}
         (\vec{y})
                    &=&
            -\Delta s\cdot
                \frac{\partial
                                      }{\partial
                          {z}}\,
  \hat{\cal{H}}_{L}^{(2)}(x,p_{x},z,p_{z},\sigma,p_{\sigma})
\nonumber   \\
                    &=&
                     0\ ;
\\      \nonumber
 \\
    F_{5}
         (\vec{y})
                    &=&
            +\Delta s\cdot
                \frac{\partial
                                      }{\partial
                          {p_{\sigma}}}\,
  \hat{\cal{H}}_{L}^{(2)}(x,p_{x},z,p_{z},\sigma,p_{\sigma})
\nonumber   \\
                    &=&
         -\frac{1}{2}\,
          \frac{
                       K_{x}(s_{0})
               \cdot\Delta s
                                                   }
               {
                \left[1+
                   f(p_{\sigma})\right]
                                }
          \cdot
          \frac{
                     f'(p_{\sigma})
                                    }
               {
                \left[1+
                   f(p_{\sigma})\right]
                                }
                                   \cdot x
                     \cdot
                p_{x}^{2}\ ;
\nonumber   \\
                    &=&
      -\frac{1}{2}\,A
          \cdot
          \frac{
                     f'(p_{\sigma})
                                    }
               {
                \left[1+
                   f(p_{\sigma})\right]
                                }
                     \cdot
                        [y_{1}\,y_{2}^{2}]\ ;
\\      \nonumber
 \\
    F_{6}
         (\vec{y})
                    &=&
            -\Delta s\cdot
                \frac{\partial
                                      }{\partial
                          {\sigma}}\,
  \hat{\cal{H}}_{L}^{(2)}(x,p_{x},z,p_{z},\sigma,p_{\sigma})
\nonumber   \\
                    &=&
      0
\end{eqnarray*}
whereby we have written for abbreviation\,:
\begin{eqnarray}
           A
              &=&
      \frac{
    K_{x}(s_{0})
               \cdot\Delta s
                                 }
               {
                \left[1+
                   f(\hat{p}_{\sigma})\right]
                                }\ .
\end{eqnarray}

Thus\,:
\begin{eqnarray}
       \hat{D}
        &=&
    F_{1}
        (\vec{\hat{y}})
         \cdot
                \frac{\partial\ }
                     {\partial{\hat{y}_{1}}}
          +
    F_{2}
        (\vec{\hat{y}})
         \cdot
                \frac{\partial\ }
                     {\partial{\hat{y}_{2}}}
          +
    F_{5}
        (\vec{\hat{y}})
         \cdot
                \frac{\partial\ }
                     {\partial{\hat{y}_{5}}}
\nonumber   \\
\nonumber   \\
        &=&
    A\cdot
   \left\{
      \left[\hat{y}_{1}\,\hat{y}_{2}\right]
         \cdot
                \frac{\partial\ }
                     {\partial{\hat{y}_{1}}}
          -
     \frac{1}{2}\,
           \hat{y}_{2}^{2}
         \cdot
                \frac{\partial\ }
                     {\partial{\hat{y}_{2}}}
          -
          \frac{1}{2}\,
          \frac{
                     f'(p_{\sigma})
                                    }
               {
                \left[1+
                   f(p_{\sigma})\right]
                                }
         \cdot
      \left[\hat{y}_{1}\,\hat{y}_{2}^{\,2}\right]
           \hat{y}_{2}^{2}
         \cdot
                \frac{\partial\ }
                     {\partial{\hat{y}_{5}}}
                   \right\}
\end{eqnarray}
and
\begin{eqnarray}
       \hat{D}\,
       \vec{\hat{y}}
        &=&
                 \left( \begin{array}{c}
                   F_{1}(\vec{\hat{y}})  \\
                   F_{2}(\vec{\hat{y}})  \\
                   0                     \\
                   0                     \\
                   F_{5}(\vec{\hat{y}})  \\
                   0
              \end{array}
       \right)\ .
\end{eqnarray}

We then obtain\,:
\\
\vspace*{0.5ex}

a) For\, $x$\,:
\begin{eqnarray*}
       \hat{D}\,
        \hat{x}
        &=&
       F_{1}(\vec{\hat{y}})
       \ =\
         A
                \cdot
            \left[
                   \hat{y}_{1}\,
                   \hat{y}_{2}
                         \right]\ ;
\\
\\
       \hat{D}^{2}\,
        \hat{x}
        &=&
       \hat{D}\,
       F_{1}(\vec{\hat{y}})
\\
        &=&
 \left\{
    F_{1}
        (\vec{\hat{y}})
         \cdot
                \frac{\partial\ }
                     {\partial{\hat{y}_{1}}}
          +
    F_{2}
        (\vec{\hat{y}})
         \cdot
                \frac{\partial\ }
                     {\partial{\hat{y}_{2}}}
                           \right\}\,
       F_{1}(\vec{\hat{y}})
\\
        &=&
    A^{2}\cdot
   \left\{
      \left[\hat{y}_{1}\,\hat{y}_{2}\right]
         \cdot
                \frac{\partial\ }
                     {\partial{\hat{y}_{1}}}
          -
     \frac{1}{2}\,
           \hat{y}_{2}^{2}
         \cdot
                \frac{\partial\ }
                     {\partial{\hat{y}_{2}}}
                   \right\}
            \left[
                   \hat{y}_{1}\,
                   \hat{y}_{2}
                         \right]
\\
        &=&
    A^{2}\cdot
   \left\{
            \hat{y}_{1}\cdot\hat{y}_{2}^{2}
          -
     \frac{1}{2}\,
           \hat{y}_{2}^{2}
         \cdot
              \hat{y}_{1}
                   \right\}
\\
        &=&
    A^{2}\cdot
    \frac{1}{2}\,
   \left\{
            \hat{y}_{1}\cdot\hat{y}_{2}^{2}
                   \right\}
\\
        &=&
    A^{2}\cdot
    \frac{1}{2}\,
                 \hat{y}_{1}\,
           \hat{y}_{2}^{2}\ ;
\\
\\
       \hat{D}^{3}\,
        \hat{x}
        &=&
    A^{2}\cdot
 \left\{
    F_{1}
        (\vec{\hat{y}})
         \cdot
                \frac{\partial\ }
                     {\partial{\hat{y}_{1}}}
          +
    F_{2}
        (\vec{\hat{y}})
         \cdot
                \frac{\partial\ }
                     {\partial{\hat{y}_{2}}}
                           \right\}\,
   \left\{
    \frac{1}{2}\,
                 \hat{y}_{1}\,
           \hat{y}_{2}^{2}
                   \right\}
\\
        &=&
    A^{3}\cdot
   \left\{
      \left[\hat{y}_{1}\,\hat{y}_{2}\right]
         \cdot
                \frac{\partial\ }
                     {\partial{\hat{y}_{1}}}
          -
     \frac{1}{2}\,
           \hat{y}_{2}^{2}
         \cdot
                \frac{\partial\ }
                     {\partial{\hat{y}_{2}}}
                   \right\}\,
   \left\{
    \frac{1}{2}\,
                 \hat{y}_{1}\,
           \hat{y}_{2}^{2}
                   \right\}
\\
        &=&
    A^{3}\cdot\frac{1}{2}\,
   \left\{
      \left[\hat{y}_{1}\,\hat{y}_{2}\right]
           \hat{y}_{2}^{2}
          -
     \frac{1}{2}\,
           \hat{y}_{2}^{2}
         \cdot
      2\,\left[\hat{y}_{1}\,\hat{y}_{2}\right]
                      \right\}
\\
\\
        &=&
         0\ ;
\\
\\
    \Longrightarrow\ \
       \hat{D}^{\nu}\,
        \hat{x}
        &=&
      0\ \
       \mbox{for}\ \
       \nu\;>\;2\ .
\end{eqnarray*}
Thus\,:
\begin{eqnarray}
  \left\{
     \exp{[\hat{D}]}
                \right\}
      \hat{x}
         &\equiv&
        \sum_{n\,=\,0}^{\infty}\,
     \frac{1}{n!}\cdot
           \hat{D}^{n}\,
      \hat{x}
\nonumber   \\
         &=&
      \hat{x}
         +
       \hat{D}\,
        \hat{x}
         +
       \frac{1}{2}\,
       \hat{D}^{2}\,
        \hat{x}
\nonumber   \\
         &=&
      \hat{y}_{1}
         +
         A
                \cdot
            \left[
                   \hat{y}_{1}\,
                   \hat{y}_{2}
                         \right]
         +
    \frac{1}{2}
         \cdot
    \frac{A^{2}}{2}\,
                 \hat{y}_{1}\,
           \hat{y}_{2}^{2}
\nonumber   \\
         &=&
      \hat{y}_{1}
      \cdot
    \left\{1+
             A
                \cdot
                   \hat{y}_{2}
         +
    \frac{1}{4}\,
    A^{2}\cdot
           \hat{y}_{2}^{2}
                   \right\}
\nonumber   \\
         &\equiv&
      \hat{x}
      \cdot
    \left\{1+
             A
                \cdot
                   \hat{p}_{x}
         +
    \frac{1}{4}\,
    A^{2}\cdot
           \hat{p}_{x}^{2}
                   \right\}\ .
\end{eqnarray}
\\
\vspace*{0.5ex}

b) For\, $p_{x}$\,:
\begin{eqnarray*}
       \hat{D}\,
        \hat{p}_{x}
        &=&
       F_{2}(\vec{\hat{y}})
       \ =\
      -\frac{A}{2}
                \cdot
           \hat{y}_{2}^{2}\ ;
\\
\\
       \hat{D}^{2}\,
        \hat{p}_{x}
        &=&
       \hat{D}\,
       F_{2}(\vec{\hat{y}})
\\
        &=&
 \left\{
    F_{2}
        (\vec{\hat{y}})
         \cdot
                \frac{\partial\ }
                     {\partial{\hat{y}_{2}}}
                           \right\}\,
       F_{2}(\vec{\hat{y}})
\\
        &=&
    \left(
      -\frac{A}{2}
             \right)^{2}
   \left\{
           \hat{y}_{2}^{2}
         \cdot
                \frac{\partial\ }
                     {\partial{\hat{y}_{2}}}
                   \right\}
           \hat{y}_{2}^{2}
\\
        &=&
    \left(
      -\frac{A}{2}
             \right)^{2}
         \cdot
   \left\{
           \hat{y}_{2}^{2}
         \cdot
            2\,\hat{y}_{2}
                   \right\}
\\
        &=&
    \left(
      -\frac{A}{2}
             \right)^{2}
         \cdot
       2\,\hat{y}_{2}^{3}
\\
\\
       \hat{D}^{3}\,
        \hat{p}_{x}
        &=&
    \left(
      -\frac{A}{2}
             \right)^{2}
         \cdot
      {2}\cdot
 \left\{
    F_{2}
        (\vec{\hat{y}})
         \cdot
                \frac{\partial\ }
                     {\partial{\hat{y}_{2}}}
                           \right\}\,
                 \hat{y}_{2}^{3}
\\
        &=&
    \left(
      -\frac{A}{2}
             \right)^{3}
         \cdot
      {2}\cdot
   \left\{
           \hat{y}_{2}^{2}
         \cdot
                \frac{\partial\ }
                     {\partial{\hat{y}_{2}}}
                   \right\}\,
           \hat{y}_{2}^{3}
\\
        &=&
    \left(
      -\frac{A}{2}
             \right)^{3}
         \cdot
      {2}\cdot
      {3}\cdot
           \hat{y}_{2}^{4}\ ;
\end{eqnarray*}
Thus\,:
\begin{eqnarray*}
       \hat{D}^{n}\,
        \hat{x}
           &=&
    \left(
      -\frac{A}{2}
             \right)^{n}
         \cdot
      (n!)
         \cdot
           \hat{y}_{2}^{n+1}\ ;
\end{eqnarray*}
\begin{eqnarray}
    \Longrightarrow\ \
  \left\{
     \exp{[\hat{D}]}
                \right\}
      \hat{p}_{x}
         &=&
        \sum_{n\,=\,0}^{\infty}\,
     \frac{1}{n!}\cdot
           \hat{D}^{n}\,
      \hat{x}
\nonumber   \\
         &=&
        \sum_{n\,=\,0}^{\infty}\,
    \left(
      -\frac{A}{2}
             \right)^{n}
         \cdot
           \hat{y}_{2}^{n+1}
\nonumber   \\
         &=&
\frac{
              \hat{y}_{2}
             }
     {1+\frac{A}{2}\,
              \hat{y}_{2}
             }
\nonumber   \\
         &\equiv&
\frac{
              \hat{p}_{x}
             }
     {1+\frac{A}{2}\,
              \hat{p}_{x}
             }\ .
\end{eqnarray}
\\
\vspace*{0.5ex}

c) For\, $z$\,:
\begin{eqnarray*}
       \hat{D}\,
        \hat{z}
        &=&
       0\ ;
\\
\\
    \Longrightarrow\ \
       \hat{D}^{\nu}\,
        \hat{z}
        &=&
      0\ \
       \mbox{for}\ \
       \nu\;>\;0\ .
\end{eqnarray*}
Thus\,:
\begin{eqnarray}
  \left\{
     \exp{[\hat{D}]}
                \right\}
      \hat{z}
         &=&
      \hat{z}\ .
\end{eqnarray}
\\
\vspace*{0.5ex}

d) For\, $p_{z}$\,:
\begin{eqnarray*}
       \hat{D}\,
        \hat{p}_{z}
        &=&
       0\ ;
\\
\\
    \Longrightarrow\ \
       \hat{D}^{\nu}\,
        \hat{p}_{z}
        &=&
      0\ \
       \mbox{for}\ \
       \nu\;>\;0\ .
\end{eqnarray*}
Thus\,:
\begin{eqnarray}
  \left\{
     \exp{[\hat{D}]}
                \right\}
        \hat{p}_{z}
         &=&
        \hat{p}_{z}\ .
\end{eqnarray}
\\
\vspace*{0.5ex}

e) For\, $\sigma$\,:
\begin{eqnarray*}
       \hat{D}\,
        \hat{\sigma}
        &=&
       F_{5}(\vec{\hat{y}})
       \ =\
         -\frac{1}{2}\,
         A\cdot
          \frac{
                     f'(\hat{p}_{\sigma})
                                    }
               {
                \left[1+
                   f(\hat{p}_{\sigma})\right]
                                }
                     \cdot
                [\hat{y}_{1}\,
                 \hat{y}_{2}^{\,2}]\ ;
\\
\\
       \hat{D}^{2}\,
        \hat{\sigma}
        &=&
       \hat{D}\,
       F_{5}(\vec{\hat{y}})
\\
        &=&
 \left\{
    F_{1}
        (\vec{\hat{y}})
         \cdot
                \frac{\partial\ }
                     {\partial{\hat{y}_{1}}}
          +
    F_{2}
        (\vec{\hat{y}})
         \cdot
                \frac{\partial\ }
                     {\partial{\hat{y}_{2}}}
                           \right\}\,
       F_{5}(\vec{\hat{y}})
\\
        &=&
         -\frac{1}{2}\,
         A^{2}\cdot
          \frac{
                     f'(\hat{p}_{\sigma})
                                    }
               {
                \left[1+
                   f(\hat{p}_{\sigma})\right]
                                }
         \cdot
   \left\{
      \left[\hat{y}_{1}\,\hat{y}_{2}\right]
         \cdot
                \frac{\partial\ }
                     {\partial{\hat{y}_{1}}}
          -
     \frac{1}{2}\,
           \hat{y}_{2}^{2}
         \cdot
                \frac{\partial\ }
                     {\partial{\hat{y}_{2}}}
                   \right\}
            \left[
                   \hat{y}_{1}\,
                   \hat{y}_{2}^{2}
                         \right]
\\
        &=&
         -\frac{1}{2}\,
         A^{2}\cdot
          \frac{
                     f'(\hat{p}_{\sigma})
                                    }
               {
                \left[1+
                   f(\hat{p}_{\sigma})\right]
                                }
         \cdot
   \left\{
          \left[\hat{y}_{1}\,\hat{y}_{2}\right]
          \cdot
          \hat{y}_{2}^{2}
          -
     \frac{1}{2}\,
           \hat{y}_{2}^{2}
         \cdot 2\,
          \left[\hat{y}_{1}\,\hat{y}_{2}\right]
                   \right\}
\\
        &=&
         0\ ;
\\
\\
    \Longrightarrow\ \
       \hat{D}^{\nu}\,
        \hat{\sigma}
        &=&
      0\ \
       \mbox{for}\ \
       \nu\;>\;1\ .
\end{eqnarray*}
Thus\,:
\begin{eqnarray}
  \left\{
     \exp{[\hat{D}]}
                \right\}
        \hat{\sigma}
         &\equiv&
        \sum_{n\,=\,0}^{\infty}\,
     \frac{1}{n!}\cdot
           \hat{D}^{n}\,
        \hat{\sigma}
\nonumber   \\
         &=&
        \hat{\sigma}
         +
       \hat{D}\,
        \hat{\sigma}
\nonumber   \\
         &=&
      \hat{y}_{5}
         -
          \frac{1}{2}\,
         A\cdot
          \frac{
                     f'(\hat{p}_{\sigma})
                                    }
               {
                \left[1+
                   f(\hat{p}_{\sigma})\right]
                                }
                     \cdot
                [\hat{y}_{1}\,
                 \hat{y}_{2}^{\,2}]
\nonumber   \\
         &\equiv&
      \hat{\sigma}
         -
          \frac{1}{2}\,
         A\cdot
          \frac{
                     f'(\hat{p}_{\sigma})
                                    }
               {
                \left[1+
                   f(\hat{p}_{\sigma})\right]
                                }
                     \cdot
                 \hat{x}\,
                 \hat{p}_{x}^{\,2}\ .
\end{eqnarray}
\\
\vspace*{0.5ex}

f) For\, $p_{\sigma}$\,:
\begin{eqnarray*}
       \hat{D}\,
        \hat{p}_{\sigma}
        &=&
       0\ ;
\\
\\
    \Longrightarrow\ \
       \hat{D}^{\nu}\,
        \hat{p}_{\sigma}
        &=&
      0\ \
       \mbox{for}\ \
       \nu\;>\;0\ .
\end{eqnarray*}
Thus\,:
\begin{eqnarray}
  \left\{
     \exp{[\hat{D}]}
                \right\}
        \hat{p}_{\sigma}
         &=&
        \hat{p}_{\sigma}\ .
\end{eqnarray}
\bigskip

{}From eqns. (B.20\,--\,25)
we finally get\,:
\setcounter{INDEX}{1}
\begin{eqnarray}
        x^{f}
             &=&
          {x}^{i}
      \cdot
    \left\{1+
             A
                \cdot
                       {p}_{x}^{\,i}
         +
    \frac{1}{4}\,
    A^{2}\cdot
           (p_{x}^{\,i})^{2}
                   \right\}
\nonumber    \\
             &=&
          {x}^{i}
      \cdot
    \left\{1+
             \frac{1}{2}\,
             A
                \cdot
                       {p}_{x}^{\,i}
                   \right\}^{2}\ ;
 \\   \nonumber
 \\
   \addtocounter{equation}{-1}
\addtocounter{INDEX}{1}
        p_{x}^{\,f}
             &=&
\frac{
             {p}_{x}^{\,i}
             }
     {1+\frac{A}{2}\,
                {p}_{x}^{\,i}
             }\ ;
 \\   \nonumber
 \\
   \addtocounter{equation}{-1}
\addtocounter{INDEX}{1}
        z^{f}
             &=&
          z^{i}\ ;
 \\   \nonumber
 \\
   \addtocounter{equation}{-1}
\addtocounter{INDEX}{1}
        p_{z}^{\,f}
             &=&
        p_{z}^{\,i}\ ;
 \\   \nonumber
 \\
   \addtocounter{equation}{-1}
\addtocounter{INDEX}{1}
          \sigma^{f}
             &=&
          \sigma^{i}
         -
          \frac{1}{2}\,
         A\cdot
          \frac{
                     f'({p}_{\sigma}^{\,i})
                                    }
               {
                \left[1+
                   f({p}_{\sigma}^{\,i})\right]
                                }
                     \cdot
                   x^{i}
                     \cdot
                   ({p}_{x}^{\,i})^{\,2}\ ;
 \\   \nonumber
 \\
   \addtocounter{equation}{-1}
\addtocounter{INDEX}{1}
          p_{\sigma}^{\,f}
             &=&
          p_{\sigma}^{\,i}
\end{eqnarray}
\setcounter{INDEX}{0}
with\,
       $A$\,
given by (B.17).
\\
\\
\\
\underline{Remarks:}
                    \newline

\vspace{-0.3ex}
1) The thin\,-\,lens transport map
resulting from eqn. (B.5)
reads as\,:
\begin{eqnarray*}
         T_{L}
              &=&
               \exp{
                    \left[
                            {\hat{D}^{(1)}}
                             \right]
                               }
              \cdot
               \exp{
                    \left[
                            {\hat{D}^{(2)}}
                             \right]
                               }
\end{eqnarray*}
whereby\,
        ${\hat{D}^{(1)}}$\,
corresponds to the component\,
    $\hat{\cal{H}}_{L}^{(1)}$\,
and
        ${\hat{D}^{(2)}}$\,
to\,
    $\hat{\cal{H}}_{L}^{(2)}$.
One could also use
the more symmetric
composition\,:
\begin{eqnarray*}
         T_{L}
              &=&
               \exp{
                    \left[
                          \frac{1}{2}\,
                            {\hat{D}^{(1)}}
                             \right]
                               }
              \cdot
               \exp{
                    \left[
                            {\hat{D}^{(2)}}
                             \right]
                               }
              \cdot
               \exp{
                    \left[
                          \frac{1}{2}\,
                            {\hat{D}^{(1)}}
                             \right]
                               }\ .
\end{eqnarray*}
\\

2) As mentioned in the remark
at the end of section 3.2.2,
one can calculate the transport map\,
       $\underline{T}_{L}$\,
alternatively
by solving the differential equation (3.30).
To illustrate this method
we use the lens corresponding to
    $\hat{\cal{H}}_{L}^{(2)}$.
In this case the differential equations (3.30)
take the form\,:
\setcounter{INDEX}{1}
\begin{eqnarray}
        x'
             &=&
   \frac{A}{\Delta s}
      \cdot
    x\cdot p_{x}\ ;
 \\   \nonumber
 \\
   \addtocounter{equation}{-1}
\addtocounter{INDEX}{1}
        p_{x}\,'
             &=&
        -\frac{1}{2}
              \cdot
   \frac{A}{\Delta s}
              \cdot
             {p}_{x}^{2}\ ;
 \\   \nonumber
 \\
   \addtocounter{equation}{-1}
\addtocounter{INDEX}{1}
        z'
             &=&
          0\ ;
 \\   \nonumber
 \\
   \addtocounter{equation}{-1}
\addtocounter{INDEX}{1}
        p_{z}\,'
             &=&
        0\ ;
 \\   \nonumber
 \\
   \addtocounter{equation}{-1}
\addtocounter{INDEX}{1}
          \sigma'
             &=&
          -
          \frac{1}{2}
              \cdot
   \frac{A}{\Delta s}
          \cdot
          \frac{
                     f'({p}_{\sigma})
                                    }
               {
                \left[1+
                   f({p}_{\sigma})\right]
                                }
                     \cdot
                   x
                     \cdot
                    {p}_{x}^{\,2}\ ;
 \\   \nonumber
 \\
   \addtocounter{equation}{-1}
\addtocounter{INDEX}{1}
          p_{\sigma}\,'
             &=&
          0\ ,
\end{eqnarray}
\setcounter{INDEX}{0}
which are solved by\,:
\begin{eqnarray}
        x(s)
             &=&
        x(s_{0})
      \cdot
    \left[\,
          1+
          \frac{1}{2}
              \cdot
   \frac{A}{\Delta s}
      \cdot
      p_{x}(s_{0})
      \cdot
      (s-s_{0})\,
              \right]^{2}\ ;
 \\   \nonumber
 \\
   \addtocounter{equation}{-1}
\addtocounter{INDEX}{1}
        p_{x}(s)
             &=&
        \frac{
              p_{x}(s_{0})
                           }
             {
       \displaystyle{
          1+
          \frac{1}{2}
              \cdot
   \frac{A}{\Delta s}
      \cdot
      p_{x}(s_{0})
      \cdot
      (s-s_{0})}
                           }\ ;
 \\   \nonumber
 \\
   \addtocounter{equation}{-1}
\addtocounter{INDEX}{1}
        z(s)
             &=&
        z(s_{0})\ ;
 \\   \nonumber
 \\
   \addtocounter{equation}{-1}
\addtocounter{INDEX}{1}
        p_{z}(s)
             &=&
        p_{z}(s_{0})\ ;
 \\   \nonumber
 \\
   \addtocounter{equation}{-1}
\addtocounter{INDEX}{1}
          \sigma(s)
             &=&
          \sigma(s_{0})
          -
          \frac{1}{2}
              \cdot
   \frac{A}{\Delta s}
              \cdot
          \frac{
                     f'[{p}_{\sigma}(s_{0})]
                                    }
               {
                      1+
                   f[{p}_{\sigma}(s_{0})]
                                }
                     \cdot
                   x(s_{0})
                     \cdot
                    {p}_{x}^{\,2}(s_{0})
                     \cdot
                    (s-s_{0})\ ;
 \\   \nonumber
 \\
   \addtocounter{equation}{-1}
\addtocounter{INDEX}{1}
          p_{\sigma}(s)
             &=&
          p_{\sigma}(s_{0})\ .
\end{eqnarray}
\setcounter{INDEX}{0}
Using then the relations\,:
\begin{eqnarray*}
       \vec{y}^{\,i}
         &\equiv&
       \vec{y}(s_{0})\ ;
\\
       \vec{y}^{\,f}
         &\equiv&
       \vec{y}(s_{0}+\Delta s)\ ,
\end{eqnarray*}
(see (3.31) and (3.32)\,)
one indeed regains eqns. (B.26a\,--\,f).
\bigskip
%
%

\setcounter{section}{3}
\setcounter{subsection}{0}
\setcounter{equation}{0}
\section*{Appendix C:
                      Superposition
                      of Solenoids
                      and Quadrupoles}
\addcontentsline{toc}{section}{Appendix C:
                      Superposition
                      of Solenoids
                      and Quadrupoles}
\par
\newcommand{\xr}{{\bf x},s}

\subsection{Exponentiation}

\ \ \ \ \
For a superposition of a solenoid
and a quadrupole
we have\,:
\begin{eqnarray*}
      H&\neq&0\,;\ \
      g\ \neq\ 0
\end{eqnarray*}
and
\begin{eqnarray*}
          K_{x}&=&K_{z}\ =\ N\ =\ \lambda\ =\ \mu\ =\ V
                                          \ =\ 0\ .
\end{eqnarray*}

{}From (3.16) and (3.29b) we then
obtain\,:
\begin{eqnarray}
    F_{1}
         (\vec{y})
                    &=&
     +\frac{
                H(s_{0})
              \cdot\Delta s
                            }
              {
               \left[1+
                  f(p_{\sigma})\right]
                               }
                         \cdot z\ ;
 \\       \nonumber
 \\
   \addtocounter{equation}{-1}
\addtocounter{INDEX}{1}
    F_{2}
         (\vec{y})
                        &=&
     +\frac{
                 H(s_{0})
               \cdot\Delta s
                             }
               {
                \left[1+
                   f(p_{\sigma})\right]
                                }
               \cdot
        \left[p_{z}-H(s_{0})\cdot x\right]
                -
           g\cdot x
           \cdot\Delta s\ ;
 \\       \nonumber
 \\
   \addtocounter{equation}{-1}
\addtocounter{INDEX}{1}
    F_{3}
         (\vec{y})
                   &=&
     -\frac{
                 H(s_{0})
               \cdot\Delta s
                             }
               {
                \left[1+
                   f(p_{\sigma})\right]
                                }
                         \cdot x\ ;
 \\       \nonumber
 \\
   \addtocounter{equation}{-1}
\addtocounter{INDEX}{1}
    F_{4}
         (\vec{y})
                       &=&
     -\frac{
                 H(s_{0})
               \cdot\Delta s
                             }
               {
                \left[1+
                   f(p_{\sigma})\right]
                                }
               \cdot
        \left[p_{x}+H(s_{0})\cdot z\right]
                +
           g\cdot z
           \cdot\Delta s\ ;
 \\       \nonumber
 \\
   \addtocounter{equation}{-1}
\addtocounter{INDEX}{1}
    F_{5}
         (\vec{y})
                        &=&
     -\frac{
                 H(s_{0})
               \cdot\Delta s
                             }
               {
                \left[1+
                   f(p_{\sigma})\right]
                                }
       \cdot
      \frac{
            f'(p_{\sigma})
                            }
               {
                \left[1+
                   f(p_{\sigma})\right]
                                }
      \cdot
      \left\{
          \frac{1}{2}\,
          H(s_{0})
                  \cdot[x^{2}+
                            z^{2}]
             +
              [p_{x}\cdot z-p_{z}\cdot x]
                        \right\}\ ;
 \\       \nonumber
 \\
   \addtocounter{equation}{-1}
\addtocounter{INDEX}{1}
    F_{6}
         (\vec{y})
                      &=&
                       0\ .
\end{eqnarray}
\setcounter{INDEX}{0}

Thus\,:
\begin{eqnarray}
       \hat{D}
        &=&
    F_{1}
         (\vec{\hat{y}})
         \cdot
                \frac{\partial\ }
                     {\partial{\hat{y}_{1}}}
          +
    F_{2}
         (\vec{\hat{y}})
         \cdot
                \frac{\partial\ }
                     {\partial{\hat{y}_{2}}}
          +
    F_{3}
         (\vec{\hat{y}})
         \cdot
                \frac{\partial\ }
                     {\partial{\hat{y}_{3}}}
          +
    F_{4}
         (\vec{\hat{y}})
         \cdot
                \frac{\partial\ }
                     {\partial{\hat{y}_{4}}}
          +
    F_{5}
         (\vec{\hat{y}})
         \cdot
                \frac{\partial\ }
                     {\partial{\hat{y}_{5}}}
\end{eqnarray}
and
\setcounter{INDEX}{1}
\begin{eqnarray}
       \hat{D}\,
                 \left( \begin{array}{c}
                   \hat{x}         \\
                   \hat{p}_{x}     \\
                   \hat{z}         \\
                   \hat{p}_{z}
              \end{array}
       \right)
        &=&
                 \left( \begin{array}{c}
                   F_{1}(\vec{\hat{y}})  \\
                   F_{2}(\vec{\hat{y}})  \\
                   F_{3}(\vec{\hat{y}})  \\
                   F_{4}(\vec{\hat{y}})
              \end{array}
       \right)
       \ =\
       \underline{\hat{A}}_{\,0}\,
                 \left( \begin{array}{c}
                   \hat{x}         \\
                   \hat{p}_{x}     \\
                   \hat{z}         \\
                   \hat{p}
              \end{array}
       \right)\ ;
 \\         \nonumber
 \\         \nonumber
 \\
   \addtocounter{equation}{-1}
\addtocounter{INDEX}{1}
       \hat{D}\,
       \hat{\sigma}
        &=&
       F_{5}(\vec{\hat{y}})\ ;
 \\         \nonumber
 \\
   \addtocounter{equation}{-1}
\addtocounter{INDEX}{1}
       \hat{D}\,
       \hat{p}_{\sigma}
        &=&
         0
    \ \ \Longrightarrow\ \
         \left\{
               \exp{
                    \left[
                            {\hat{D}}
                             \right]
                               }
                      \right\}\,
       \hat{p}_{\sigma}
       \ =\
       \hat{p}_{\sigma}
\end{eqnarray}
\setcounter{INDEX}{0}
with
\begin{eqnarray}
   \underline{\hat{A}}_{\,0}
               &=&
   \Delta s
       \cdot
          \frac{1}
               {
                \left[1+
                   f(\hat{p}_{\sigma})\right]
                                }
       \cdot
                 \left( \begin{array}{cccc}
                   0  &  0
                            & +H                 &  0   \\
   -[H^{2}+\hat{g}\,]     &  0
                            &  0  & +H       \\
   -H                 &  0
                            &  0  &  0   \\
                   0  &  -H
       &  -[H^{2}-\hat{g}\,]   &  0
              \end{array}
       \right)
\end{eqnarray}
and
\begin{eqnarray}
          \frac{
                \hat{g}
                              }
               {
                \left[1+
                   f(p_{\sigma})\right]
                                }
                  &=&
                   g\ .
\end{eqnarray}

We decompose\,
the matrix\,
  $\underline{\hat{A}}_{\,0}$\,
into the components\,
  $\underline{\hat{A}}_{\,01}$\,
and\,
  $\underline{\hat{A}}_{\,02}$\,:
\begin{eqnarray}
   \underline{\hat{A}}_{\,0}
          &=&
   \underline{\hat{A}}_{\,01}
           +
   \underline{\hat{A}}_{\,02}
\end{eqnarray}
with
\setcounter{INDEX}{1}
\begin{eqnarray}
   \underline{\hat{A}}_{\,01}
          &=&
   \Delta s
       \cdot
          \frac{H^{2}}
               {
                \left[1+
                   f(\hat{p}_{\sigma})\right]
                                }
       \cdot
                 \left( \begin{array}{cccc}
                   0  &  0
                            & 0                 &  0    \\
   -1                 &  0
                            &  0  & 0       \\
   0                  &  0
                            &  0  &  0    \\
                   0  &  0
       &  -1                 &  0
              \end{array}
       \right)
\end{eqnarray}
and
\begin{eqnarray}
   \addtocounter{equation}{-1}
\addtocounter{INDEX}{1}
   \underline{\hat{A}}_{\,02}
          &=&
   \Delta s
       \cdot
          \frac{H}
               {
                \left[1+
                   f(\hat{p}_{\sigma})\right]
                                }
       \cdot
                 \left( \begin{array}{cccccc}
                   0  &  0
                            & +1                 &  0    \\
   -(\hat{g}/H)       &  0
                            &  0  & +1      \\
   -1                 &  0
                            &  0  &  0    \\
                   0  &  -1
       &
   +(\hat{g}/H)       &  0
                             &  0
              \end{array}
       \right)\ .
\end{eqnarray}
\setcounter{INDEX}{0}

Since
\begin{eqnarray*}
       \hat{D}\,
   \underline{\hat{A}}_{\,0}
               &=&
   \underline{\hat{A}}_{\,0}\,
       \hat{D}
\ \ \Longrightarrow\ \
       \hat{D}^{\,\nu}\,
       \vec{\hat{y}}_{0}
       \ =\
       \underline{\hat{A}}^{\,\nu}
       \vec{\hat{y}}_{0}
\ \ \Longrightarrow\ \
         \left\{
               \exp{
                    \left[
                            {\hat{D}}
                             \right]
                               }
                      \right\}
       \vec{\hat{y}}_{0}
       \ =\
         \left\{
               \exp{
                    \left[
                  \underline{\hat{A}}_{0}
                             \right]
                               }
                      \right\}
       \vec{\hat{y}}_{0}
\end{eqnarray*}
the transfer matrix
for
\begin{eqnarray}
       \vec{y}_{0}
           &=&
                 \left( \begin{array}{c}
                       {x}         \\
                       {p}_{x}     \\
                       {z}         \\
                       {p}
              \end{array}
       \right)
\end{eqnarray}
reads as\,:
\begin{eqnarray}
   \underline{M}_{0}
               &=&
               \exp{
                    \left[
                  \underline{\hat{A}}_{\,0}
                             \right]
                               }\ .
\end{eqnarray}
\bigskip

Using the equations
\begin{eqnarray*}
   \underline{\hat{A}}_{\,01}
          \cdot
   \underline{\hat{A}}_{\,02}
            &=&
   \underline{\hat{A}}_{\,02}
          \cdot
   \underline{\hat{A}}_{\,01}
\end{eqnarray*}
we have\,:
\begin{eqnarray*}
               \exp{
                    \left[
                  \underline{\hat{A}}
                             \right]
                                    }
                     \ =\
               \exp{
                    \left[
                  \underline{\hat{A}}_{\,01}
                             \right]
                                    }
               \cdot
               \exp{
                    \left[
                  \underline{\hat{A}}_{\,02}
                             \right]
                                    }\ .
\end{eqnarray*}
Also since
\begin{eqnarray*}
    \left[
       \underline{\hat{A}}_{01}
                             \right]
                               ^{\,\nu}
               &=&
       \underline{0}\ \
       \mbox{for}\ \ \
       \nu\;>\;1
\end{eqnarray*}
we get\,:
\begin{eqnarray*}
               \exp{
                    \left[
                  \underline{\hat{A}}_{\,01}
                             \right]
                                    }
                     \ =\
                  \underline{1}
                       +
                  \underline{\hat{A}}_{\,01}\ .
\end{eqnarray*}
Because
\begin{eqnarray*}
& &
                 \left( \begin{array}{cccccc}
                   0  &  0
                            & 1                 &  0    \\
  -(\hat{g}/H)
                      &  0
                            &  0  & 1      \\
   -1                 &  0
                            &  0  &  0    \\
                   0  &  -1
       &
  +(\hat{g}/H)
                             &  0
              \end{array}
       \right)^{\,2n}
            \ =\
       (-1)^{n}
       \cdot
       \underline{1}\ ;
\\
\\
& &
                 \left( \begin{array}{cccccc}
                   0  &  0
                            & 1                 &  0    \\
  -(\hat{g}/H)
                      &  0
                            &  0  & 1      \\
   -1                 &  0
                            &  0  &  0    \\
                   0  &  -1
       &
  +(\hat{g}/H)
                             &  0
              \end{array}
       \right)^{\,2n+1}
            \ =\
       (-1)^{n}
       \cdot
                 \left( \begin{array}{cccccc}
                   0  &  0
                            & 1                 &  0    \\
  -(\hat{g}/H)
                      &  0
                            &  0  & 1      \\
   -1                 &  0
                            &  0  &  0    \\
                   0  &  -1
       &
  +(\hat{g}/H)
              \end{array}
       \right)
\end{eqnarray*}
we obtain\,:
\begin{eqnarray*}
& &
               \exp{
                    \left[
                  \underline{\hat{A}}_{\,02}
                             \right]
                                    }
                     \ =\
\\
\\
& &
        \sum_{n,\,=\,0}^{\infty}\,
     \frac{1}{(2n)!}\cdot
     (-1)^{n}\cdot
     (\Delta\Theta)^{2n}
           \cdot
                  \underline{1}
\\
\\
& &
\hspace*{1.0cm}
                       +
        \sum_{n,\,=\,0}^{\infty}\,
     \frac{1}{(2n+1)!}\cdot
     (-1)^{n}\cdot
     (\Delta\Theta)^{2n+1}
           \cdot
                 \left( \begin{array}{cccccc}
                   0  &  0
                            & +1                 &  0    \\
  -(\hat{g}/H)
                      &  0
                            &  0  & +1      \\
   -1                 &  0
                            &  0  &  0    \\
                   0  &  -1
       &
  +(\hat{g}/H)
                             &  0
              \end{array}
       \right)
\\
\\
& &
           \ =\
          \underline{1}
     \cdot
     \cos(\Delta\Theta)
                       +
                 \left( \begin{array}{cccccc}
                   0  &  0
                            & +1                 &  0    \\
  -(\hat{g}/H)
                      &  0
                            &  0  & +1      \\
   -1                 &  0
                            &  0  &  0    \\
                   0  &  -1
       &
  +(\hat{g}/H)
                             &  0
              \end{array}
       \right)
           \cdot
     \sin(\Delta\Theta)
\\
\\
& &
           \ =\
                 \left( \begin{array}{cccccc}
     \cos(\Delta\Theta)
                      &  0
                            &
                              +\sin(\Delta\Theta)
                                                 &  0    \\
  -(\hat{g}/H)\cdot
                   \sin(\Delta\Theta)
                      &
     \cos(\Delta\Theta)
                            &  0  &
                              +\sin(\Delta\Theta)
                                            \\
   -\sin(\Delta\Theta)
                      &  0
                            &
     \cos(\Delta\Theta)
                                  &  0    \\
                   0  &
                         -\sin(\Delta\Theta)
       &
  +(\hat{g}/H)\cdot
                   \sin(\Delta\Theta)
                             &
     \cos(\Delta\Theta)
              \end{array}
       \right)
\end{eqnarray*}
with
\begin{eqnarray*}
          \Delta\Theta
          &=&
          \frac{
                H(s_{0})
               \cdot\Delta s
                             }
               {
                \left[1+
                   f(\hat{p}_{\sigma})\right]
                                }\ .
\end{eqnarray*}
Then from (C.9)\,:
\begin{eqnarray}
                  \underline{M}_{0}
                      &=&
        \left[
                  \underline{1}
                       +
                  \underline{\hat{A}}_{\,01}
                        \right]
\nonumber    \\
\nonumber    \\
&\times&
                 \left( \begin{array}{cccccc}
     \cos(\Delta\Theta)
                      &  0
                            &
                              +\sin(\Delta\Theta)
                                                 &  0    \\
  -(\hat{g}/H)\cdot
                   \sin(\Delta\Theta)
                      &
     \cos(\Delta\Theta)
                            &  0  &
                              +\sin(\Delta\Theta)
                                            \\
   -\sin(\Delta\Theta)
                      &  0
                            &
     \cos(\Delta\Theta)
                                  &  0    \\
                   0  &
                         -\sin(\Delta\Theta)
       &
  +(\hat{g}/H)\cdot
                   \sin(\Delta\Theta)
                             &
     \cos(\Delta\Theta)
              \end{array}
       \right)\ .
\end{eqnarray}

For the variable\,
      $\sigma$\,
we obtain from (C.3b)\,:
\begin{eqnarray*}
   {\hat{D}}^{2}\,
    \hat{\sigma}
        &=&
           {\hat{D}}\,
       F_{5}(\vec{\hat{y}})
 \\
 \\
        &=&
  \left\{
    F_{1}
         (\vec{\hat{y}})
         \cdot
                \frac{\partial\ }
                     {\partial{\hat{y}_{1}}}
          +
    F_{2}
         (\vec{\hat{y}})
         \cdot
                \frac{\partial\ }
                     {\partial{\hat{y}_{2}}}
          +
    F_{3}
         (\vec{\hat{y}})
         \cdot
                \frac{\partial\ }
                     {\partial{\hat{y}_{3}}}
          +
    F_{4}
         (\vec{\hat{y}})
         \cdot
                \frac{\partial\ }
                     {\partial{\hat{y}_{4}}}
               \right\}\,
       F_{5}(\vec{\hat{y}})
 \\
 \\
        &=&
          \frac{
                H
               \cdot\Delta s
                             }
               {
                \left[1+
                   f(\hat{p}_{\sigma})\right]
                                }\,
    \left\{
    \hat{y}_{3}
         \cdot
                \frac{\partial\ }
                     {\partial{\hat{y}_{1}}}
          -
    \left[
          H\cdot
                 \hat{y}_{1}
                 -
                 \hat{y}_{4}
                            \right]
         \cdot
                \frac{\partial\ }
                     {\partial{\hat{y}_{2}}}
                 -
    \hat{y}_{1}
         \cdot
                \frac{\partial\ }
                     {\partial{\hat{y}_{3}}}
          -
    \left[H\cdot
                 \hat{y}_{3}
                 +
                 \hat{y}_{2}
                            \right]
         \cdot
                \frac{\partial\ }
                     {\partial{\hat{y}_{4}}}
                                           \right.
 \\
 \\
& &
\hspace*{5.0cm}
         \left.
          -
          (\hat{g}/H)\cdot
                 \hat{y}_{1}
         \cdot
                \frac{\partial\ }
                     {\partial{\hat{y}_{2}}}
          +
          (\hat{g}/H)\cdot
                 \hat{y}_{3}
         \cdot
                \frac{\partial\ }
                     {\partial{\hat{y}_{4}}}
               \right\}
 \\
 \\
& &
       \left\{
          \frac{
                (-H)
               \cdot f'(\hat{p}_{\sigma})
               \cdot\Delta s
                             }
               {
                \left[1+
                   f(\hat{p}_{\sigma})\right]^{2}
                                }\,
       \left[
              \frac{1}{2}\,H\cdot
              (
               \hat{y}_{1}^{2}
                      +
               \hat{y}_{3}^{2}
                              )
                      +
       \left(
             \hat{y}_{2}\cdot
             \hat{y}_{3}
                      -
             \hat{y}_{4}\cdot
             \hat{y}_{1}
                               \right)
                          \right]
                          \right\}
 \\
 \\
                      &=&
          \frac{
                H
               \cdot\Delta s
                             }
               {
                \left[1+
                   f(\hat{p}_{\sigma})\right]
                                }\cdot
          \frac{
                (-H)
               \cdot f'(\hat{p}_{\sigma})
               \cdot\Delta s
                             }
               {
                \left[1+
                   f(\hat{p}_{\sigma})\right]^{2}
                                }
 \\
 \\
& &
         \times\,
       \left\{
               \hat{y}_{3}
                          \cdot
              [
               H\cdot\hat{y}_{1}
                      -
               \hat{y}_{4}
                              ]
                      -
              [
               H\cdot\hat{y}_{1}
                      -
               \hat{y}_{4}
                              ]
                          \cdot
               \hat{y}_{3}
                      -
               \hat{y}_{1}
                          \cdot
              [
               H\cdot\hat{y}_{3}
                      +
               \hat{y}_{2}
                              ]
                      +
              [
               H\cdot\hat{y}_{3}
                      +
               \hat{y}_{2}
                              ]
                          \cdot
               \hat{y}_{1}
                                           \right.
 \\
 \\
& &
\hspace*{5.0cm}
         \left.
          -
    (\hat{g}/H)\cdot
          \left[
                \hat{y}_{1}\cdot\hat{y}_{3}
                           +
                \hat{y}_{3}\cdot\hat{y}_{1}
                     \right]
                          \right\}
 \\
 \\
                      &=&
        -2\,
          \frac{
                H
               \cdot\Delta s
                             }
               {
                \left[1+
                   f(\hat{p}_{\sigma})\right]
                                }\cdot
          \frac{
                (-H)
               \cdot f'(\hat{p}_{\sigma})
               \cdot\Delta s
                             }
               {
                \left[1+
                   f(\hat{p}_{\sigma})\right]^{2}
                                }
    \cdot
    (\hat{g}/H)\cdot
                \hat{y}_{1}\cdot\hat{y}_{3}
 \\
 \\
                      &=&
         2\cdot[\Delta\Theta]^{2}
          \cdot
          \frac{
                f'(\hat{p}_{\sigma})
               \cdot\Delta s
                             }
               {
                \left[1+
                   f(\hat{p}_{\sigma})\right]
                                }
    \cdot
    (\hat{g}/H)\cdot
                \hat{y}_{1}\cdot\hat{y}_{3}\ .
\end{eqnarray*}

Furthermore, using the relations\,:
\begin{eqnarray*}
       \hat{D}^{2n}\,
       [\hat{y}_{1}\,\hat{y}_{3}]
           &=&
       (-1)^{n}
               \cdot
       [2\cdot\Delta\Theta]^{2n}
               \cdot
       [\hat{y}_{1}\,\hat{y}_{3}]\ ;
\\
       \hat{D}^{2n+1}\,
       [\hat{y}_{1}\,\hat{y}_{3}]
           &=&
       (-1)^{n}
               \cdot
       [2\cdot\Delta\Theta]^{2n+1}
               \cdot
       \frac{1}{2}\,
       [\hat{y}_{3}^{2}-\hat{y}_{1}^{2}]\ ;
\end{eqnarray*}
\begin{eqnarray*}
    \Longrightarrow\ \
  \left\{
     \exp{[\hat{D}]}
                \right\}
     [\hat{y}_{1}\,\hat{y}_{3}]
         &=&
        \sum_{n\,=\,0}^{\infty}\,
     \frac{1}{n!}\cdot
           \hat{D}^{n}\,
     [\hat{y}_{1}\,\hat{y}_{3}]
\\
         &=&
        \sum_{n\,=\,0}^{\infty}\,
     \frac{1}{(2n)!}\cdot
           \hat{D}^{2n}\,
     [\hat{y}_{1}\,\hat{y}_{3}]
          +
        \sum_{n\,=\,0}^{\infty}\,
     \frac{1}{(2n+1)!}\cdot
           \hat{D}^{2n+1}\,
     [\hat{y}_{1}\,\hat{y}_{3}]
\\
         &=&
        \sum_{n\,=\,0}^{\infty}\,
     \frac{1}{(2n)!}\cdot
       (-1)^{n}
               \cdot
       [2\cdot\Delta\Theta]^{2n}
               \cdot
       [\hat{y}_{1}\,\hat{y}_{3}]
\\
         &+&
        \sum_{n\,=\,0}^{\infty}\,
     \frac{1}{(2n+1)!}\cdot
       (-1)^{n}
               \cdot
       [2\cdot\Delta\Theta]^{2n+1}
               \cdot
       \frac{1}{2}\,
       [\hat{y}_{3}^{2}-\hat{y}_{1}^{2}]
\\
         &=&
       [\hat{y}_{1}\,\hat{y}_{3}]
       \cdot
       \cos
           [2\,\Delta\Theta]
             +
       \frac{1}{2}\,
       [\hat{y}_{3}^{2}-\hat{y}_{1}^{2}]
       \cdot
       \sin
           [2\,\Delta\Theta]
\end{eqnarray*}
and
\begin{eqnarray*}
   \hat{D}\,F_{5}(\vec{\hat{y}})
          &=&
   \kappa
   \cdot
   \hat{D}^{2}
       [\hat{y}_{1}\,\hat{y}_{3}]
\end{eqnarray*}
with
\begin{eqnarray*}
   \kappa
          &=&
   -\frac{1}{2}\,
          \frac{f'({p}_{\sigma})}
               {
                \left[1+
                   f({p}_{\sigma})\right]^{2}
                                }
          \cdot
          \frac{\hat{g}}{H}
\\
          &=&
   -\frac{1}{2}\,
          \frac{f'({p}_{\sigma})}
               {
                \left[1+
                   f({p}_{\sigma})\right]
                                }
          \cdot
          \frac{g}{H}\ ,
\end{eqnarray*}
we get\,:
\begin{eqnarray}
         \left\{
               \exp{
                    \left[
                            {\hat{D}}
                             \right]
                               }
                      \right\}
       \hat{\sigma}
        &=&
        \sum_{n\,=\,0}^{\infty}\,
     \frac{1}{n!}\cdot
           \hat{D}\,
       \hat{\sigma}
\nonumber     \\
        &=&
       \hat{\sigma}
           +
           \hat{D}\,
       \hat{\sigma}
           +
        \sum_{n\,=\,2}^{\infty}\,
     \frac{1}{n!}\cdot
           \hat{D}^{n}\,
       \hat{\sigma}
\nonumber     \\
        &=&
       \hat{\sigma}
           +
       F_{5}(\vec{\hat{y}})
           +
        \sum_{n\,=\,2}^{\infty}\,
     \frac{1}{n!}\cdot
           \hat{D}^{n-1}\,
       F_{5}(\vec{\hat{y}})
\nonumber     \\
        &=&
       \hat{\sigma}
           +
       F_{5}(\vec{\hat{y}})
           +
        \sum_{n\,=\,2}^{\infty}\,
     \frac{1}{n!}\cdot
           \hat{D}^{n-2}\cdot
           \hat{D}\,
       F_{5}(\vec{\hat{y}})
\nonumber     \\
        &=&
       \hat{\sigma}
           +
       F_{5}(\vec{\hat{y}})
           +
       \kappa\cdot
        \sum_{n\,=\,2}^{\infty}\,
     \frac{1}{n!}\cdot
           \hat{D}^{n}\,
       [\hat{y}_{1}\,\hat{y}_{3}]
\nonumber     \\
        &=&
       \hat{\sigma}
           +
       F_{5}(\vec{\hat{y}})
           +
       \kappa\cdot
        \sum_{n\,=\,0}^{\infty}\,
     \frac{1}{n!}\cdot
           \hat{D}^{n}\,
       [\hat{y}_{1}\,\hat{y}_{3}]
           -
       \kappa\cdot
       (1+\hat{D})\,
       [\hat{y}_{1}\,\hat{y}_{3}]
\nonumber     \\
        &=&
       \hat{\sigma}
           +
       F_{5}(\vec{\hat{y}})
           -
       \kappa\cdot
       [\hat{y}_{1}\,\hat{y}_{3}]
           -
       \kappa\cdot
       [2\cdot\Delta\Theta]
               \cdot
       \frac{1}{2}\,
       [\hat{y}_{3}^{2}-\hat{y}_{1}^{2}]
\nonumber     \\
& &
\hspace*{2.0cm}
      +
       \kappa\cdot
       [\hat{y}_{1}\,\hat{y}_{3}]
       \cdot
       \cos
           [2\cdot\Delta\Theta]
             +
       \kappa\cdot
       \frac{1}{2}\,
       [\hat{y}_{3}^{2}-\hat{y}_{1}^{2}]
       \cdot
       \sin
           [2\cdot\Delta\Theta]
\nonumber     \\
        &=&
       \hat{\sigma}
           +
       F_{5}(\vec{\hat{y}})
           -
       \kappa\cdot
       [\hat{y}_{1}\,\hat{y}_{3}]
       \cdot
    \left\{
           1-
       \cos
           [2\cdot\Delta\Theta]
                   \right\}
\nonumber     \\
& &
\hspace*{2.0cm}
           +
       \kappa\cdot
       \frac{1}{2}\,
       [\hat{y}_{3}^{2}-\hat{y}_{1}^{2}]\cdot
    \left\{
       \sin
           [2\cdot\Delta\Theta]
            -
       [2\cdot\Delta\Theta]
                   \right\}\ .
\end{eqnarray}
\bigskip

\subsection{Thin\,-\,Lens Transport Map}
\bigskip

\ \ \ \ \
Equations (C.9), (C.11) and (C.3c)
finally lead to\,:
\setcounter{INDEX}{1}
\begin{eqnarray}
        \vec{y}_{0}^{\;f}
             &=&
        \underline{M}_{0}\,
        \vec{y}_{0}^{\;i}\ ;
 \\   \nonumber
 \\
   \addtocounter{equation}{-1}
\addtocounter{INDEX}{1}
          \sigma^{f}
             &=&
                   {\sigma}^{i}
         -\frac{f'({p}_{\sigma}^{\,i})}
               {
                \left[1+
                   f({p}_{\sigma}^{\,i})\right]
                                }
                    \cdot
                \Delta\Theta
\nonumber     \\
& &
\hspace*{1.0cm}
      \times
  \left\{
          \frac{1}{2}\,
                   H(s_{0})
                   \cdot
           \left[
                ({x}^{i})^{2}
               +({z}^{i})^{2}
                                \right]
                   +
    \left[
    {p}_{x}^{\,i}
                   \cdot
            {z}^{i}
                 -
    {p}_{z}^{\,i}
                   \cdot
            {x}^{i}
                      \right]
                 \right\}
\nonumber    \\
\nonumber    \\
& &
\hspace*{1.0cm}
           -
       \kappa\cdot
       [
            {x}^{i}\cdot
            {z}^{i}
                                ]
       \cdot
    \left\{
           1-
       \cos
           [2\cdot\Delta\Theta]
                   \right\}
\nonumber    \\
\nonumber    \\
& &
\hspace*{1.0cm}
           +
       \kappa\cdot
       \frac{1}{2}\,
       [
           ({z}^{i})^{2}
            -
           ({x}^{i})^{2}
                                       ]
               \cdot
    \left\{
       \sin
           [2\cdot\Delta\Theta]
            -
       [2\cdot\Delta\Theta]
                   \right\}\ ;
 \\
   \addtocounter{equation}{-1}
\addtocounter{INDEX}{1}
          p_{\sigma}^{\,f}
             &=&
          p_{\sigma}^{\,i}
\end{eqnarray}
with\,
\begin{eqnarray}
\addtocounter{INDEX}{1}
   \addtocounter{equation}{-1}
      \Delta\Theta
           &=&
      \frac{
               H(s_{0})\cdot\Delta s
             }
               {
                \left[1+
                   f({p}_{\sigma}^{\,i})\right]
                                }
\\
\addtocounter{INDEX}{1}
   \addtocounter{equation}{-1}
   \kappa
          &=&
   -\frac{1}{2}\,
          \frac{f'({p}_{\sigma}^{\,i})}
               {
                \left[1+
                   f({p}_{\sigma}^{\,i})\right]
                                }
          \cdot
          \frac{g}{H}\ ,
\end{eqnarray}
and
       $\underline{M}_{0}$\,
given by (C.10),
where the matrix\,
  $\underline{\hat{A}}_{\,01}$\,
appearing in (C.10) takes the form\,:
\begin{eqnarray}
\addtocounter{INDEX}{1}
   \addtocounter{equation}{-1}
   \underline{\hat{A}}_{\,01}
          &=&
   \Delta\Theta
       \cdot
          H(s_{0})
       \cdot
                 \left( \begin{array}{cccc}
                   0  &  0
                            & 0                 &  0    \\
   -1                 &  0
                            &  0  & 0       \\
   0                  &  0
                            &  0  &  0    \\
                   0  &  0
       &  -1                 &  0
              \end{array}
       \right)
\end{eqnarray}
\setcounter{INDEX}{0}
(see eqn. (C.7a)\,).
\bigskip

Equation (C.12) contains as special cases
the transport maps of a simple solenoid
and a simple quadrupole
already derived
in section 4.
\bigskip
\end{document}